\documentclass[letterpaper,english,iop]{emulateapj}
\usepackage[T1]{fontenc}
\usepackage{graphicx}
\usepackage{wasysym}

\makeatletter

\usepackage{apjfonts}
\usepackage[hyperfootnotes=false,colorlinks=true,citecolor=blue,urlcolor=blue,linkcolor=blue]{hyperref}

\shorttitle{Dust Attenuation Curves}
\shortauthors{SEON \& DRAINE}

\makeatother

\usepackage{babel}
\begin{document}

\title{Radiative Transfer Model of Dust Attenuation Curves\\in Clumpy,
Galactic Environments}

\author{Kwang-Il Seon\altaffilmark{1,2,3} and Bruce T. Draine\altaffilmark{2}}

\altaffiltext{1}{Korea Astronomy and Space Science Institute, Daejeon, 305-348, Korea; kiseon@kasi.re.kr; seon@princeton.edu}
\altaffiltext{2}{Department of Astrophysical Sciences, Princeton University, Princeton, NJ, 08544, USA}
\altaffiltext{3}{Astronomy and Space Science Major, Korea University of Science and Technology, Daejeon, 305-350, Korea}
\begin{abstract}
The attenuation of starlight by dust in galactic environments is investigated
through models of radiative transfer in a spherical, clumpy interstellar
medium (ISM). The photon sources are uniformly distributed within
a spherical subregion. Extinction properties for Milky Way (MW), Large
Magellanic Cloud (LMC), and Small Magellanic Cloud (SMC) dust types
are considered. It is illustrated that the attenuation curves are
primarily determined by the wavelength dependence of absorption rather
than by the underlying extinction (absorption+scattering) curve; in
other words, the observationally derived attenuation curves are not
able to constrain a unique extinction curve unless the absorption
or scattering efficiency is additionally specified. Attenuation curves
consistent with the ``Calzetti attenuation curve'' are found by assuming
the silicate-carbonaceous dust model for the MW, but with the 2175\AA\
absorption bump suppressed or absent. The discrepancy between our
results and previous work that claimed the SMC-type dust to be the
most likely origin of the Calzetti curve is ascribed to the difference
in adopted albedos; this study uses the theoretically calculated albedos
whereas the previous ones adopted empirically derived albedos from
observations of reflection nebulae. It is also found that the model
attenuation curves calculated with the MW dust are well represented
by a modified Calzetti curve with a varying slope and UV bump strength.
The strong correlation between the slope and UV bump strength, with
steeper curves having stronger bumps, as found in star-forming galaxies
at $0.5<z<2.0$, is well reproduced by our models if the abundance
of the UV bump carriers or PAHs is assumed to be 30\% or 40\% of that
of the MW-dust. The trend is explained by radiative transfer effects
which lead to shallower attenuation curves with weaker UV bumps as
the ISM is more clumpy and dustier. We also argue that at least some
of the \emph{IUE} local starburst galaxies may have a UV bump feature
in their attenuation curves, albeit much weaker than that of the MW
extinction curve.
\end{abstract}


\section{INTRODUCTION}

Spectral energy distributions (SEDs) of galaxies provide crucial information
on stellar populations, metal content and star formation history \citep[e.g.,][]{1998ARA&A..36..189K,2013ARA&A..51..393C}.
However, internal dust absorbs and scatters starlight, thereby hampering
our ability to directly measure the intrinsic SEDs of galaxies, especially
at ultraviolet (UV) wavelengths. Quantifying the dust attenuation
effect is, therefore, essential in deriving the intrinsic properties
of galaxies from the observed SEDs. It also helps us understand the
nature of dust grains and the star/dust geometry.

The dust attenuation for extended objects, such as external galaxies,
should be distinguished from the extinction for individual point sources.
The extinction represents the loss of starlight out of the observer's
line of sight due to both absorption and scattering. Extinction curves
as a function of wavelength have been measured only for the Milky
Way (MW; \citealt{1999PASP..111...63F}), the Large Magellanic Cloud
(LMC; \citealt{1999ApJ...515..128M}), the Small Magellanic Cloud
(SMC; \citealt{1998ApJ...500..816G}), and M31 \citep{1996ApJ...471..203B,2015ApJ...815...14C},
where individual stars are resolved, and for galaxies in front of
QSOs and GRBs \citep{2006ApJS..166..443E,2009ApJ...697.1725E}, and
for a few cases of foreground/background galaxies \citep[e.g.,][]{2013MNRAS.433...47H,2013A&A...556A..42H}.
The attenuation, on the contrary, is defined as net loss of starlight
due to complex radiative transfer effects of the underlying dust extinction
properties and spatial distributions of dust and stars. The attenuation
is sometimes referred to as the effective extinction.

\begin{figure*}[t]
\begin{centering}
\medskip{}
\includegraphics[clip]{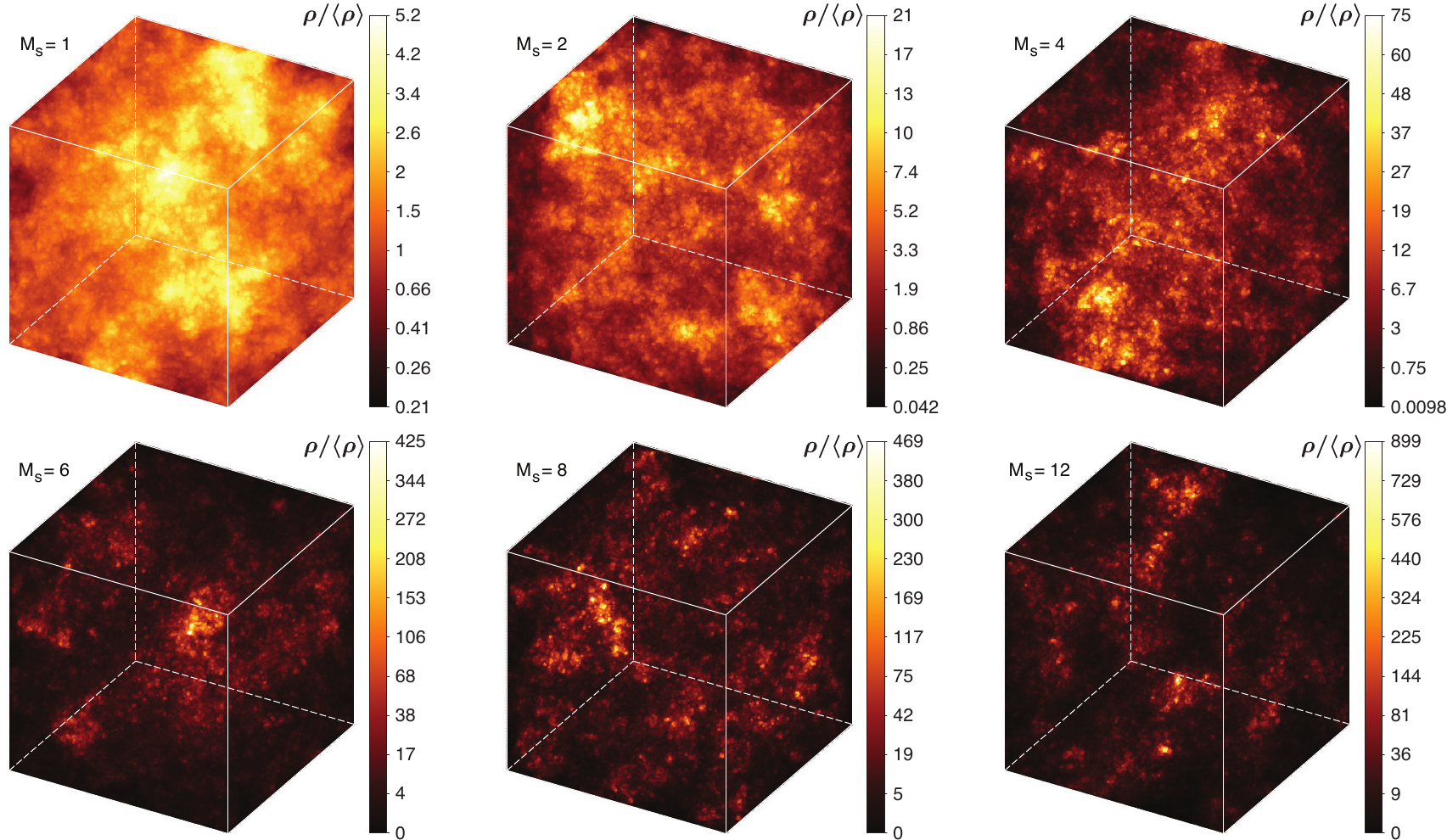}\medskip{}
\par\end{centering}
\caption{\label{fig1}Some realizations of lognormal density fields for sonic
Mach numbers $M_{{\rm s}}$ = 1, 2, 4, 6, 8, 12. The color bars show
$\rho/\left\langle \rho\right\rangle $, where $\left\langle \rho\right\rangle $
is the mean density.}
\medskip{}
\end{figure*}

To quantify the attenuation effect, \citet{1994ApJ...429..172K} estimated
a dust attenuation curve from the UV to the near-infrared (NIR) for
two local starburst galaxies, which were supposed to share the same
intrinsic properties, by dividing the spectrum of the more reddened
galaxy with that of the less reddened galaxy. By applying a similar
method to a sample of 39 local starburst and blue compact galaxies,
\citet{1994ApJ...429..582C,2000ApJ...533..682C} derived a mean attenuation
curve, which is now referred to as the Calzetti attenuation curve
or the Calzetti law. More recently, \citet{2016ApJ...818...13B} applied
the method of \citet{1994ApJ...429..582C,2000ApJ...533..682C} to
a sample of $\sim10,000$ local ($z\lesssim0.1$) star-forming galaxies
obtained from \emph{GALEX} and \emph{SDSS} and found that an attenuation
curve derived from the sample is consistent with the Calzetti curve
for local starburst galaxies. Attenuation curves covering the far-UV
wavelength shortward of 0.12$\mu$m were also derived for local starburst
galaxies \citep{2002ApJS..140..303L} and for galaxies at $z\sim3$
\citep{2016arXiv160600434R}; the far-UV attenuation curves are similar
to the Calzetti curve at $\lambda\gtrsim0.12\mu{\rm m}$, but shallower
than the extrapolation of the Calzetti curve at $\lambda\lesssim0.12$
$\mu$m. \citet{2000ApJ...539..718C} found that an attenuation curve
proportional to $\lambda^{-0.7}$ reproduces the relation between
the ratio of far-infrared (FIR) to UV luminosities and the UV spectral
slope observed in local starburst galaxies. The Calzetti curve somewhat
resembles a power law $\lambda^{-0.7}$ \citep[e.g., ][]{2003MNRAS.341...33K}.
These attenuation curves have been the most commonly adopted for modeling
the stellar populations of high-redshift galaxies. The most prominent
characteristics of both attenuation curves are the absence of the
UV absorption bump at 2175\AA\, which is commonly visible in the
MW and LMC, and a grayer (shallower) slope than the average MW extinction
curve.

The absence of a UV absorption bump in the Calzetti attenuation curve
motivated the radiative transfer studies of \citet{1997ApJ...487..625G}
and \citet{2000ApJ...528..799W} to investigate the nature of dust
grains in local starburst galaxies. By varying the underlying extinction
curve and star/dust geometries, they concluded that the dust in local
starburst galaxies has an SMC-like extinction curve lacking a 2175\AA\
bump. Using the same model, \citet{1999ASPC..193..517G} also showed
that color-color plots, suitable to probe the presence/lack of the
2175\AA\ bump, of starburst galaxies at $z=1-3$ in the Hubble Deep
Fields (HDFs) are consistent with SMC-like dust. The Far-UV (FUV)
slope (observed $G-\mathcal{R}$ color) of a large sample of Lyman
break galaxies (LBGs) at $z\sim3$ was also found to be consistent
with a SMC-like extinction curve \citep{2003ApJ...587..533V}. For
a sample of $\sim$1000 galaxies with UV to IR photometry and optical
spectroscopy, \citet{2007ApJS..173..392J} derived mean, low-resolution
attenuation curves which are consistent both with a $\lambda^{-0.7}$
power law and with the Calzetti curve. \citet{2010MNRAS.404..247C}
found no evidence for the presence of a UV bump using the $B-R$ color
of star-forming galaxies at $0.6<z<1.4$.

However, there is ample evidence for the presence of a UV bump at
2175\AA, though not as strong as for the MW, not only in nearby but
also in high redshift star-forming galaxies. Variations in the attenuation
curve have also been observed. \citet{2005MNRAS.360.1413B} claimed
that their UV- and FIR-selected galaxy samples have attenuation curves
with a UV bump. \citet{2005A&A...444..137N} and \citet{2007A&A...472..455N,2009A&A...507.1793N}
also found strong evidence for a UV bump of moderate strength in one
third of a sample of massive star-forming galaxies at $1<z<2.5$.
By analyzing UV and optical colors of disk dominated galaxies at $0.01<z<0.05$,
\citet{2010ApJ...718..184C} speculated that a strong UV bump could
be present in their extinction curves. \citet{2011MNRAS.417.1760W}
found evidence for the presence of a UV bump in local star-forming
galaxies. \citet{2011A&A...533A..93B,2012A&A...545A.141B} used a
modified Calzetti curve, which was originally proposed by \citet{2009A&A...499...69N}
to allow for variation in the UV bump strength and the attenuation
curve slope, for galaxies at $1<z<2.2$ and found evidence of a UV
bump with strength $\sim$ 35\% that of MW. They also showed that
the resulting attenuation curves are on average slightly steeper than
the Calzetti curve. \citet{2013ApJ...775L..16K} also applied the
modified Calzetti curve to composite SEDs constructed from a survey
of $0.5<z<2.0$ galaxies and found not only evidence for the presence
of a UV bump but also a strong correlation between the attenuation
curve slope and the UV bump strength \citep[see also][]{2011ApJ...743..168K}.
\citet{2015ApJ...804..149S} found that an attenuation curve steeper
than the Calzetti curve is needed to reproduce the observed IR/UV
ratios of galaxies younger than 100 Myr. \citet{2015ApJ...806..259R}
analyzed the dust attenuation curve of $z=1.4-2.6$ galaxies and found
marginal evidence for excess absorption at 2175\AA. High redshift
galaxies at $z=2-6.5$, for which the UV and optical SEDs are dominated
by OB stars, were also shown to have a clear UV bump feature \citep{2015ApJ...800..108S}.
\citet{2015ApJ...814..162Z} found that the average attenuation curve
in a sample of high redshift ($1.90<z<2.35$) galaxies is similar
to the Calzetti curve. However, it was also found that the attenuation
curve slope at UV is shallower over the galaxy mass range $7.2\lesssim\log M/M_{\odot}\lesssim10.2$
and the UV slope steepens as the galaxy mass increases. \citet{2015arXiv151205396S}
analyzed a sample of IR-luminous galaxies at $z\sim1.5-3$ using a
Bayesian analysis and found considerable diversity in the attenuation
curves; galaxies with high color excess have shallower attenuation
curves and those with low color excess have steeper attenuation curves.

Interestingly, it was also demonstrated that adding a UV bump to the
Calzetti curve significantly improves the agreement between photometric
and spectroscopic redshifts at least for some galaxies \citep{2001A&A...380..425M,2009ApJ...690.1236I}.
Detailed analyses of individual galaxies also revealed the presence
of a UV bump. The UV to NIR SED of the central region of the M33 nucleus
was found to be consistent with MW-type dust with a strong UV bump
\citep{1999ApJ...519..165G}. \citet{2011AJ....141..205H} analyzed
medium-band UV photometry from the \emph{Swift} UV/Optical Telescope
(UVOT) and found evidence for a prominent UV bump in star-forming
regions of M81 and Holmberg IX. \citet{2015arXiv150406635H} used
\emph{Swift} UVOT to measure the attenuation curve for star-forming
regions in the SMC and M33 and found that both the slope of the curve
and the strength of the 2175\AA\ bump vary across both galaxies.
\citet{2015ApJ...815...14C} found strong 2175\AA\ extinction bumps
for 4 sightlines in M31.

In understanding these observational results, radiative transfer studies
play an essential role. The main question in these studies would be
whether the differences between the attenuation curves for starburst
and/or normal star forming galaxies and the MW extinction curve are
caused by intrinsic differences in dust properties or can be explained
by radiative transfer effects. \citet{1997ApJ...487..625G} and \citet{2000ApJ...528..799W}
performed Monte Carlo radiative transfer calculations by assuming
three geometrical configurations, filled with either homogeneous or
two-phase clumpy density distributions, and two types of interstellar
dust properties similar to those of the MW and the SMC. Their models
reproduced not only the absence of a UV bump but also the overall
shape of the Calzetti attenuation curve, provided that one adopts
the SMC-type dust and a clumpy shell-type dust distribution surrounding
the starbursts. They also argued that the absence of a 2175 \AA\
absorption feature in the Calzetti curve can only be explained with
dust that lacks the 2175\AA\ feature in its extinction curve. It
was also noted that there is no unique attenuation curve. By adopting
a dusty foreground screen with a lognormal density distribution and
the MW extinction curve, \citet{2003ApJ...599L..21F} and \citet{2005ApJ...619..340F}
provided a physical explanation on the shallow shape of the Calzetti
curve. Later, \citet{2011A&A...533A.117F} explained the lack of a
2175\AA\ bump by assuming that below a certain critical column density
the UV bump carriers are destroyed by strong radiation fields.

In contrast, \citet{2000ApJ...542..710G} and \citet{2007MNRAS.375..640P}
claimed that the lack of a UV bump in the Calzetti curve is caused
by the age-dependent extinction in which newly formed young stars
are strongly attenuated by dense molecular clouds while old stars
are less attenuated by the diffuse ISM. \citet{2005MNRAS.359..171I}
solved the radiative transfer equation with multiple anisotropic scattering
through the clumpy ISM by using the ``mega-grain'' approximation,
treating clumps as large dust grains but with the effective absorption
and scattering coefficients calculated from the properties of the
dust of which the clumps are composed. He found that the age-dependent
attenuation gives rise to a much steeper attenuation curve for normal
star forming galaxies than the Calzetti curve. The Calzetti curve
was interpreted by \citet{2005MNRAS.359..171I} as a special case
at high optical depth. We note, however, that his attenuation curves
at high optical depth obtained using the MW extinction curve still
show the UV bump while those calculated using the SMC extinction curve
are consistent with the Calzetti curve. \citet{2004ApJ...617.1022P}
found that for highly inclined, weak bulge galaxies the 2175 \AA{}
feature was very weak in the global attenuation curves.

The theoretical models have supplied useful insights on the attenuation
curves in star-forming and starburst galaxies. However, the origin
of variations in the observed attenuation curve, such as found in
\citet{2013ApJ...775L..16K}, is still poorly understood. It is also
not well known what processes determine the strength of the UV bump.
In this paper, we therefore aimed to investigate what is the primary
determinant in shaping the attenuation curve. The present study is
based on radiative transfer calculations which employ dust density
distributions appropriate for a turbulent ISM. In Section \ref{sec:2},
we present a description of the dust density model and the radiative
transfer model. The main results are presented in Section \ref{sec:3},
in which we describe radiative transfer effects needed to understand
the attenuation curve, and show that the wavelength dependence of
the absorption efficiency (hereafter, absorption curve) is the most
important factor in determining the shape of the attenuation curve.
Our results are then compared with a strong correlation between the
attenuation curve slope and the UV bump strength found by \citet{2013ApJ...775L..16K}
and with a correlation between the slope and the $E(B-V)$ color excess
found by \citet{2015arXiv151205396S}. Radiative transfer models in
which starlight originates only from locally dense regions are also
briefly described. In Section \ref{sec:4}, correlation of the observed
attenuation curves with galaxy properties are discussed. We also discuss
the UV bump-less attenuation curve in scattering-dominant geometry
and present hints on the existence of a weak UV bump feature in local
starburst galaxies. The results are summarized in Section \ref{sec:5}.

\section{MODEL}

\label{sec:2}

\subsection{Turbulent Medium}

\label{subsec:2.1}

We first present a brief introduction to the ISM density distribution
and then detail the algorithm that is employed to simulate the density
structure in this study. The ISM is known to be clumpy and turbulent,
with a hierarchical density structure that is often treated as scale-free,
and/or fractal \citep[e.g.,][]{2004ARA&A..42..211E}. To represent
a clumpy medium, a two- or multi-phase medium has been adopted in
some radiative transfer models \citep{1996ApJ...463..681W,1997ApJ...481..809W,2000ApJ...528..799W,2000MNRAS.311..601B,2012MNRAS.420.2756S,2012ApJ...758..109S}.
However, these models ignored the hierarchical structure of the ISM.
\citet{1997ApJ...477..196E} proposed a more complex algorithm to
mimic hierarchically clumped clouds, which has been used for radiative
transfer models in the dusty or photoionized ISM \citep{2002ApJ...574..812M,2015MNRAS.447..559B}.
However, this model has also a critical limitation that results in
a large number of unrealistic empty holes.

The density field in the turbulent ISM can be represented by a power
spectral density (PSD) in the Fourier domain and a probability density
function (PDF) in real space. The PSDs of turbulent media are well
represented by a power-law function in the inertial range over which
viscous effects are not important \citep[e.g.,][]{2004ApJ...604L..49P,2005ApJ...630L..45K,2006ApJ...638L..25K}.
The PDFs of the three-dimensional (3D) density of the turbulent ISM
are known to be close to lognormal both in numerical simulations and
in observations \citep{1994ApJ...423..681V,1997ApJ...474..730P,2000ApJ...535..869K,2001ApJ...546..980O,2008MNRAS.390L..19B,2008A&A...489..143L,2009ApJ...703.1159S,2010MNRAS.406.1350F,2011ApJS..196...15S,2012ApJ...755L..19B,2013ApJ...772...57S,2015MNRAS.448.2469B}.
Therefore, the statistical properties of a density field in the turbulent
ISM can be characterized by two parameters: a standard deviation ($\sigma_{\ln\rho}$)
of the logarithm of the density and a power-law index ($\gamma$)
of the density power spectrum \citep{2012ApJ...761L..17S}.

The two parameters should be related to the sonic Mach number $M_{{\rm s}}$
of the turbulent ISM. \citet{2009ApJ...703.1159S} combined numerical
simulation results of \citet{2004ApJ...604L..49P}, \citet{2005ApJ...630L..45K},
and \citet{2006ApJ...638L..25K}, finding a relationship between $M_{{\rm s}}$
and $\gamma$,
\begin{equation}
\gamma=3.81M_{{\rm s}}^{-0.16},\label{eq:1}
\end{equation}
which is applicable to isothermal hydrodynamic turbulence. In the
isothermal hydrodynamic regimes, the variance of the log-density is
found to be related to $M_{{\rm s}}$ according to 
\begin{equation}
\sigma_{\ln\rho}^{2}=\ln(1+b^{2}M_{{\rm s}}^{2}),\label{eq:2}
\end{equation}
where the proportional constant $b$ is $\sim0.4$ for natural turbulence
forcing mode \citep[e.g.][]{2008ApJ...688L..79F,2010A&A...512A..81F}.
More complex relations between the density variance and Mach number
have been proposed for isothermal, magnetized gas or for the ISM in
non-isothermal regime \citep{2012MNRAS.423.2680M,2015arXiv150404370N}.
However, the simple relation in Equation (\ref{eq:2}) is sufficient
for the present purpose that pursues general properties of radiative
transfer effects. Given a Mach number, the standard deviation and
power spectral index of the lognormal density are determined by the
above two equations.

\begin{figure}[t]
\begin{centering}
\medskip{}
\includegraphics[clip,scale=0.8]{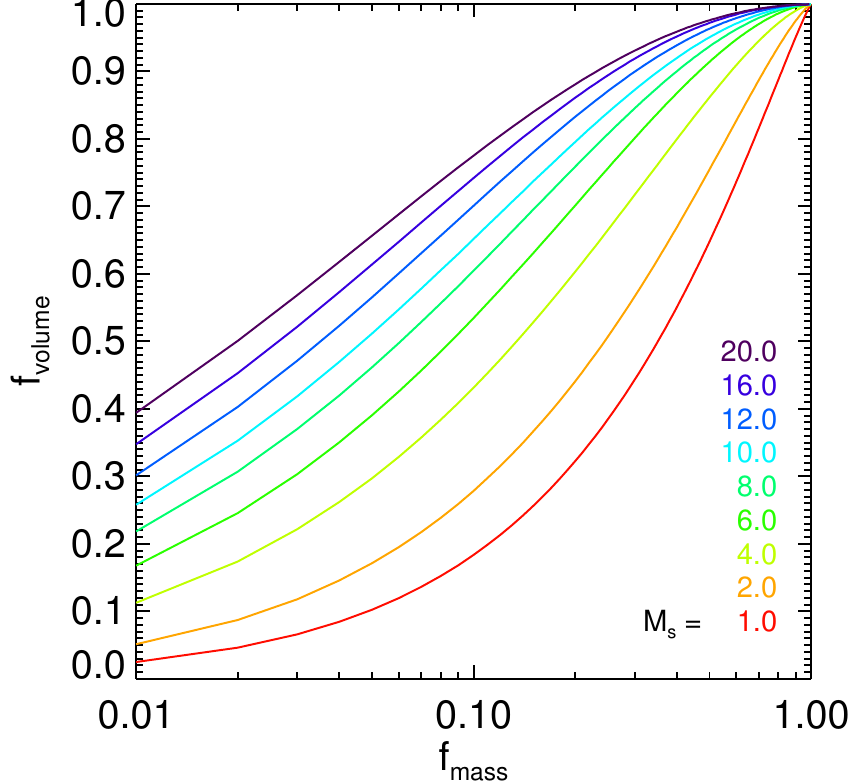}
\par\end{centering}
\caption{\label{fig2}Volume fraction vs mass fraction for the density fields
with Mach numbers of $M_{s}=$ 1, 2, 4, 6, 8, 10, 12, 16, and 20.}

\begin{centering}
\medskip{}
\par\end{centering}
\medskip{}
\end{figure}

\begin{figure}[t]
\begin{centering}
\includegraphics[clip,scale=0.51]{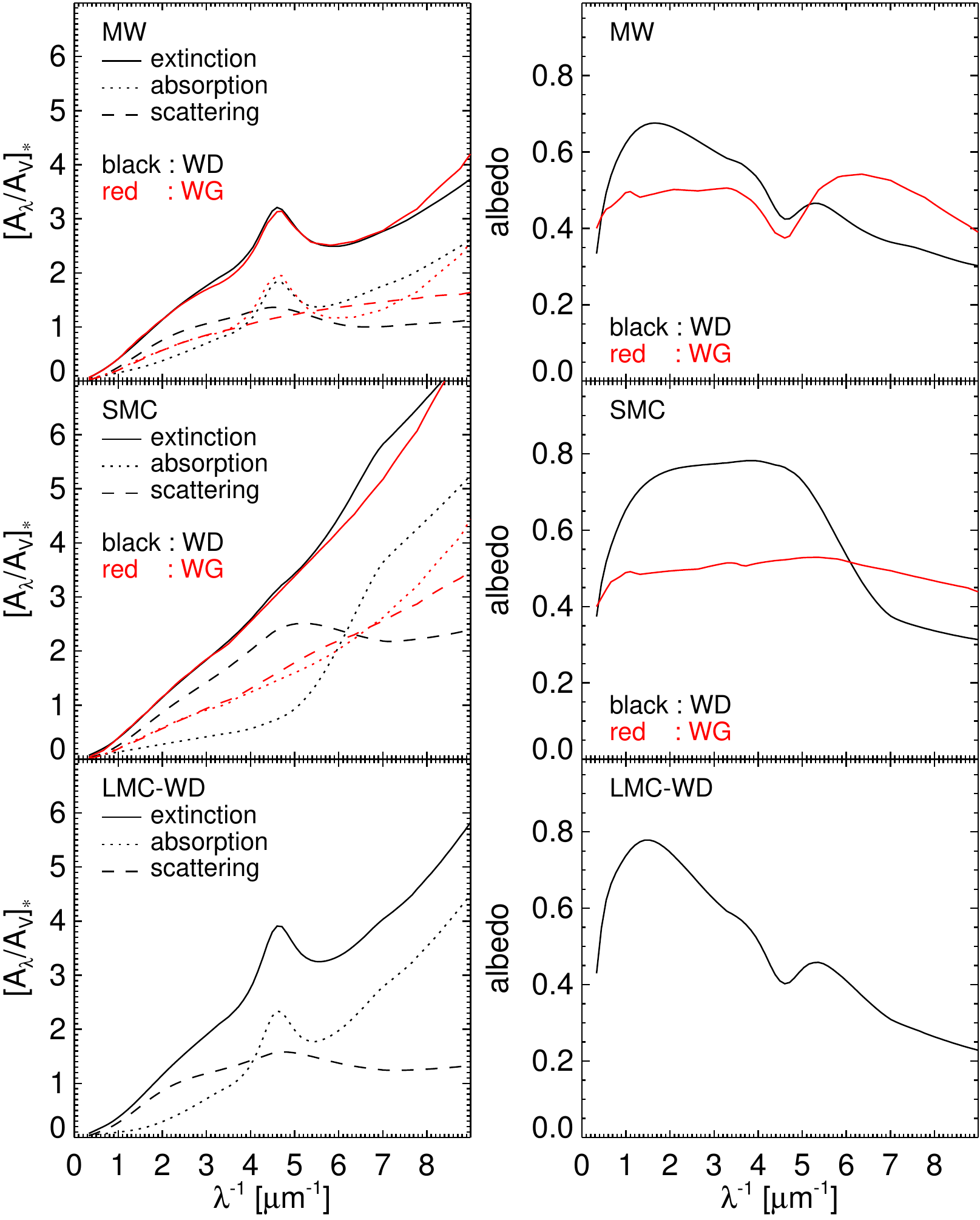}
\par\end{centering}
\caption{\label{fig3}Comparison of the extinction and albedo curves of \citet{2001ApJ...548..296W}
and \citet{2000ApJ...528..799W}. In left panels, the extinction,
absorption and scattering curves are denoted in solid, dotted and
dashed lines, respectively. In top and middle panels, the MW and SMC
curves of \citet{2001ApJ...548..296W} and \citet{2000ApJ...528..799W}
are denoted in black and red lines, respectively. The LMC curves of
\citet{2001ApJ...548..296W} are also shown in bottom panels.}
\medskip{}
\end{figure}

A Gaussian random field $\rho_{g}(\mathbf{x})$ for a given power-law
index $\gamma_{g}$ is generated by first assigning 3D Fourier coefficients
$s(\mathbf{k})=\left|s(\mathbf{k})\right|e^{i\phi(\mathbf{k})}$ of
$\rho_{g}(\mathbf{x})$ defined over the 3D wavenumber grid $\mathbf{k}=(k_{x,}k_{y},k_{z})$
using a prescription similar to those described in \citet{1998A&A...336..697S}
and \citet{2002ApJ...564..773E}. The Fourier coefficients $s(\mathbf{k})$
are related to the random field $\rho_{g}(\mathbf{x})$ by 
\begin{eqnarray}
\rho_{g}(n_{x},n_{y},n_{z}) & = & \sum_{k_{x}=0}^{N_{x}-1}\sum_{k_{y}=0}^{N_{y}-1}\sum_{k_{z}=0}^{N_{z}-1}e^{2\pi ik_{x}n_{x}/N_{x}}e^{2\pi ik_{y}n_{y}/N_{y}}e^{2\pi ik_{z}n_{z}/N_{z}}\nonumber \\
 &  & \times s(k_{x},k_{y},k_{z}),
\end{eqnarray}
where $N_{x}$, $N_{y}$, and $N_{z}$ are numbers of elements in
each dimension, and $(n_{x},n_{y},n_{z})$ and $(k_{x},k_{y},k_{z})$
are indexes in real space and wavenumber space, respectively, ranging
from 0 to $N_{x}-1$, $N_{y}-1$, and $N_{z}-1$. The amplitudes $\left|s(\mathbf{k})\right|$
are taken to be Gaussian random variables with zero mean and a variance
of $\left|\mathbf{k}\right|^{-\gamma_{g}}$. Mean value of zero in
$\rho_{g}(\mathbf{x})$ is achieved by setting $s(0,0,0)=0$. To obtain
unit dispersion, the amplitudes are normalized by the sum $\sum_{\mathbf{k}}\left|s(\mathbf{k})\right|^{2}$
over the wavenumber grid. The phases $\phi(\mathbf{k})$ are randomly
selected in the range of 0 to $2\pi$. Because the density field should
be real valued, the Fourier coefficients must satisfy $s(\mathbf{k})=s^{*}(-\mathbf{k})$.
This results in the condition for the phases to be odd: $\phi(\mathbf{k})=-\phi(-\mathbf{k})$,
which can be easily realized by first generating a set of phases $\chi(\mathbf{k})$
and then taking $\phi(\mathbf{k})=\chi(\mathbf{k})-\chi(-\mathbf{k})$.
The real space counterparts are then calculated using the fast Fourier
transform library FFTW \citep{Frigo:cp}; the resulting density field
$\rho_{g}(\mathbf{x})$ is a Gaussian random field with zero mean
and unit dispersion \citep{Voss:1988:FNC:61153.61155}.

The Gaussian random field is multiplied by the standard deviation
$\sigma_{\ln\rho}$ of the logarithmic density, exponentiated, and
then multiplied by a normalization constant $\rho_{0}$ adjusted to
yield a given mean density; this results in the log-normal density
field $\rho(\mathbf{x})=\rho_{0}\exp\left[\sigma_{\ln\rho}\rho_{{\rm g}}(\mathbf{x})\right]$.
However, the power-law spectral index of the resulting lognormal density
field ($\gamma$) is different from that of the input Gaussian field
($\gamma_{g}$). The relationship between the power spectral indexes
of the lognormal density field and the Gaussian field was derived
in \citet{2012ApJ...761L..17S}. The power spectral index of $\rho(\mathbf{x})$
was obtained by least-squares fit over the range of $3\le\left|\mathbf{k}\right|\le35$.
This relationship, together with Equations (\ref{eq:1}) and (\ref{eq:2}),
was used to obtain a random realization of the lognormal density field
with the spectral index $\gamma$ and standard deviation $\sigma_{\ln\rho}$
appropriate for a given Mach number. In this way, ten random realizations
of the lognormal density distribution with box size of $N^{3}=128^{3}$
were generated for each Mach number of $M_{{\rm s}}$ = 1, 2, 4, 6,
8, 10, 12, 16, and 20. Some realizations of the lognormal density
field are shown in Figure \ref{fig1}. Because of the lognormal property
of the density fields, high-density cells are mostly contained in
a relatively small fraction of the volume. Figure \ref{fig2} shows
variations of the fraction of volume $f_{{\rm volume}}$ occupied
by a mass fraction $f_{{\rm mass}}$, when the mass fraction is measured
from the lowest density cell, for each Mach number. The results presented
here are results averaged over the ten realizations for each Mach
number.

\begin{figure}[t]
\begin{centering}
\includegraphics[clip,scale=0.9]{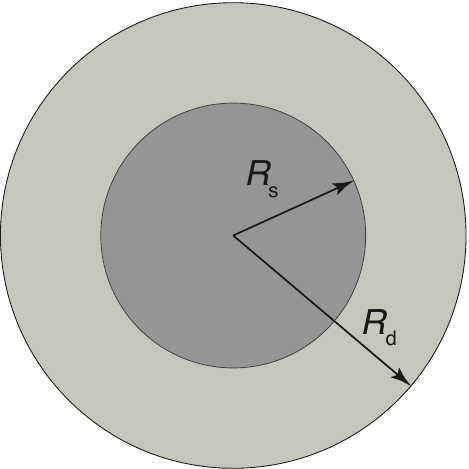}
\par\end{centering}
\begin{centering}
\medskip{}
\par\end{centering}
\caption{\label{fig4}Geometry used in the present study. $R_{{\rm s}}$ and
$R_{{\rm d}}$ represent the radii of the stellar and dust distributions,
respectively.}

\medskip{}

\medskip{}
\end{figure}

\subsection{Dust Extinction Curves}

\label{subsec:2.2}

To examine which type of underlying dust model best reproduces the
Calzetti attenuation curve and other attenuation curves inferred from
star-forming galaxies, five dust types are considered: three theoretical
dust models adopted from \citet{2001ApJ...548..296W} and \citet{2003ApJ...598.1017D}
for the MW, LMC, and SMC bar and two empirical models for the MW and
SMC from \citet{2000ApJ...528..799W}. The dust models of \citet{2001ApJ...548..296W}
will be referred to as MW-WD, LMC-WD and SMC-WD and the dust models
of \citet{2000ApJ...528..799W} as MW-WG and SMC-WG. As will be discussed
later, the resulting attenuation curve primarily depends on the underlying
absorption curve (wavelength dependence of absorption), rather than
on the extinction (absorption+scattering) curve. The wavelength dependences
of albedo ($a$) and scattering phase function asymmetry factor ($g$)
were calculated using Mie theory and the size distributions of carbonaceous
and silicate grain populations obtained in \citet{2001ApJ...548..296W}.
It should be noted that the albedo and scattering phase factor for
the MW-WG dust were empirically determined based on the observations
of reflection nebulae which usually have higher values of $R_{V}=A_{V}/E(B-V)$
than 3.1. The scattering phase function for the SMC-WG dust is the
same as that of the MW-WG dust. The albedo for the SMC-WG dust was
obtained by modifying that of the MW-WG to reflect the lack of a 2175\AA\
absorption bump in the SMC extinction curve.

Figure \ref{fig3} compares the wavelength dependences of extinction,
absorption and scattering cross-sections normalized by the extinction
values at $V$-band for the MW- and SMC-type dusts of \citet{2001ApJ...548..296W}
and \citet{2000ApJ...528..799W}. The extinction, absorption and scattering
cross-sections for the LMC-type dust of \citet{2001ApJ...548..296W}
are also shown in the bottom panels. The albedos of the MW-WG and
SMC-WG dust types remain more or less constant except for the absorption
feature at $\sim$ 2175\AA\ in the MW-WG dust. However, the albedo
of the MW-WD dust gradually decreases at $\lambda^{-1}>1.5$ $\mu$m$^{-1}$
and that of the SMC-WD dust decreases at $\lambda^{-1}>5$ $\mu$m$^{-1}$.
This results in a gradual decrease of the scattering cross-section
of the MW-WD dust at $\lambda^{-1}>4.5$ $\mu$m$^{-1}$; in contrast,
the scattering cross-section of the MW-WG dust gradually increases
as wavelength decreases. Similar trends are also found for the SMC-type
dust models. These properties will be further discussed with respect
to the fraction of scattered light in Section \ref{subsec:3.1}.

\begin{figure*}[t]
\begin{centering}
\includegraphics[clip,scale=0.73]{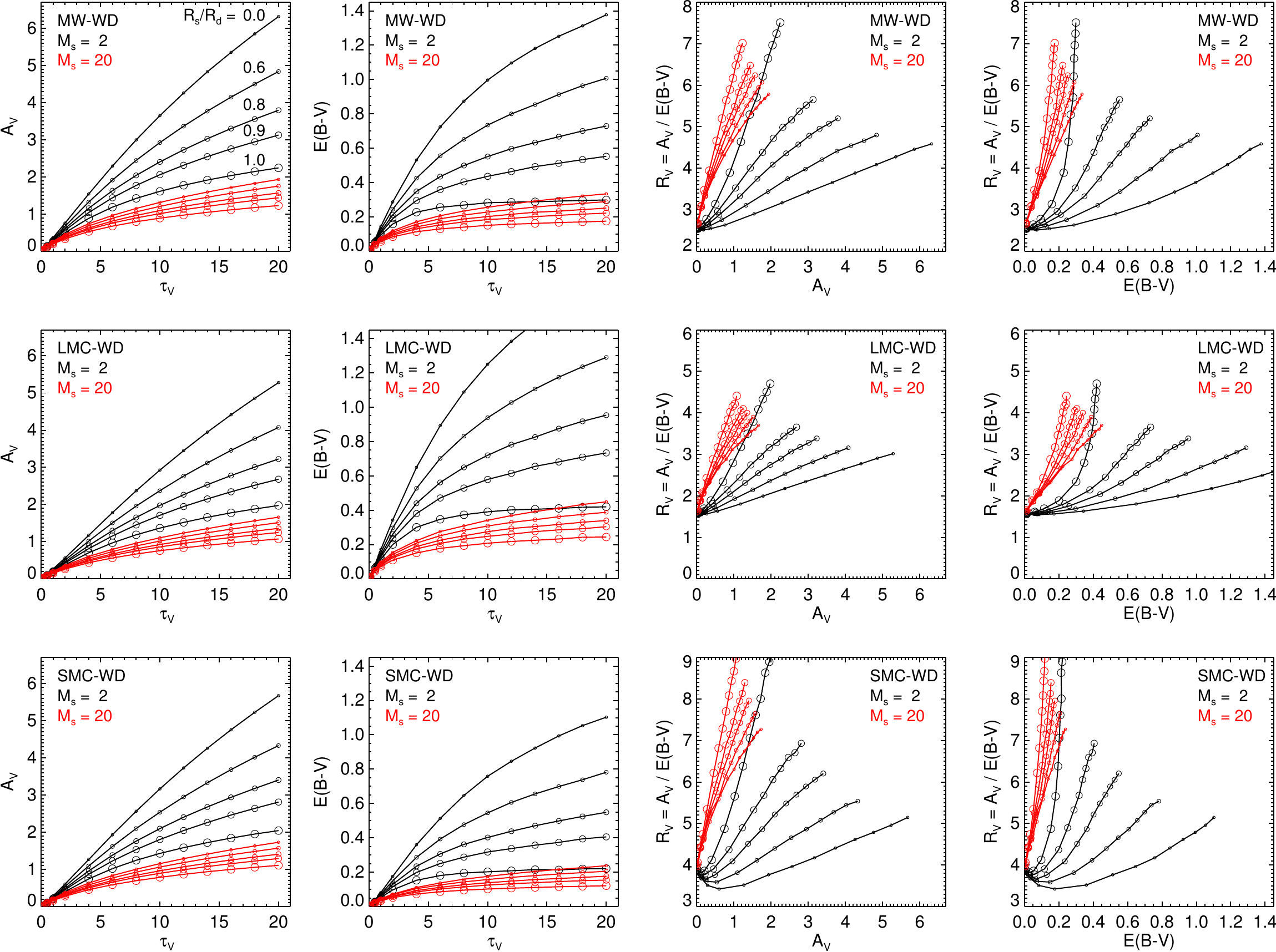}
\par\end{centering}
\caption{\label{fig5} (first and second columns) Attenuation $A_{V}$ and
color excess $E(B-V)$ as a function of homogeneous optical depth
$\tau_{V}$. (third column) Total to selective attenuation $R_{V}$
as a function of $A_{V}$. (fourth column) $R_{V}$ as a function
of the $E(B-V)$ color excess. First, second, and third rows represent
the results obtained for the MW, LMC and SMC dust types of \citet{2001ApJ...548..296W},
respectively. The homogeneous optical depth $\tau_{V}$ increases
from 0.1 up to 20. Symbol size is proportional to the size of stellar
distribution, corresponding to $R_{{\rm s}}/R_{{\rm d}}=0.0$, 0.6,
0.8, 0.9, and 1.0. Black and red colors denote the sonic Mach numbers
of $M_{{\rm s}}=2$ and 20, respectively.}

\centering{}\medskip{}
\end{figure*}

\subsection{Radiative Transfer Code}

\label{subsec:2.3}

For the radiative transfer simulation in turbulent media with lognormal
density distributions, a 3D Monte Carlo radiative transfer code MoCafe
\textbf{(Mo}nte \textbf{Ca}rlo radiative trans\textbf{fe}r) was used
\citep{2014ApJ...785L..18S,2015JKAS...48...57S}. The basic Monte
Carlo radiative transfer algorithms have been detailed by many authors
\citep{2001ApJ...551..269G,2011ApJS..196...22B,2013ARA&A..51...63S}.
Here, the major techniques implemented in the code are briefly described.
The forced first scattering technique was adopted for efficiency at
low optical depths, as in most Monte Carlo radiative transfer codes
\citep{cashwell1959pratical}. A forced scattering technique, as implemented
in \citet{2011ApJS..196...22B}, in which photon packets are always
forced to be scattered until the weight of photon packet decreases
down to a minimum value was also implemented and confirmed to yield
results that are consistent with the forced first scattering technique.
At high optical depths, the scattering scheme of \citet{2011ApJS..196...22B}
is faster, but underestimates the output flux unless a sufficiently
small minimum weight is adopted. The scheme of \citet{2011ApJS..196...22B}
adopting a very small minimum weight is in fact equivalent to the
forced first scattering scheme in speed.

MoCafe employs a fast voxel traversal algorithm for ray tracing, developed
for computer graphics rendering by \citet{Amanatides_1987}. The voxel
traversal algorithm dramatically increases the speed of radiative
transfer calculation compared to the conventional methods of \citet{1999A&A...349..839W}
or \citet{2001A&A...379..336K}. The algorithm also provides a method
to accurately track the paths of photon packets, which is crucial
for simulations in highly inhomogeneous media, even with single precision
floating-point numbers. The code uses the KISS (for Keep It Simple
Stupid) random number generator of \citet{marsaglia1993kiss}, which
is up to $\sim5$ times faster and has a much longer period than the
popular ``ran2'' generator given in \citet{1992nrfa.book.....P}.
A new fast algorithm to accelerate the simulation at high optical
depths was also developed; the algorithm will be published in a separate
paper.

Radiative transfer calculations were performed for 44 wavelengths
defined in a wavelength range of $\lambda^{-1}=0.3-8.9$ $\mu$m$^{-1}$
with a bin size of $\Delta\lambda^{-1}=0.2$ $\mu$m$^{-1}$. The
wavelength of a photon packet is selected among the wavelengths so
that the input SED is constant. The number of photon packages for
each wavelength was varied between $10^{5}$ and $10^{7}$ depending
on the optical depth of the system. Photon packages leaving the system
were recorded for the output SED $f_{\lambda}^{{\rm esc}}$ as a function
of wavelength. The attenuation optical depth as a function of wavelength
is then calculated by

\begin{equation}
\tau_{\lambda}^{{\rm att}}=-\ln\left(f_{\lambda}^{{\rm esc}}\right).\label{eq:3}
\end{equation}
The attenuation optical depth is always smaller than the extinction
optical depth, which is defined by the center-to-edge optical depth
of a cloud, because the dust albedo is always nonzero and the stars
are distributed throughout the extended volume. Total attenuation
at a wavelength $\lambda$ is defined by $A_{\lambda}=(2.5\log_{10}e)\tau_{\lambda}^{{\rm att}}=1.086\tau_{\lambda}^{{\rm att}}$
as analogous to the total extinction. Attenuation color excess $E(B-V)$
is defined by $A_{B}-A_{V}$ as for the extinction color excess. In
this paper, extinction-related quantities are denoted inside the square
brackets $[]_{\ast}$ with an asterisk to distinguish from those for
attenuation; for instance, the total extinction is denoted by $[A_{\lambda}]_{\ast}$.

Scattering directions of photons were randomly determined to follow
a numerical phase function calculated by the Mie theory for the dust
models of \citet{2001ApJ...548..296W} or a Henyey-Greenstein phase
function, using the algorithm of \citet{1977ApJS...35....1W}, for
the dust models of \citet{2000ApJ...528..799W} who do not provide
a size distribution of dust grains.

\subsection{Geometry}

\label{subsec:2.4}

In the present study, both the turbulent dusty medium and the stellar
sources were assumed to be spherically distributed; radii for stellar
and dust distributions are defined by $R_{{\rm s}}$ and $R_{{\rm d}}$,
respectively, as shown in Figure \ref{fig4}. Although the dust distribution
is highly inhomogeneous, photon sources were uniformly distributed
within the sphere, unless otherwise specified. The dust density was
set to zero for $r>R_{{\rm d}}$. The radial extents of the stellar
source were varied within a range of $0\le R_{{\rm s}}/R_{{\rm d}}\le1$;
the case of $R_{{\rm s}}=0$ represents a compact OB association surrounded
by a cloud, while the case of $R_{{\rm s}}/R_{{\rm d}}=1$ corresponds
to the ``dusty'' configuration of \citet{2000ApJ...528..799W} in
which photon sources are uniformly distributed over the spherical
dusty medium.

The optical depth of the medium is measured from the center to the
cloud surface at the $V$-band wavelength ($\lambda=0.55$ $\mu$m).
As in \citet{2000ApJ...528..799W}, the homogeneous radial optical
depth $\tau_{V}$ at $V$-band for a lognormal density cloud is defined
by the center-to-edge optical depth of a cloud with a constant density
and radius $R_{{\rm d}}$, but with the same dust mass as the lognormal
density cloud. For each Mach number, $\tau_{V}$ was varied from 0.1
to 20. To compare with the results of \citet{2000ApJ...528..799W},
radiative transfer calculations were also performed for the clumpy,
``shell'' geometry, in which stars are distributed only within $R_{{\rm s}}/R_{{\rm d}}\le0.3$
while dust grains are only located outside of the stellar distribution
radius. Ten realizations of the clumpy density distribution according
to \citet{2000ApJ...528..799W} were produced and the radiative transfer
calculations were averaged to remove any possible peculiarity due
to a specific realization, as for the case of the turbulent density
distribution. The results presented in this paper are averages over
all directions.

\begin{figure*}[t]
\begin{centering}
\includegraphics[clip,scale=0.51]{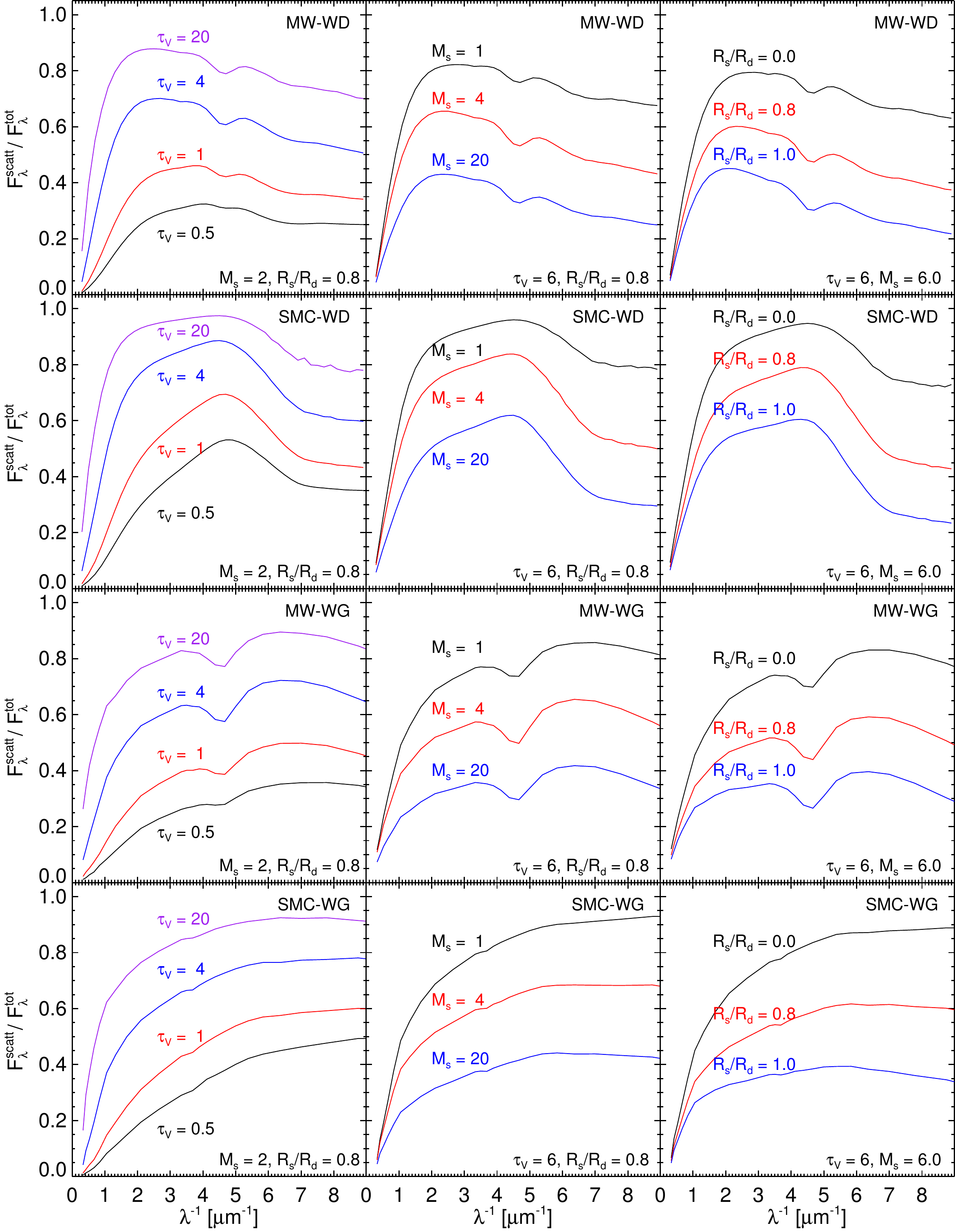}
\par\end{centering}
\caption{\label{fig6}Fraction of scattered light. Left, middle and right columns
show variations of the relative SEDs as optical depth, Mach number
and stellar distribution size vary, respectively. First and second
rows were calculated with the MW- and SMC-type extinction curves of
\citet{2001ApJ...548..296W}, respectively. Third and fourth rows
were calculated assuming the MW- and SMC-type extinction curves of
\citet{2000ApJ...528..799W}, respectively.}

\medskip{}
\end{figure*}

\begin{figure}[t]
\begin{centering}
\includegraphics[clip,scale=0.51]{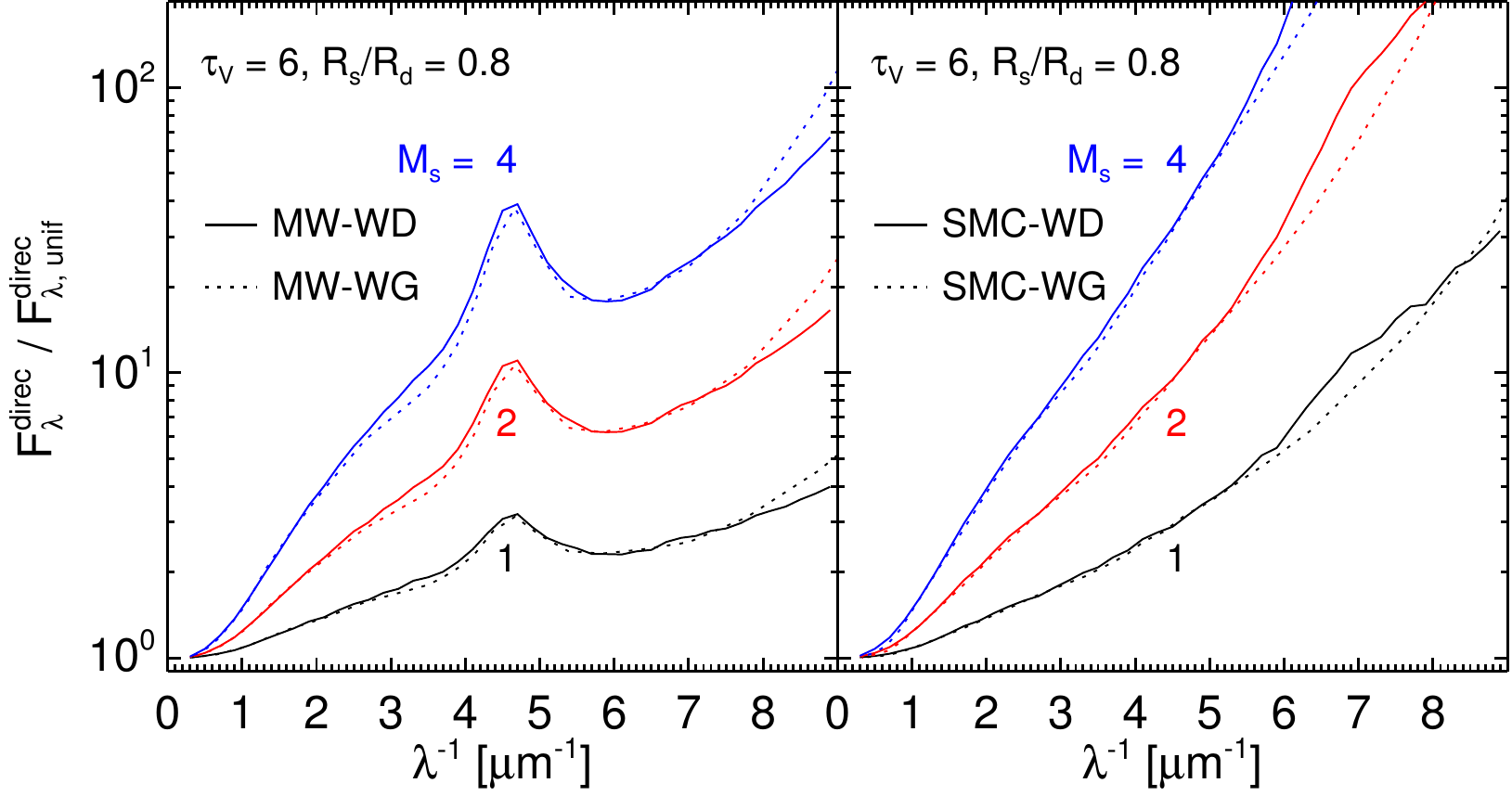}
\par\end{centering}
\caption{\label{fig7}Ratio of direct starlight escaping from a clumpy medium
($F_{\lambda}^{{\rm direc}}$) to that from a uniform medium ($F_{\lambda,{\rm unif}}^{{\rm direc}}$).
The relative SED is shown for $M_{{\rm s}}=1$ (black), 2 (red), and
4 (blue). Left and right panels were calculated with the MW- and SMC-type
extinction curves, respectively. Dust types of \citet{2001ApJ...548..296W}
and \citet{2000ApJ...528..799W} are denoted as solid and dotted lines,
respectively.}

\medskip{}
\end{figure}

\begin{figure}[tp]
\begin{centering}
\medskip{}
\includegraphics[clip,scale=0.92]{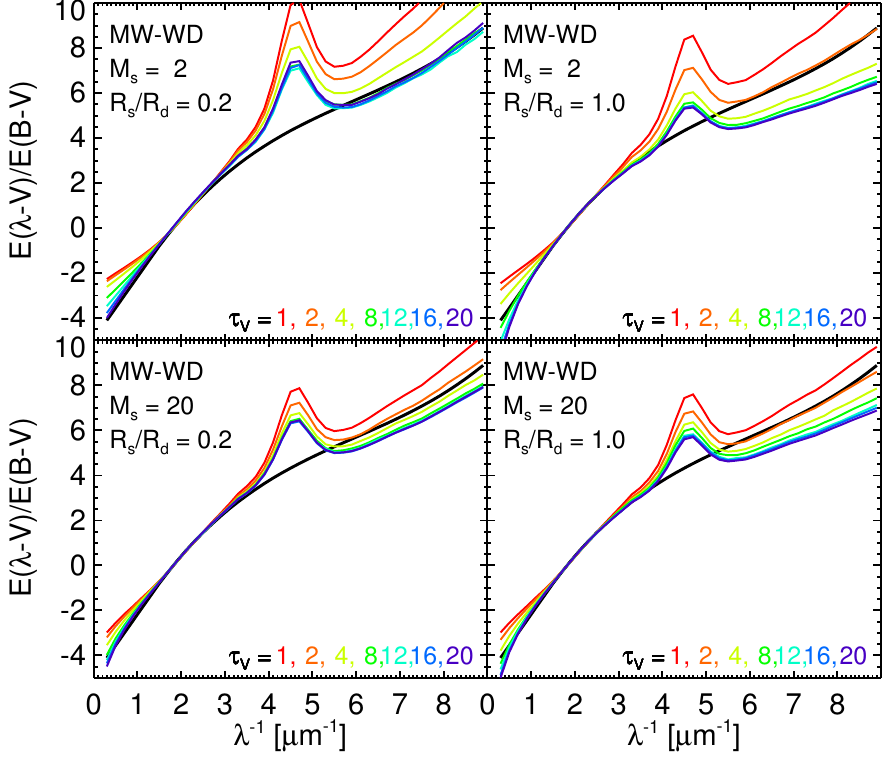}\medskip{}
\par\end{centering}
\begin{centering}
\includegraphics[clip,scale=0.92]{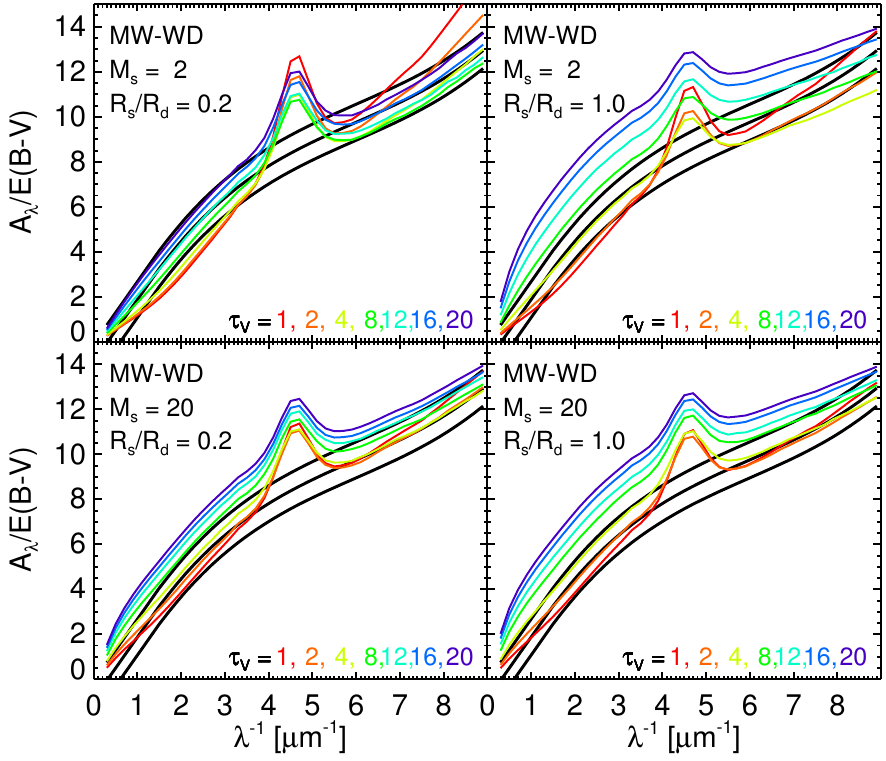}\medskip{}
\par\end{centering}
\begin{centering}
\includegraphics[clip,scale=0.92]{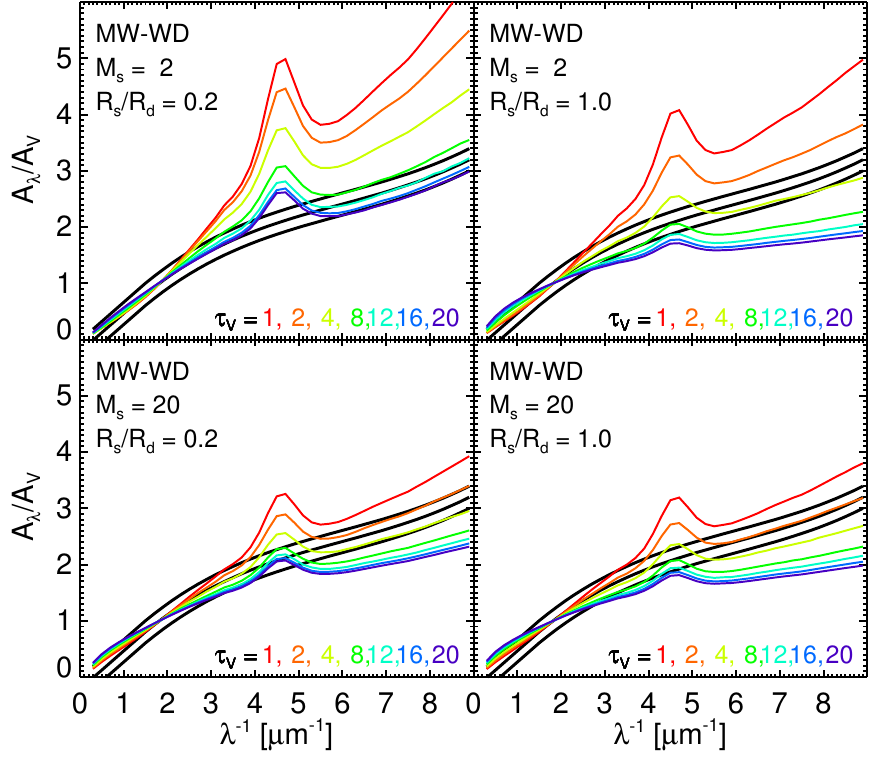}
\par\end{centering}
\caption{\label{fig8}Attenuation curves for the MW dust of \citet{2001ApJ...548..296W}
are plotted in three forms: (top) $E(\lambda-V)/E(B-V)$, (middle)
$A_{\lambda}/E(B-V)$, (bottom) and $A_{\lambda}/A_{V}$. Black curves
denote the Calzetti curve and its uncertainty boundaries. The upper
and lower bounds of the Calzetti curve in middle and bottom panels
are obtained using the error range of $R_{V}$. Note that no uncertainty
of the Calzetti curve in the $E(\lambda-V)/E(B-V)$ form (shown in
top panels) are provided.}
\end{figure}

\begin{figure}[tp]
\begin{centering}
\medskip{}
\includegraphics[clip,scale=0.92]{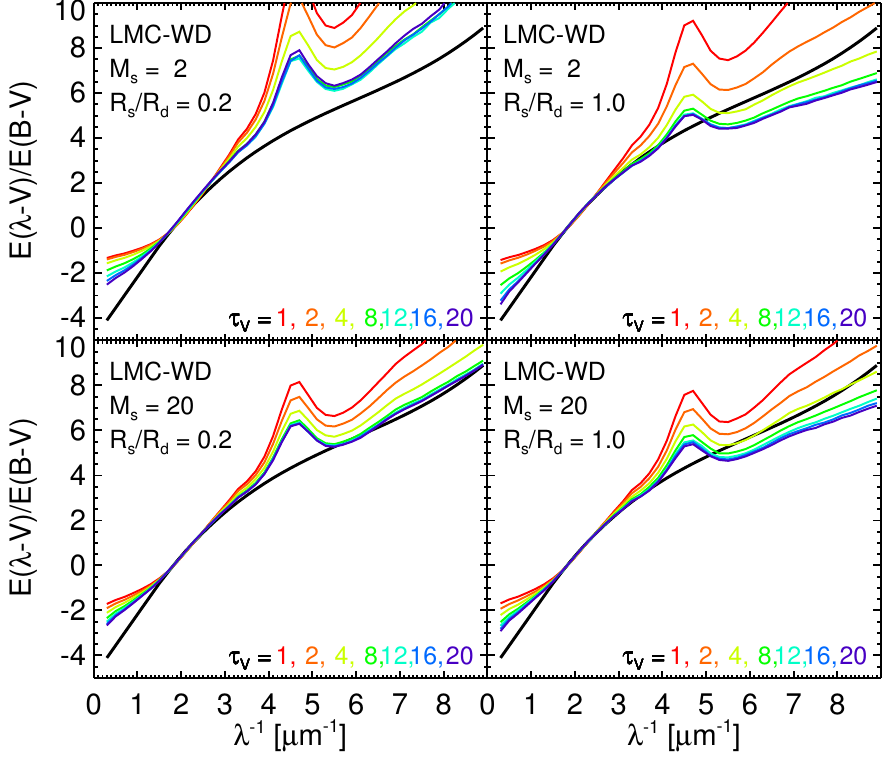}\medskip{}
\par\end{centering}
\begin{centering}
\includegraphics[clip,scale=0.92]{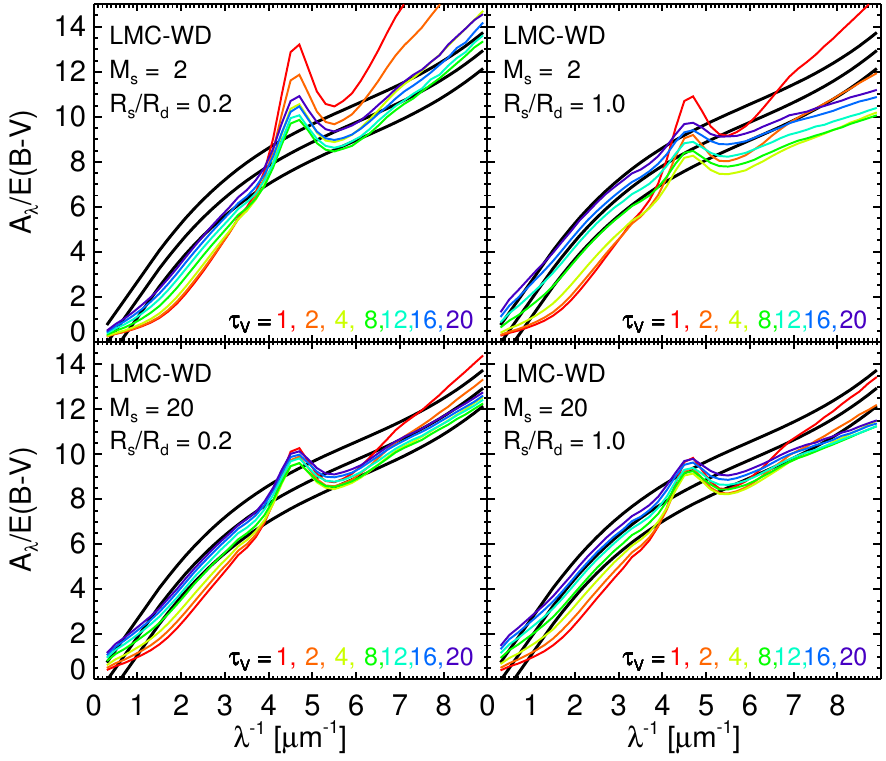}\medskip{}
\par\end{centering}
\begin{centering}
\includegraphics[clip,scale=0.92]{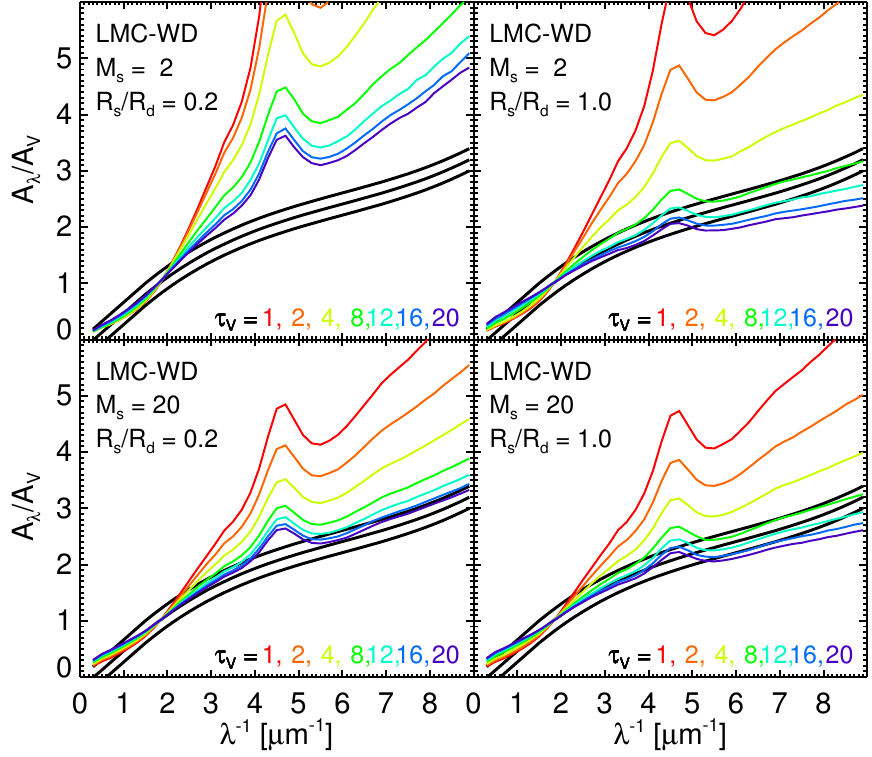}
\par\end{centering}
\caption{\label{fig9}Attenuation curves for the LMC dust of \citet{2001ApJ...548..296W}
are plotted in three forms: (top) $E(\lambda-V)/E(B-V)$, (middle)
$A_{\lambda}/E(B-V)$, (bottom) and $A_{\lambda}/A_{V}$. Black curves
denote the Calzetti curve and its uncertainty boundaries.}
\end{figure}

\begin{figure}[tp]
\begin{centering}
\medskip{}
\includegraphics[clip,scale=0.92]{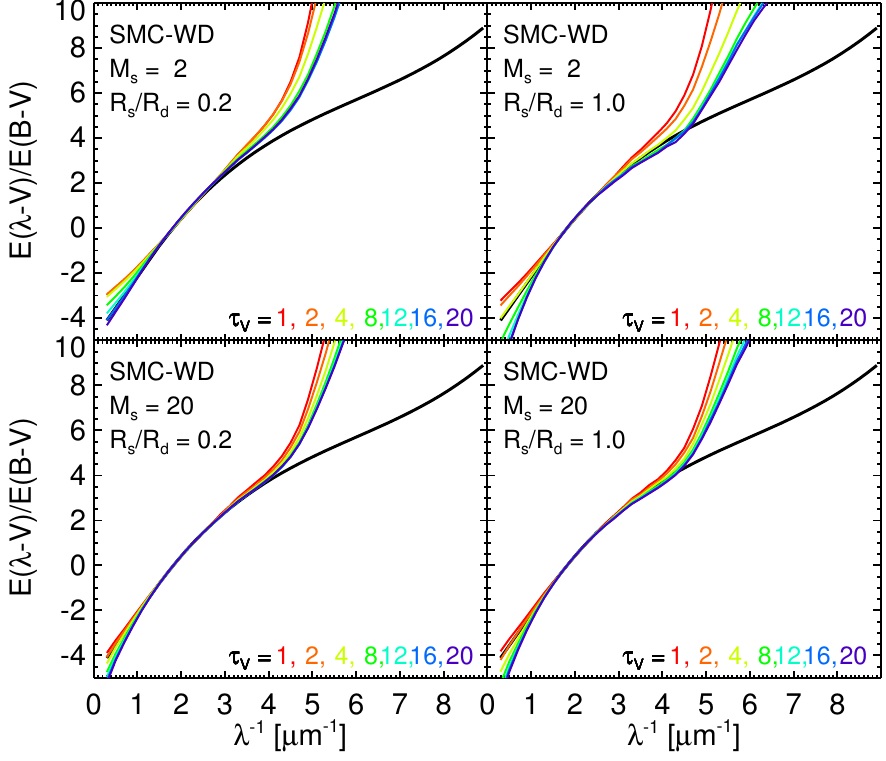}\medskip{}
\par\end{centering}
\begin{centering}
\includegraphics[clip,scale=0.92]{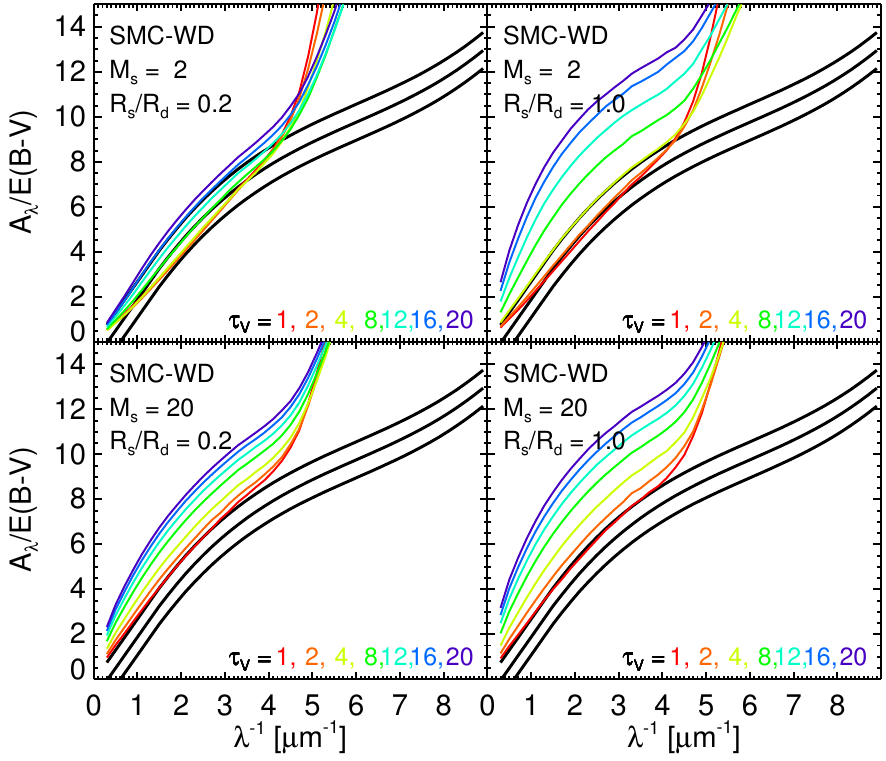}\medskip{}
\par\end{centering}
\begin{centering}
\includegraphics[clip,scale=0.92]{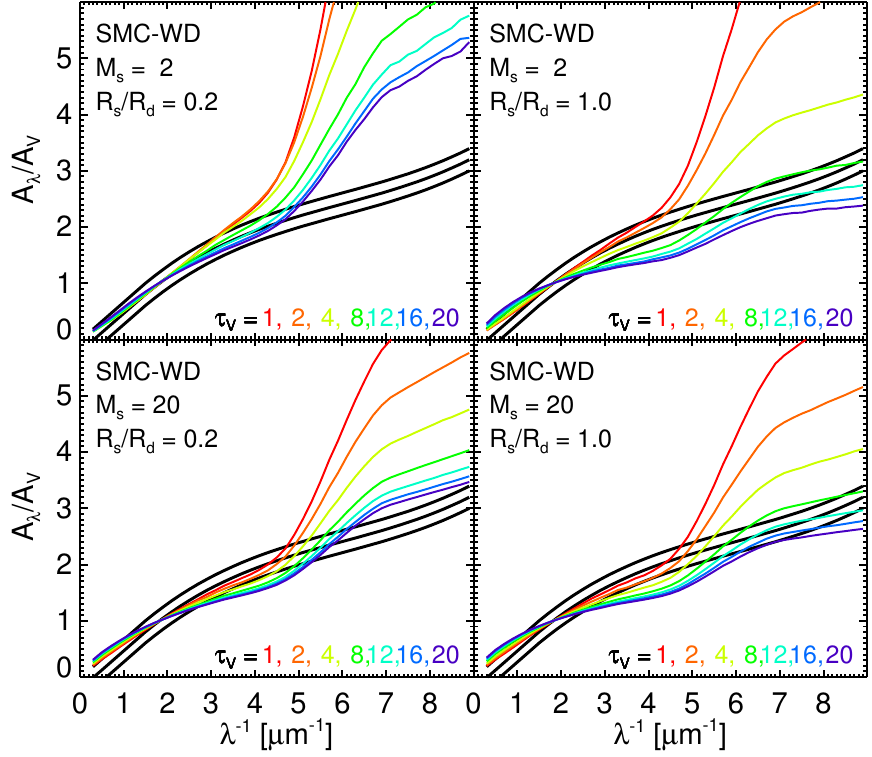}
\par\end{centering}
\caption{\label{fig10}Attenuation curves for the SMC dust of \citet{2001ApJ...548..296W}
are plotted in three forms: (top) $E(\lambda-V)/E(B-V)$, (middle)
$A_{\lambda}/E(B-V)$, (bottom) and $A_{\lambda}/A_{V}$. Black curves
denote the Calzetti curve and its uncertainty boundaries.}
\end{figure}

\section{RESULTS}

\label{sec:3}

\subsection{Radiative Transfer Effects}

\label{subsec:3.1}

In dealing with an extinction toward a point source, the amount of
total extinction $[A_{\lambda}]_{\ast}$ at a wavelength $\lambda$
as well as the color excess $[E(B-V)]_{\ast}\equiv[A_{B}-A_{V}]_{\ast}$
are linearly proportional to dust column density (or optical depth
at a particular wavelength) unless the physical properties of dust
grains are changed; thus, the total to selective extinction $[R_{V}]_{\ast}\equiv[A_{V}/E(B-V)]_{\ast}$
remains constant as optical depth varies. However, for spatially extended
sources, the linearity of total attenuation $A_{\lambda}$, defined
analogous to total extinction, on the amount of dust is no longer
valid even in a uniform medium \citep{1992ApJ...393..611W,2000ApJ...528..799W}.
Consequently, the color excess $E(B-V)$ is a non-linear function
of $A_{V}$; the total to selective attenuation $R_{V}$ is also not
a constant.

The first and second columns of Figure \ref{fig5} show $A_{V}$ and
$E(B-V)$, respectively, calculated as a function of the homogeneous
radial optical depth $\tau_{V}$ for the MW-WD, LMC-WD and SMC-WD
dust types. The total attenuation $A_{V}$ and color excess $E(B-V)$
decrease as Mach number and/or the stellar distribution size increases
for a given $\tau_{V}$. The MW-WD model shows the highest total attenuation
$A_{V}$ for a given $R_{{\rm s}}/R_{{\rm d}}$, $\tau_{V}$, and
$M_{{\rm s}}$ whereas the color excess $E(B-V)$ is highest for the
LMC-WD model. Both the total attenuation and color excess are increasing
functions of the homogeneous optical depth $\tau_{V}$. However, the
increase rapidly slows down. Especially at high Mach number ($M_{{\rm s}}\gtrsim6$)
and for extended stellar distribution ($R_{{\rm s}}/R_{{\rm d}}\gtrsim0.8$),
$E(B-V)$ starts to saturate at $\tau_{V}\sim5$, indicating that
the actual amount of dust cannot be determined by the color excess
measurement. For $M_{{\rm s}}\apprge6$ and $R_{{\rm s}}/R_{{\rm d}}\gtrsim0.8$,
$E(B-V)\lesssim0.4$ for the MW-WD model. As noted in \citet{2000ApJ...528..799W},
$E(B-V)$ of LBGs crowds around values of $\sim0.2$, rarely exceeding
0.4 \citep{1998AJ....115.1329S,1999ApJ...519....1S,2001ApJ...562...95S,2008ApJS..175...48R}.
The results in Figure \ref{fig5} suggest that the LBGs are highly
turbulent or clumpy with stars spatially mixed well with dust. As
will be shown in next section, the saturation of $E(B-V)$ is accompanied
by an attenuation curve that is grayer or shallower than the underlying
extinction curve.

Variation of $R_{V}$ is also shown as a function of $A_{V}$ and
$E(B-V)$ in the third and fourth columns of Figure \ref{fig5}. In
general, $R_{V}$ increases with increase of $M_{{\rm s}}$, $R_{{\rm s}}/R_{{\rm d}}$
and $\tau_{V}$. The ratio $R_{V}$ monotonically increases with $A_{V}$
except for the SMC-WD models with low $M_{{\rm s}}$ values, in which
$R_{V}$ decreases down to a minimum at $A_{V}\sim0.6$ and then increases
monotonically. For the models with high $R_{{\rm s}}/R_{{\rm d}}$
values, $R_{V}$ is not well constrained as a function of $E(B-V)$
because of the saturation of $E(B-V)$. This also indicates that the
absolute amount of dust is not well determined by the color excess
$E(B-V)$.

The saturation in $E(B-V)$ is due to three reasons: (1) dust density
distribution, (2) scattering and (3) free escape of starlight through
relatively low column density paths. The saturation of $E(B-V)$ indicates
that $E(B-V)$ is not a good proxy to accurately measure dust attenuation
\citep{2000ApJ...528..799W}; even a small measurement error in $E(B-V)$
would give rise to a large error in attenuation.

We also note that scattered light partially compensates extinction
\textendash{} the integrated light consists of scattered light and
direct starlight. Figure \ref{fig6} shows the SEDs of scattered light
relative to total light for various models with the MW and SMC dust
types. The fraction of scattered light depends on dust scattering
properties, wavelength, star/dust geometry, as well as total dust
column density. Multiple scattering becomes increasingly important
as $\tau_{V}$ increases while the direct starlight decreases; hence,
the fraction of scattered light increases with $\tau_{V}$, as shown
in the first column of Figure \ref{fig6}. For the parameters ($M_{{\rm s}}=2$
and $R_{{\rm s}}/R_{{\rm d}}=0.8$) in column one, scattered light
dominates the output SEDs at $\tau_{V}\gtrsim2$. The multiple scattering
effect becomes less important with increasing $M_{{\rm s}}$ and $R_{{\rm s}}/R_{{\rm d}}$,
as shown in the second and third columns of Figure \ref{fig6}. As
$M_{{\rm s}}$ increases, the density contrast between high- and low-density
cells and the volume filling fraction of low density cells both increase.
This allows more starlight to escape relatively easily. For large
source distributions (higher $R_{{\rm s}}/R_{{\rm d}}$), photons
originating from locations close to the boundary can also freely escape.
Thus, the fraction of scattered light anti-correlates with $M_{{\rm s}}$
and $R_{{\rm s}}/R_{{\rm d}}$.

It has been known that the scattered light is relatively ``bluer''
than direct starlight and thus partially compensates for reddening
by extinction \citep{1992ApJ...393..611W,2000ApJ...528..799W}. However,
the ``blueing'' effect of scattered light is true only in optical
and NIR wavelengths ($\lambda^{-1}\lesssim2.5$ $\mu$m$^{-1}$).
At UV wavelengths ($\lambda^{-1}\gtrsim2.5$ $\mu$m$^{-1}$), scattered
light can be either bluer or redder than direct starlight depending
on the adopted dust type, as shown in Figure \ref{fig6}. For the
two dust types of \citet{2000ApJ...528..799W}, scattered light is
bluer than direct starlight in most of the wavelength range except
at $\lambda^{-1}\gtrsim7$ $\mu$m$^{-1}$ in the MW-WG dust models.
On the other hand, scattered light is redder at $\lambda^{-1}\gtrsim2$
$\mu$m$^{-1}$ for the MW-WD dust and at $\lambda^{-1}\gtrsim5$
$\mu$m$^{-1}$ for the SMC-WD dust; therefore, scattering gives a
reddening effect at UV wavelengths for the MW-WD and SMC-WD dust types.
A ``reddening'' or ``blueing'' effect similar to that for the MW-WD
dust type was also found for the LMC-WD dust type.

The ``reddening'' or ``blueing'' effect of scattering can be understood
by examining the wavelength dependence of albedo of each dust type
shown in Figure \ref{fig3}. From the figure, it can be immediately
recognized that the shape of the normalized SED of scattered light
is mainly determined, especially at short wavelengths, by the wavelength
dependence of albedo. At the longest wavelengths ($\lambda^{-1}<1$
$\mu$m$^{-1}$), starlight escapes relatively freely from the dusty
media because of low optical depth. Therefore, the scattered fraction
is relatively low and approximately proportional to optical depth.
Increase of optical depth with $\lambda^{-1}$ gives rise to a gradual
increase of the scattered fraction with $\lambda^{-1}$. However,
as $\lambda^{-1}$ further increases, optical depth becomes much larger
than unity and most of the total output SED originates near the surface
of last scattering; thus, the fraction of scattered light is roughly
proportional to albedo and the shape of the normalized SED of scattered
light resembles the albedo curve at shorter wavelengths. As $\tau_{V}$
is increased, the overall shape of the normalized SED of scattered
light becomes closer to the albedo curve. For models with $\tau_{V}\gtrsim4$,
the normalized SEDs of scattered light are very close to the albedo
function at $\lambda^{-1}\gtrsim2$ $\mu$m$^{-1}$; therefore, whether
scattering yields either ``reddening'' or ``blueing'' effect at $\lambda^{-1}\gtrsim2$
$\mu$m$^{-1}$ is primarily determined by the shape of the adopted
albedo curve. In an observational study of the FUV ($5.7\ \mu{\rm m}^{-1}<\lambda^{-1}<7.4\ \mu{\rm m}^{-1}$)
continuum background, \citet{2011ApJS..196...15S} found that the
scattered FUV continuum background is relatively redder than the direct
stellar spectrum and attributed the result to the increase of scattering
albedo with the wavelength; therefore, the diffuse FUV background
observation supports the wavelength dependence of albedo of the MW-WD
dust type. However, we note that scattering is more efficient at $B$-band
wavelength than at $V$-band wavelength for all of the dust types
considered in the present study.

The saturation of $E(B-V)$ is also observed for higher Mach numbers
(i.e., higher density contrasts) and/or more extended stellar distributions.
As already noted, increase in $M_{{\rm s}}$ or $R_{{\rm s}}/R_{{\rm d}}$
for a given $\tau_{V}$ allows more direct starlight to escape freely
without any interaction with dust grains. The middle and right columns
of Figure \ref{fig6} show the increase of the direct starlight component
with increasing $M_{{\rm s}}$ and $R_{{\rm s}}/R_{{\rm d}}$, respectively.
The contribution of direct starlight makes not only $E(B-V)$ saturate
but also attenuation curves grayer.

The above three effects play complex roles in forming attenuation
curves, which are in general grayer than the underlying extinction
curve. Direct starlight escaping from a clumpy medium is bluer than
that from a uniform medium (see Figure \ref{fig7}). This is because
a clumpier medium yields a smaller effective optical depth and the
effect becomes more important as optical depth increases with decreasing
wavelength. In NIR and optical wavelengths, scattering provides even
more blueing effect in the output SED regardless of the dust type.
However, at UV wavelengths, scattering can play a opposite role for
some dust types (e.g., MW-WD or SMC-WD).

\subsection{Attenuation Curves}

\label{subsec:3.2}

Before we discuss attenuation curves, it is first noted that \citet{1994ApJ...429..582C}
originally derived a selective attenuation curve relative to a zero
point at $V$-band (0.55 $\mu$m) and normalized to $E(B-V)$, equivalent
to $E(\lambda-V)/E(B-V)$, but not a total attenuation. Since they
averaged spectra of sample galaxies at different distances and with
different luminosities to derive an average attenuation curve for
starburst galaxies, they had to arbitrarily impose a zero point at
a conventional wavelength, e.g., 0.55 $\mu$m. The zero point was
found to be $R_{V}=A_{V}/E(B-V)=4.88\pm0.98$ by comparing the attenuated
stellar luminosity with the re-processed FIR emission \citep{1997AJ....113..162C}.
Later, the zero point was revised to be $R_{V}=4.05\pm0.80$ in \citet{2000ApJ...533..682C},
providing the final Calzetti curve in a piece-wise functional form
of $A_{\lambda}/E(B-V)\equiv E(\lambda-V)/E(B-V)+R_{V}$.

Note that it was implicitly assumed that (1) $E(B-V)$ is a proxy
for the dust attenuation and (2) $R_{V}$ is independent of $E(B-V)$
in deriving the zero-point. However, for fixed dust properties the
``observed'' total to color excess ratio $R_{V}$ is no longer a constant,
as shown in Figure \ref{fig5}. The ratio $R_{V}$ increases with
the optical depth, Mach number, and the source size. As noted in \citet{2000ApJ...528..799W},
the attenuation curve of galaxies is not unique but shows large variations.
This is because neither $A_{V}$ nor $E(B-V)$ are linear functions
of $\tau_{V}$, and $R_{V}$ is not a constant. The non-constancy
of $R_{V}$ may give an impression that a calculated $A_{\lambda}/E(B-V)$
curve is somewhat different from the Calzetti curve even when the
$E(\lambda-V)/E(B-V)$ curve seems to accord well with that of Calzetti.
It should also be noted that attenuation correction in fitting galactic
SEDs is done by using a total attenuation curve $A_{\lambda}$ normalized
to a total attenuation value at a particular wavelength, such as $A_{\lambda}/A_{V}$
\citep{2009A&A...499...69N,2011A&A...533A..93B,2013ApJ...775L..16K}
or $A_{\lambda}/A_{1300}$ \citep{2015ApJ...800..108S}. However,
both the non-linear relation between $A_{V}$ and $E(B-V)$ and the
non-constancy of $R_{V}$ would also yield a large deviation of the
calculated $A_{\lambda}/A_{V}$ curves from that of Calzetti. As will
be discussed in Section \ref{subsec:4.3}, the variability in the
attenuation curves for the local starburst galaxies is indeed significant.

\begin{figure*}[t]
\begin{centering}
\includegraphics[clip,scale=0.92]{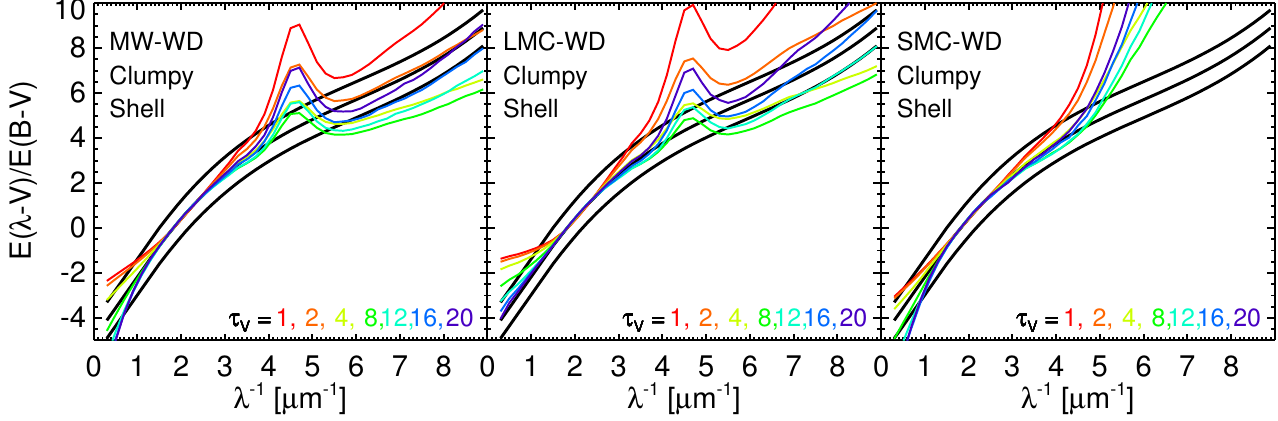}\medskip{}
\par\end{centering}
\begin{centering}
\includegraphics[clip,scale=0.92]{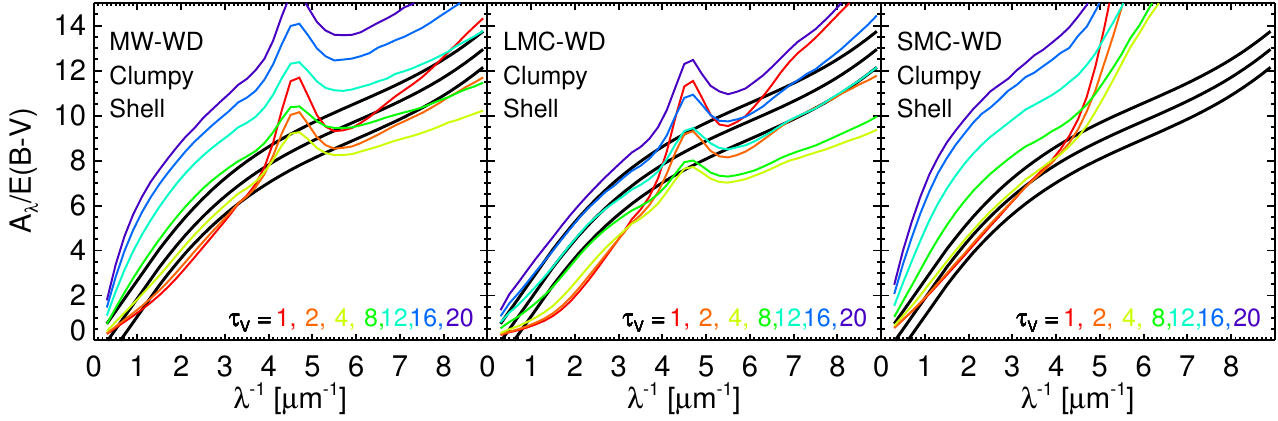}\medskip{}
\par\end{centering}
\begin{centering}
\includegraphics[clip,scale=0.92]{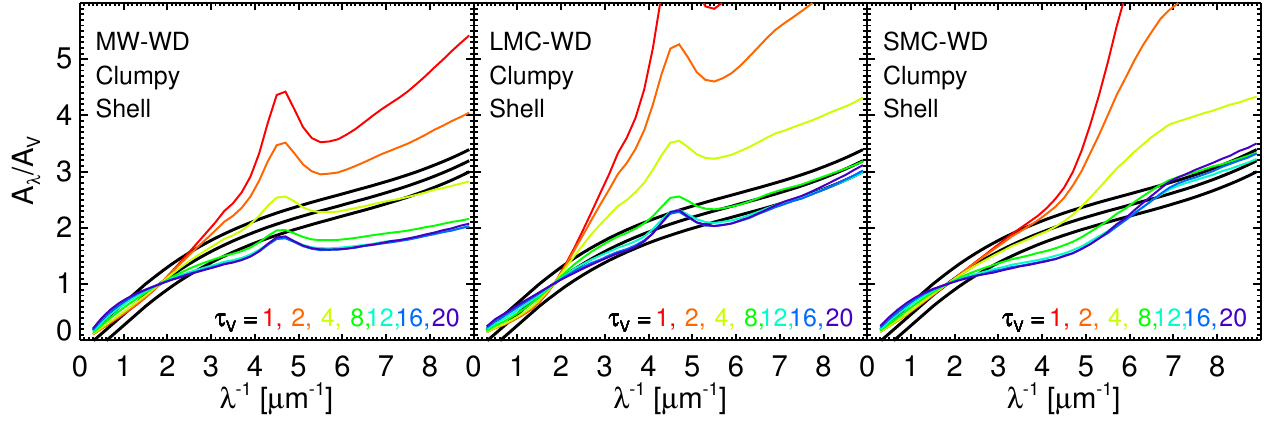}
\par\end{centering}
\caption{\label{fig11}Attenuation curves for (left) the MW, (middle) LMC,
and (right) SMC dust types of \citet{2001ApJ...548..296W} in the
clumpy media with the shell geometry of \citet{2000ApJ...528..799W}.
Three black curves are the Calzetti curve and its upper and lower
boundaries.}

\medskip{}
\end{figure*}

We therefore compare the calculated attenuation curves in three different
forms: $E(\lambda-V)/E(B-V)$, $A_{\lambda}/E(B-V)$ and $A_{\lambda}/A_{V}$.
Figures \ref{fig8}, \ref{fig9} and \ref{fig10} show the calculated
attenuation curves as functions of inverse wavelength for the MW-WD,
LMC-WD and SMC-WD dust types, respectively. In each figure, results
are shown for Mach numbers $M_{{\rm s}}=$ 2 and 20, and the stellar
distribution size of $R_{{\rm s}}/R_{{\rm d}}=0.2$ and 1.0. In each
panel of the figures, the homogeneous optical depth $\tau_{V}$ is
varied from 1 to 20 and the variation is denoted in colors. The Calzetti
curve is also shown in black for comparison. The upper and lower black
curves in the $A_{\lambda}/E(B-V)$ and $A_{\lambda}/A_{V}$ curves
denote the uncertainty boundaries of the Calzetti curve estimated
from the uncertainty range of $R_{V}=A_{V}/E(B-V)=4.05\pm0.80$, as
derived in \citet{2000ApJ...533..682C}. Note, however, that no error
range in the $E(\lambda-V)/E(B-V)$ curve was provided in \citet{2000ApJ...533..682C}.

In Figures \ref{fig8}, \ref{fig9} and \ref{fig10}, we first note
that, regardless of the adopted dust type, the attenuation curve becomes
grayer as the homogeneous optical depth $\tau_{V}$, the Mach number
$M_{{\rm s}}$, and the size of the source distribution $R_{{\rm s}}/R_{{\rm d}}$
each increase. The shallow slopes shown in the models, especially
with large optical depths, are attributable to both the turbulent
density structure and scattering effects, as discussed in Section
\ref{subsec:3.1}. The shape of the attenuation curve is least affected
by varying $R_{{\rm s}}/R_{{\rm d}}$, compared to the effects due
to changes of $\tau_{V}$ and $M_{{\rm s}}$; especially at the highest
Mach number ($M_{{\rm s}}=20$), the variation of the curve due to
the $R_{{\rm s}}/R_{{\rm d}}$ change is marginal. The largest variations
of the curves are produced by changing $\tau_{V}$. These trends are
further discussed in Section \ref{subsec:3.6}. It is also noted that
the shallower curves have weaker UV bump features. Similar trends
have been found in \citet{1992ApJ...393..611W} and \citet{2000ApJ...528..799W}.

The most important result from the figures is that the MW-WD model
calculations are in general consistent with the Calzetti curve \emph{except}
for the presence of a UV absorption feature. This is true especially
for higher $M_{{\rm s}}$ and $R_{{\rm s}}/R_{{\rm d}}$. The $E(\lambda-V)/E(B-V)$
and $A_{\lambda}/A_{V}$ curves averaged over a wide range of $\tau_{V}$
for each set of $M_{{\rm s}}$ and $R_{{\rm s}}/R_{{\rm d}}$ coincide
well with the Calzetti curve, provided that $M_{{\rm s}}\gtrsim4$
or $R_{{\rm s}}/R_{{\rm d}}\gtrsim0.2$ \emph{and} the UV bump is
removed. It will be shown later that the Calzetti attenuation curve
is very well reproduced by MW-WD dust type with the UV bump component
suppressed or entirely removed. Except for the UV bump, the average
attenuation curve for $R_{{\rm s}}/R_{{\rm d}}=1$ and $\tau_{V}\sim2-4$
is found to accord well with the Calzetti curve regardless with Mach
number. For small values of the parameters $M_{{\rm s}}$, $R_{{\rm s}}/R_{{\rm d}}$,
and $\tau_{V}$, the attenuation curves appear to be slightly steeper
than the Calzetti curve. On the other hand, the attenuation curves
are relatively shallower at large values of $\tau_{V}$. The total
to selective attenuation ratio $R_{V}$ for high values of $M_{{\rm s}}$
and $R_{{\rm s}}/R_{{\rm d}}$ tends to be slightly higher than that
of the Calzetti curve, as shown in the $A_{\lambda}/E(B-V)$ curves.

For the LMC-WD dust type, the overall shape of the $A_{\lambda}/E(B-V)$
curves, except the UV bump feature, for $M_{{\rm s}}=20$, $\tau_{V}\gtrsim4$
or $M_{{\rm s}}=2$, $\tau_{V}\gtrsim16$ seem to accord with the
Calzetti curve while $A_{\lambda}/A_{V}$ curves show much steeper
slopes except for the highest optical depths ($\tau_{V}\gtrsim8$)
and the star/dust geometries with $R_{{\rm s}}/R_{{\rm d}}>0.8$.
The $E(\lambda-V)/E(B-V)$ curve at $\lambda^{-1}>\lambda_{V}^{-1}$
is also consistent with the Calzetti curve for large $M_{{\rm s}}$
and $R_{{\rm s}}/R_{{\rm d}}$ values. We note that disregarding the
UV bump the calculated attenuation curves for the MW-WD dust were
consistent with the Calzetti curve over a wide range of parameters,
whereas the attenuation curves obtained with the LMC-WD dust accorded
within a more limited parameter range. We also note that $R_{V}$
is always lower for the LMC-WD dust than for the MW-WD dust, as shown
in $A_{\lambda}/E(B-V)$ curves. In contrast to the MW-WD and LMC-WD
dust types, the attenuation curves calculated with the SMC-WD dust
are generally not consistent with the Calzetti curve in all three
forms. 

\begin{figure}[t]
\begin{centering}
\medskip{}
\includegraphics[clip,scale=0.92]{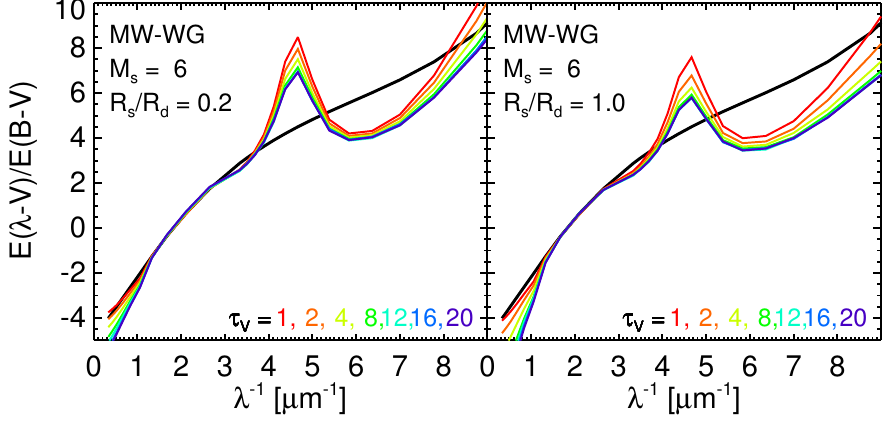}\medskip{}
\par\end{centering}
\begin{centering}
\includegraphics[clip,scale=0.92]{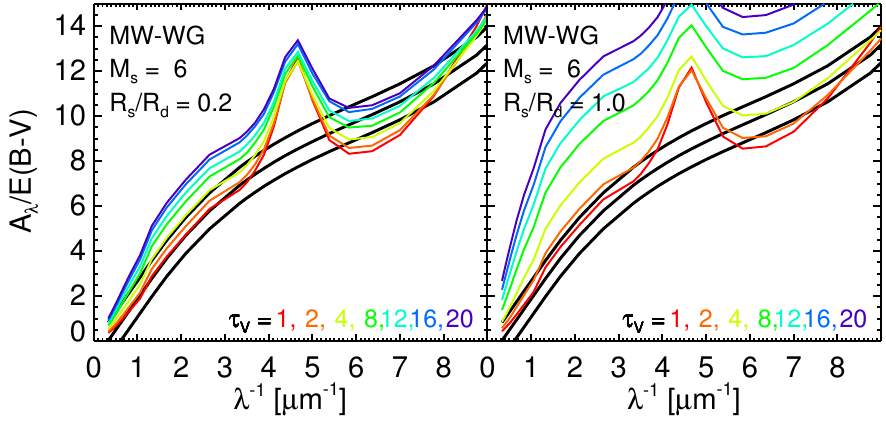}\medskip{}
\par\end{centering}
\begin{centering}
\includegraphics[clip,scale=0.92]{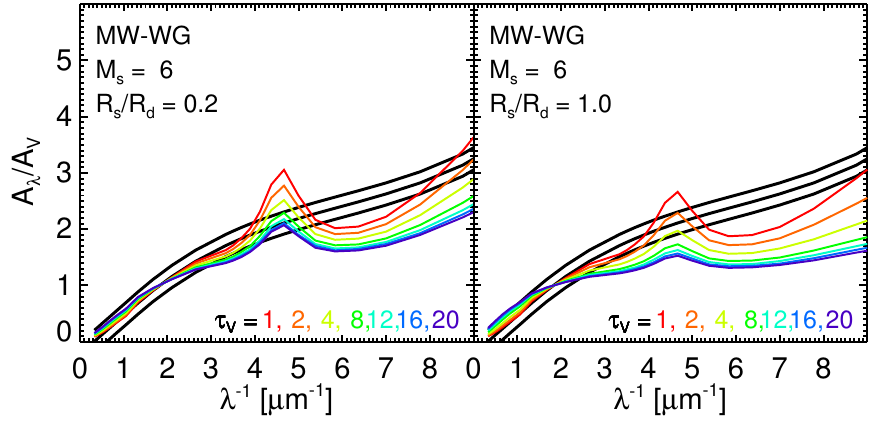}
\par\end{centering}
\caption{\label{fig12}Attenuation curves for the MW dust of \citet{2000ApJ...528..799W}
in clumpy media with $M_{{\rm s}}=6$ are plotted in three forms:
(top) $E(\lambda-V)/E(B-V)$, (middle) $A_{\lambda}/E(B-V)$, and
(bottom) $A_{\lambda}/A_{V}$. Black curves denote the Calzetti curve
and its uncertainty boundaries.}
\medskip{}
\end{figure}

\begin{figure}[t]
\begin{centering}
\medskip{}
\includegraphics[clip,scale=0.92]{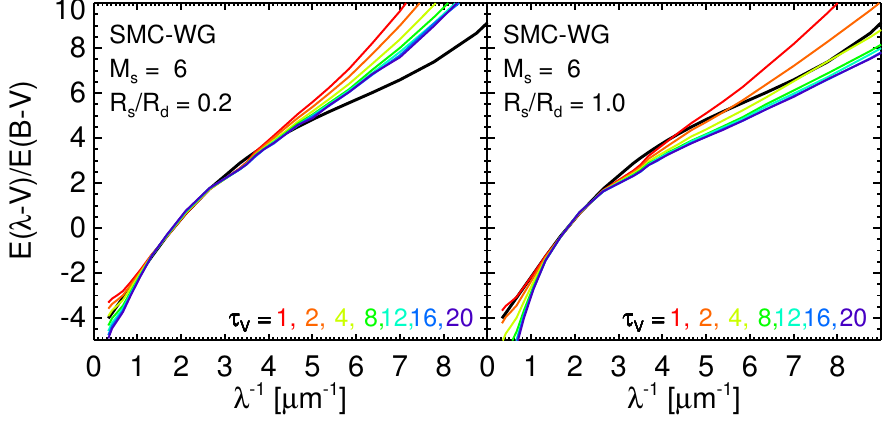}\medskip{}
\par\end{centering}
\begin{centering}
\includegraphics[clip,scale=0.92]{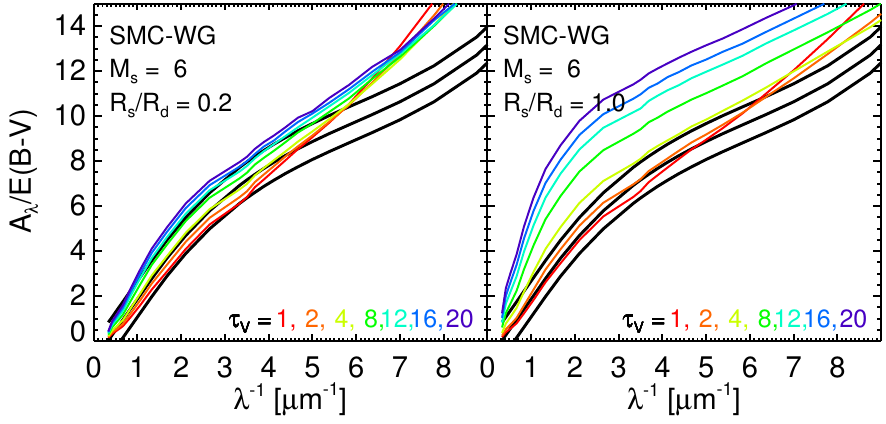}\medskip{}
\par\end{centering}
\begin{centering}
\includegraphics[clip,scale=0.92]{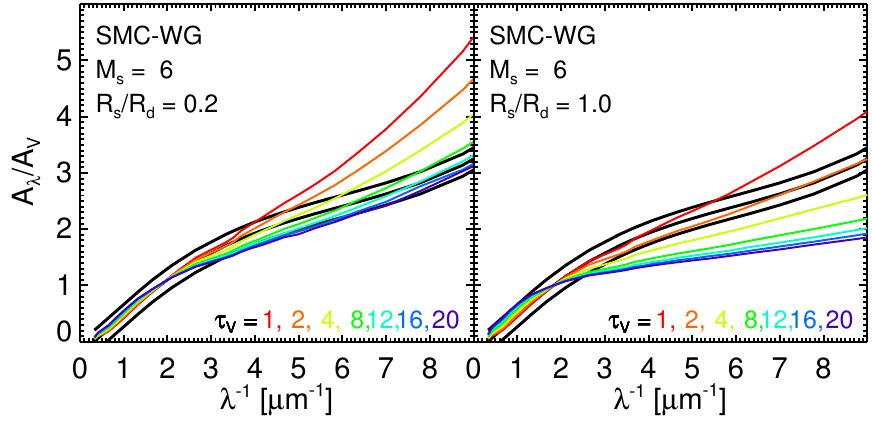}
\par\end{centering}
\caption{\label{fig13}Attenuation curves for the SMC dust of \citet{2000ApJ...528..799W}
in clumpy media with $M_{{\rm s}}=6$ are plotted in three forms:
(top) $E(\lambda-V)/E(B-V)$, (middle) $A_{\lambda}/E(B-V)$, and
(bottom) $A_{\lambda}/A_{V}$. Black curves denote the Calzetti curve
and its uncertainty boundaries.}
\end{figure}

\begin{figure}[t]
\begin{centering}
\medskip{}
\includegraphics[clip,scale=0.57]{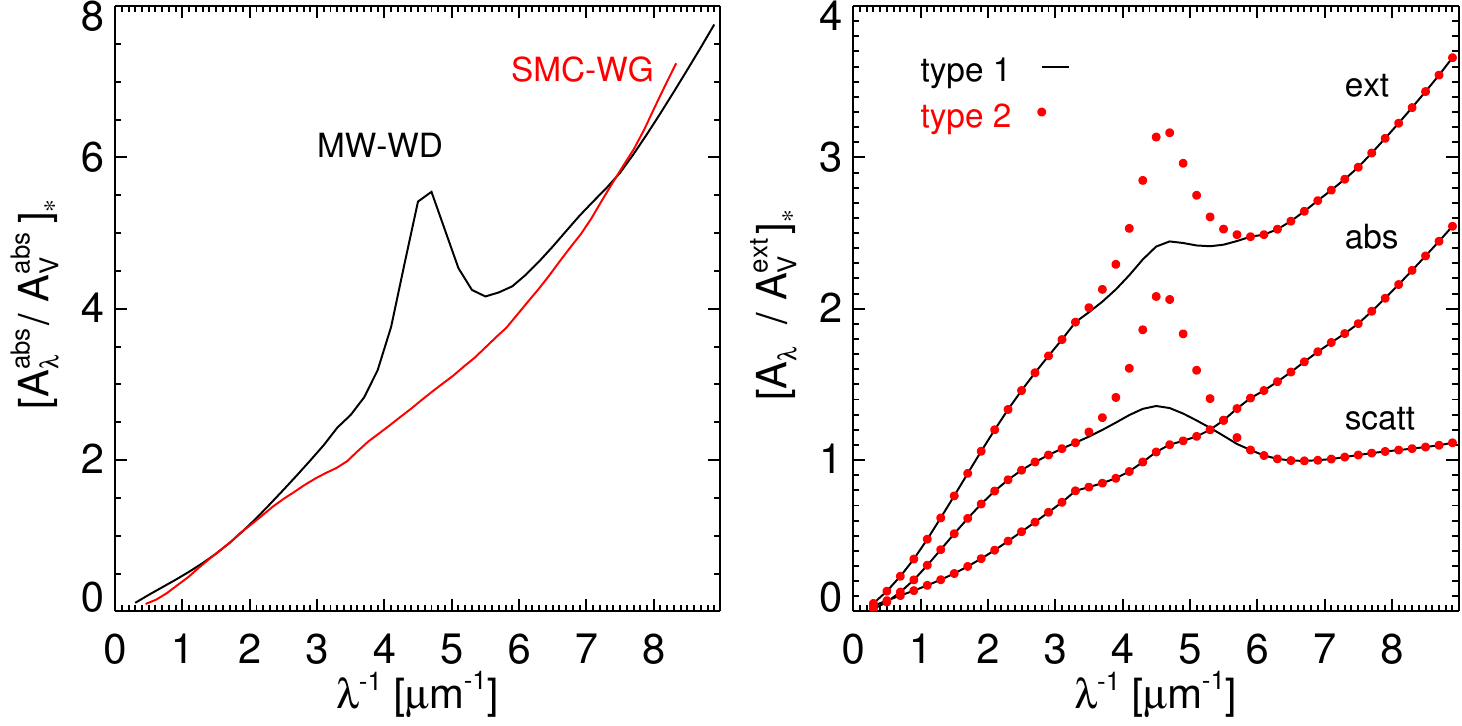}
\par\end{centering}
\caption{\label{fig14}(left) Comparison of the absorption curves of the MW-WD
dust \citep{2001ApJ...548..296W} and the SMC-WG dust \citep{2000ApJ...528..799W}.
(right) Comparison of two hypothetical dust types. The UV absorption
bump is removed from the extinction curve in type 1 (black lines).
In type 2 (red circles), the bump feature is attributed to scattering.
The scattering phase function for the added scattering was assumed
to be the same as that of the MW-WD dust.}
\medskip{}
\end{figure}

\begin{figure}[t]
\begin{centering}
\medskip{}
\includegraphics[clip,scale=0.92]{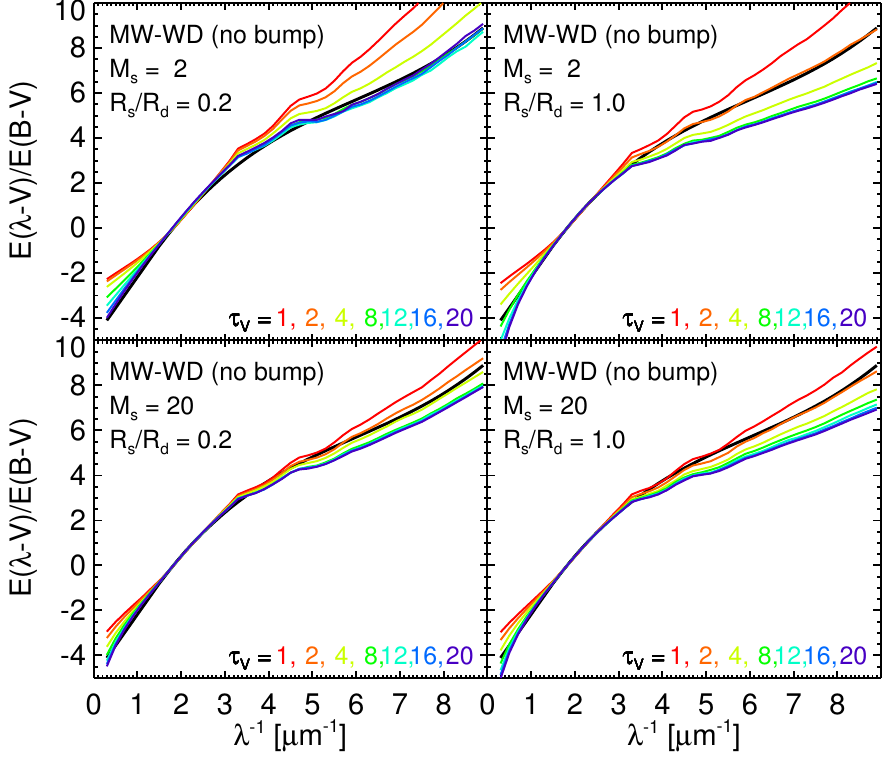}\medskip{}
\par\end{centering}
\begin{centering}
\includegraphics[clip,scale=0.92]{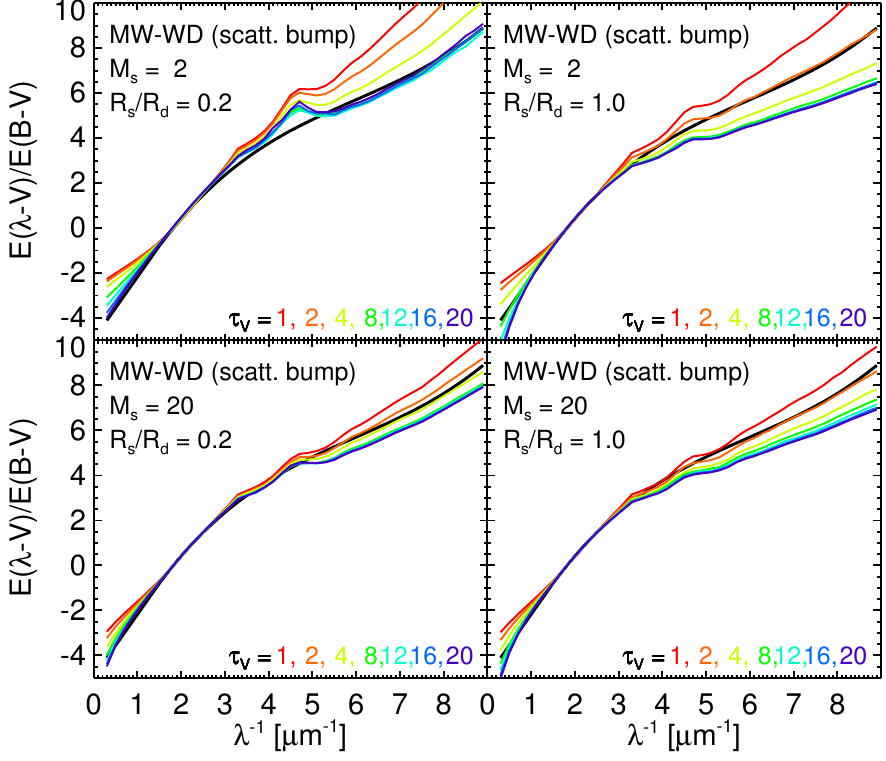}
\par\end{centering}
\caption{\label{fig15}Attenuation curves for two hypothetical dust types.
In top four panels, the UV bump was removed from the MW-WD extinction
curve (type 1). In bottom four panels, the UV absorption bump feature
was replaced by a scattering feature with the same strength (type
2). The Calzetti curve is also shown in black.}
\medskip{}
\end{figure}

\begin{figure}[t]
\begin{centering}
\medskip{}
\includegraphics[clip,scale=0.5]{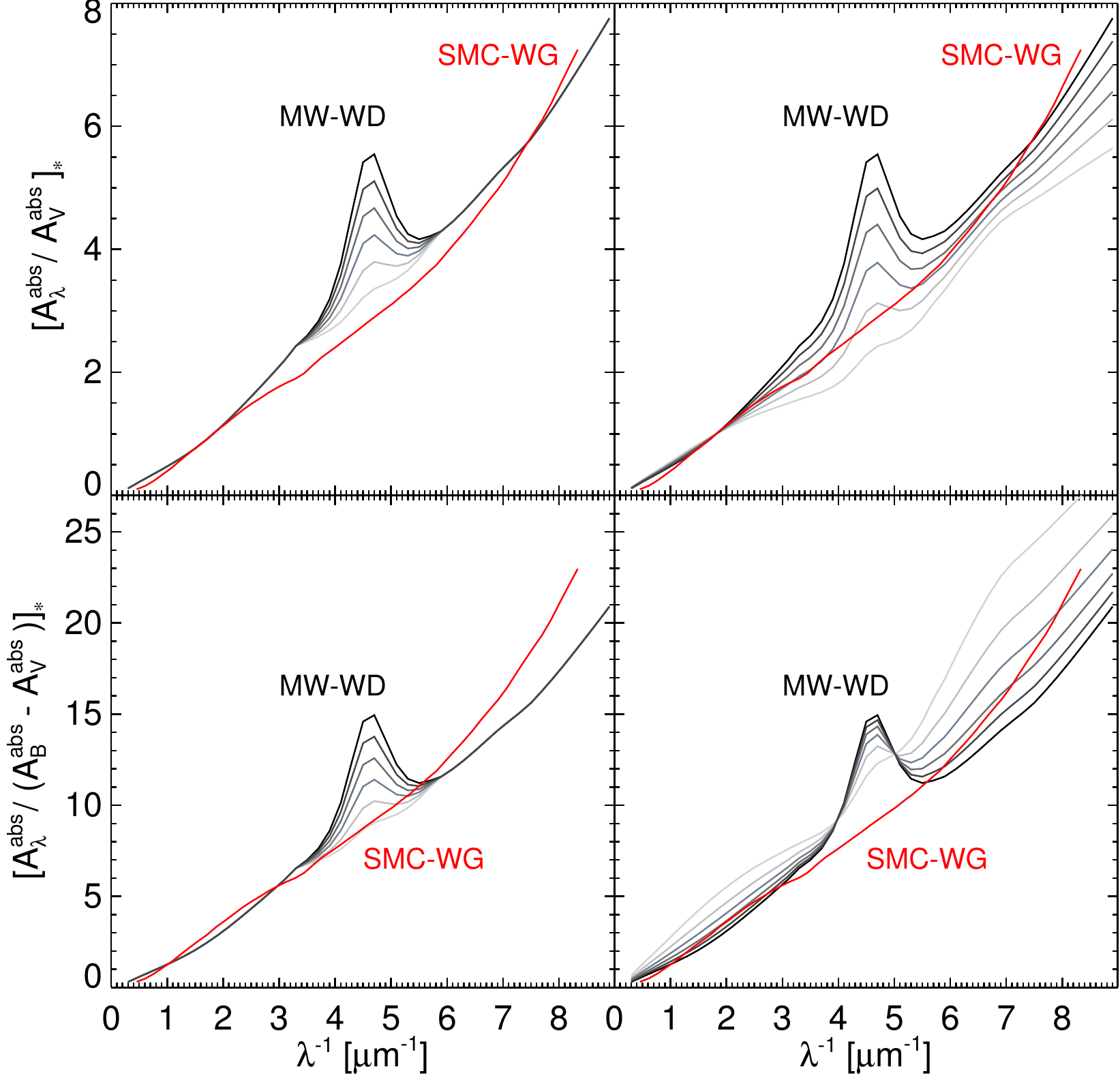}
\par\end{centering}
\caption{\label{fig16}Absorption curves with lower (left) UV bump and (right)
PAH features are shown in grayscale. Strengths of the features were
reduced from 0.8 to 0.0 in steps of 0.2 relative to those of the MW-WD
dust. The absorption curves are normalized by (top) $[A_{V}^{{\rm abs}}]_{\ast}$
and (bottom) $[A_{B}^{{\rm abs}}-A_{V}^{{\rm abs}}]_{\ast}$. The
absorption curve of the SMC-WG dust is also shown in red.}
\medskip{}
\end{figure}

\begin{figure}[tp]
\begin{centering}
\medskip{}
\includegraphics[clip,scale=0.92]{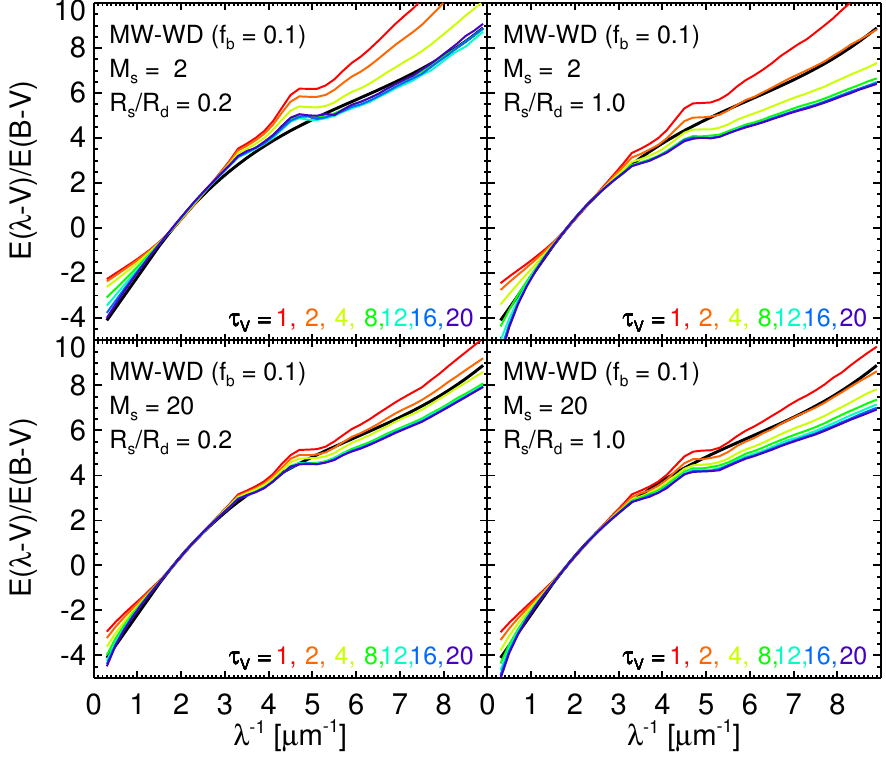}\medskip{}
\includegraphics[clip,scale=0.92]{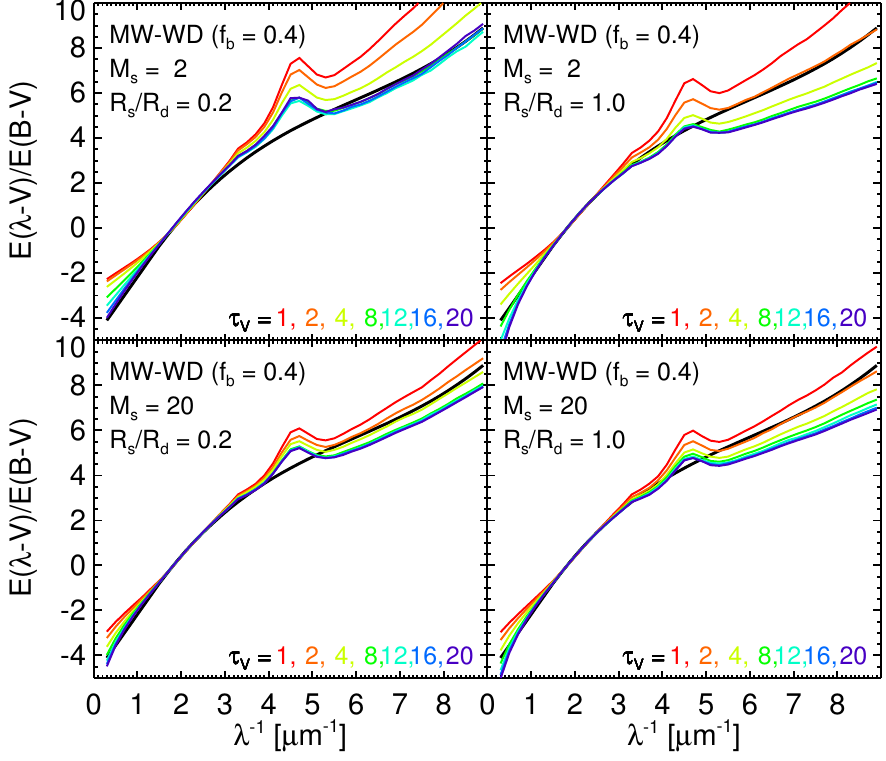}
\par\end{centering}
\caption{\label{fig17}Attenuation curves for two dust types with UV bump strengths
of (top) 0.1 and (bottom) 0.4 relative to that of the MW-WD. The Calzetti
curve is also shown in black.}
\medskip{}

\medskip{}
\end{figure}

\subsection{What Determines the Shape of Attenuation Curve?}

\label{subsec:3.3}

The above result is in contrast to the conclusion of \citet{2000ApJ...528..799W},
in which they claimed that SMC-type dust in a clumpy shell geometry
can reproduce the Calzetti curve. In this section, we will identify
the primary cause of the discrepancy. The discrepancy must be due
to differences in either geometry/density distribution or wavelength
dependence of albedo, or both. We have assumed a log-normal density
distribution while \citet{2000ApJ...528..799W} assumed a two-phase
density model. \citet{2000ApJ...528..799W} assumed a shell geometry
with stars extending from the center to 0.3 of the system radius and
dust extending from 0.3 to 1 of the system radius. In the present
study, on the other hand, stars were uniformly distributed within
a radius of $R_{{\rm s}}$. The albedo curves in the present study
were calculated using the Mie scattering theory according to the carbonaceous
and silicate dust grain models of \citet{2001ApJ...548..296W} and
\citet{2003ApJ...598.1017D}, whereas \citet{2000ApJ...528..799W}
adopted empirically determined albedo curves estimated from comparisons
of observational data of reflection nebulae with radiative transfer
models.

First, we examine the geometrical effects. Figure \ref{fig11} shows
the results obtained for the clumpy shell geometry of \citet{2000ApJ...528..799W},
but adopting the three dust types of \citet{2001ApJ...548..296W}.
The resulting attenuation curves for all three dust types are very
similar to those calculated with the turbulent media with $R_{{\rm s}}/R_{{\rm d}}\sim1$
shown in Figures \ref{fig8}, \ref{fig9} and \ref{fig10}. Therefore,
the difference in the adopted geometry and density distribution is
evidently not the main reason for the discrepancy. The other possibility
would be differences in the albedo curves. Figures \ref{fig12} and
\ref{fig13} show the attenuation curves calculated for MW-WG and
SMC-WG dust while the turbulent density distribution with $M_{{\rm s}}=6$
is adopted. For the MW-WG dust type model (Figure \ref{fig12}), curvatures
of the attenuation curves at short wavelengths of $\lambda<0.2$ $\mu$m
significantly depart not only from those obtained with the MW-WD dust,
as shown in Figure \ref{fig8}, but also from that of the Calzetti
curve. On the other hand, the overall shape of the $E(\lambda-V)/E(B-V)$
and $A_{\lambda}/A_{V}$ curves for the SMC-WG dust type (Figure \ref{fig13}),
especially with $R_{{\rm s}}/R_{{\rm d}}\sim1.0$, are consistent
with the Calzetti curve, although the $A_{\lambda}/E(B-V)$ curves
are slightly high. These results are in good agreement with the conclusion
of \citet{2000ApJ...528..799W}. We also confirmed that the attenuation
curves of \citet{2000ApJ...528..799W}\footnote{http://dirty.as.arizona.edu/dirty\_data/SGE/clumpyII/clumpyII.html}
are well reproduced when the MW-WG or SMC-WG dust and the clump shell
geometry are assumed. Therefore, the discrepancy between our results
and those of \citet{2000ApJ...528..799W} is mainly attributed to
the difference in the adopted albedo curves. In fact, \citet{2006MNRAS.370..380I}
found that wavelength dependence of the scattering albedo of the underlying
extinction curve plays a crucial role in shaping attenuation curves.
We also note that \citet{2005MNRAS.359..171I} adopted the MW-WG and
SMC-WG dust to model the attenuation curve of an edge-on galaxy using
the mega-grain approximation and a plane-parallel geometry and found
that the attenuation curves calculated with the SMC-WG dust are consistent
with the Calzetti curve (see the right panel of Figure 13 in \citet{2005MNRAS.359..171I}).

Now, it will be shown that the shape of the attenuation curve in radiative
transfer models is in fact primarily determined by the underlying
absorption curve rather than by the extinction curve. It is worthwhile
to imagine the case of perfect forward scattering, in which the scattering
asymmetry parameter $g\equiv\left\langle \cos\theta\right\rangle =1$.
In this extreme case, the scattered light will always propagate along
the initial path, equivalent to the case of no scattering. Then, the
emergent light is simply proportional to $\exp\left(-\tau_{{\rm abs}}\right)$,
where $\tau_{{\rm abs}}$ is the ``absorption'' optical depth along
the photon direction. In other words, the resulting attenuation curve
is completely determined by the absorption curve rather than by the
extinction ($\equiv$ absorption + scattering) curve. At optical and
UV wavelengths, the scattering is quite forward-directed \citep[e.g.,][]{1992ApJ...395L...5W,2003ApJ...598.1017D,2004ASPC..309...77G}.
Therefore, the attenuation curve should be mainly determined by the
shape of the adopted absorption curve. In the left panel of Figure
\ref{fig14}, the absorption curves for the MW-WD dust and the SMC-WG
dust are compared. It is clear that the two absorption curves are
very similar, except for the presence of the UV bump feature in the
MW-WD dust, while their extinction curves are significantly different
(see Figure \ref{fig3}). Therefore, if absorption is the dominant
factor, both the MW-WD dust with no UV bump and the SMC-WG dust should
equally well reproduce the Calzetti curve.

To further demonstrate our conclusion that the attenuation curve is
primarily determined by the shape of the underlying absorption curve,
we will compare two hypothetical dust types based on the MW-WD dust,
as shown in the right panel of Figure \ref{fig14}. In the first dust
type, the Drude profile for the 2175\AA\ UV bump ($3.5\le\lambda^{-1}\le5.7$
$\mu$m$^{-1}$) feature, as described in \citet{1986ApJ...307..286F,2007ApJ...663..320F},
was removed from the absorption curve. However, the scattering curve
remained the same as that of the MW-WD dust. Therefore, the extinction
curve near the UV bump is different from that of the MW dust type
(black solid lines in the right panel of Figure \ref{fig14}). In
the second dust type, the UV bump was assumed to be due to scattering
instead of absorption while keeping the extinction curve the same
as the original MW-WD extinction curve (red circles in the figure).
In other words, the Drude profile for the UV bump was removed from
the absorption curve of the MW-WD dust and the removed profile was
added to the scattering curve. The scattering phase function was kept
the same as that of the MW-WD dust. The extinction curves of these
two dust types have different shapes in the wavelength range of the
UV bump feature, but their absorption curves are exactly the same.

\begin{figure}[tp]
\begin{centering}
\medskip{}
\includegraphics[clip,scale=0.92]{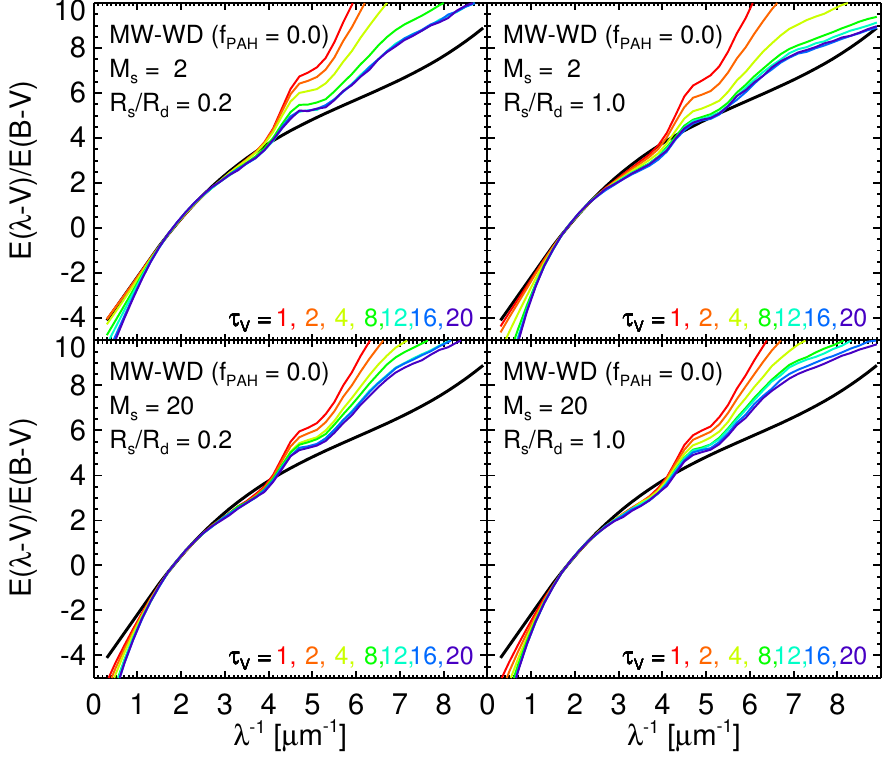}\medskip{}
\par\end{centering}
\begin{centering}
\includegraphics[clip,scale=0.92]{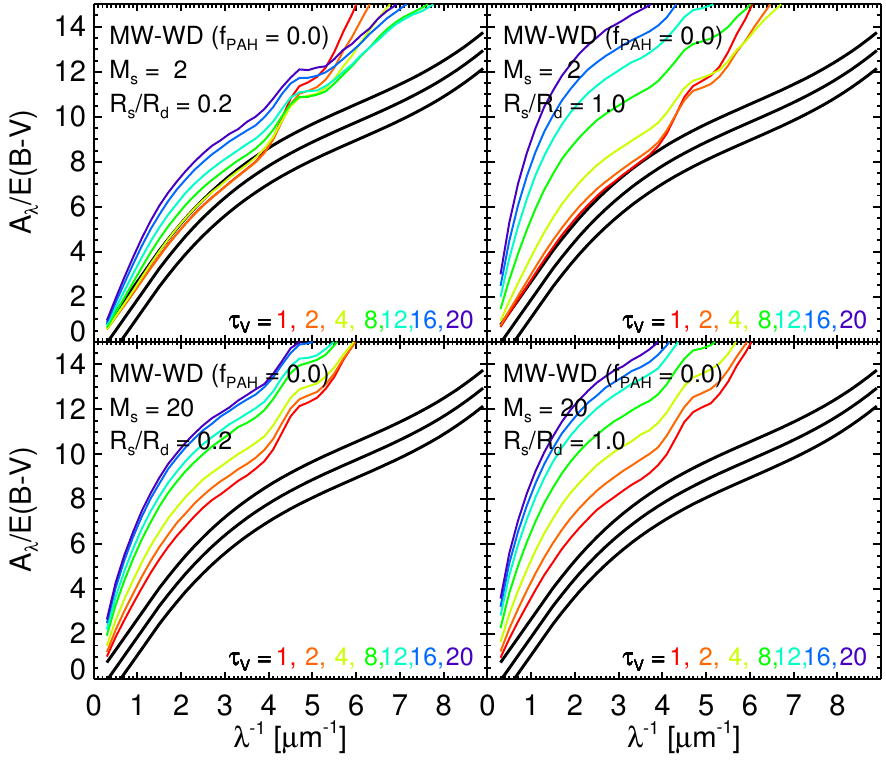}\medskip{}
\par\end{centering}
\begin{centering}
\includegraphics[clip,scale=0.92]{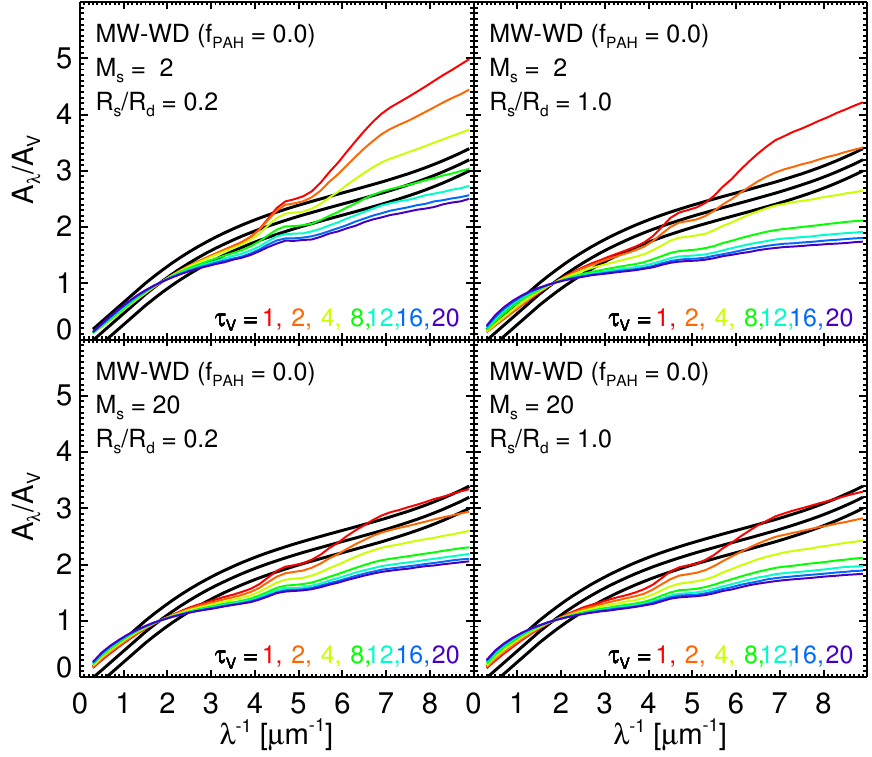}
\par\end{centering}
\caption{\label{fig18}Attenuation curves for a dust type with no PAH absorption
(UV bump + FUV extinction rise). The curves are shown in three forms:
(top) $E(\lambda-V)/E(B-V)$, (middle) $A_{\lambda}/E(B-V)$, and
(bottom) $A_{\lambda}/A_{V}$. The Calzetti curve and its uncertainty
range are also shown in black.}
\end{figure}

\begin{figure}[tp]
\begin{centering}
\medskip{}
\includegraphics[clip,scale=0.92]{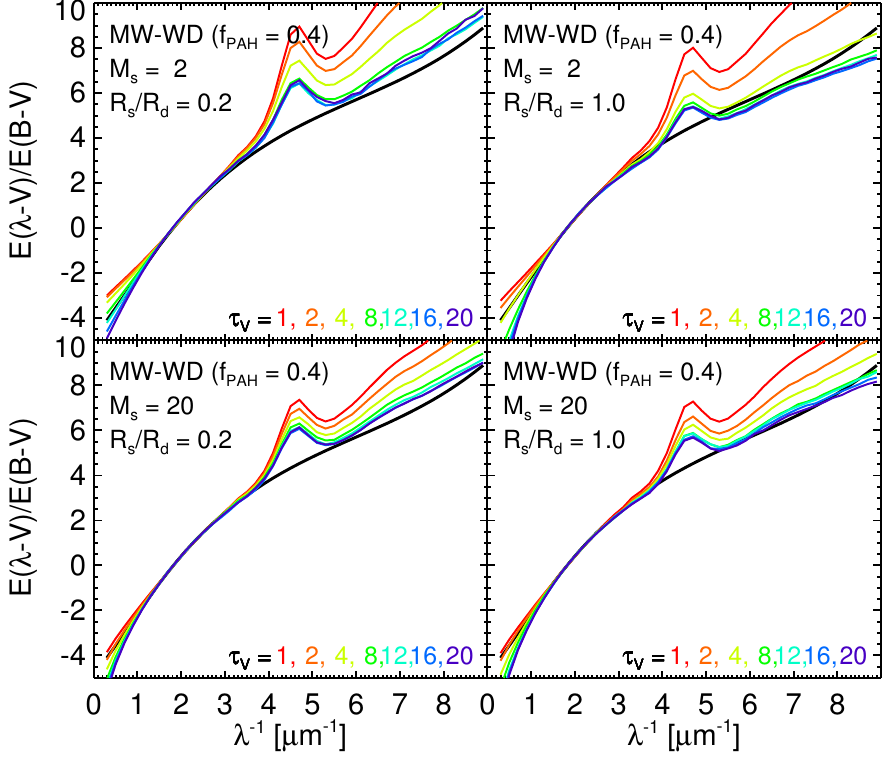}\medskip{}
\par\end{centering}
\begin{centering}
\includegraphics[clip,scale=0.92]{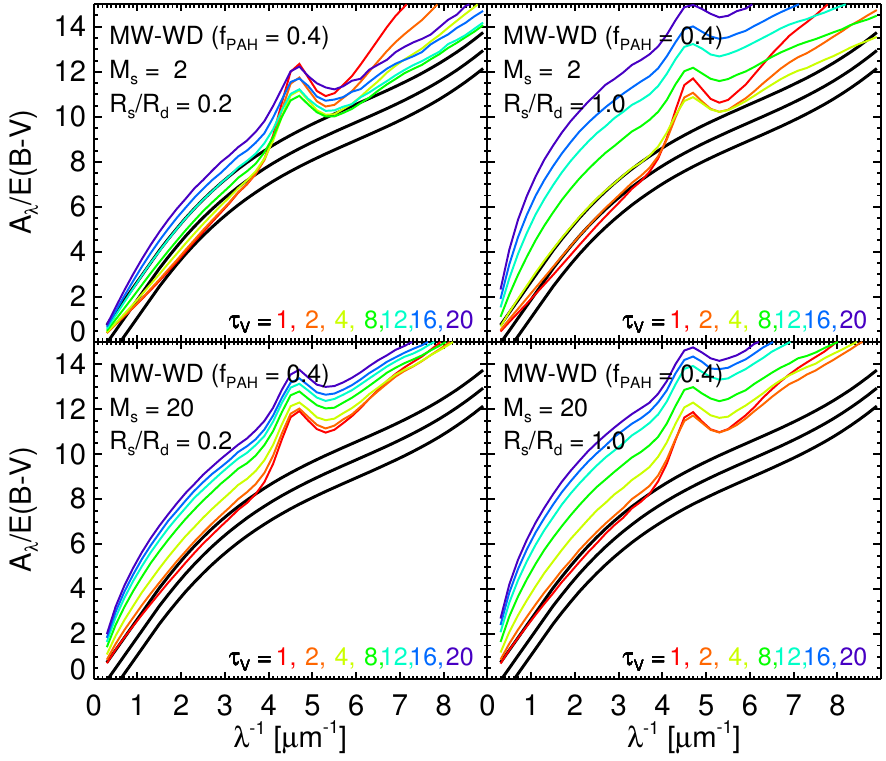}\medskip{}
\par\end{centering}
\begin{centering}
\includegraphics[clip,scale=0.92]{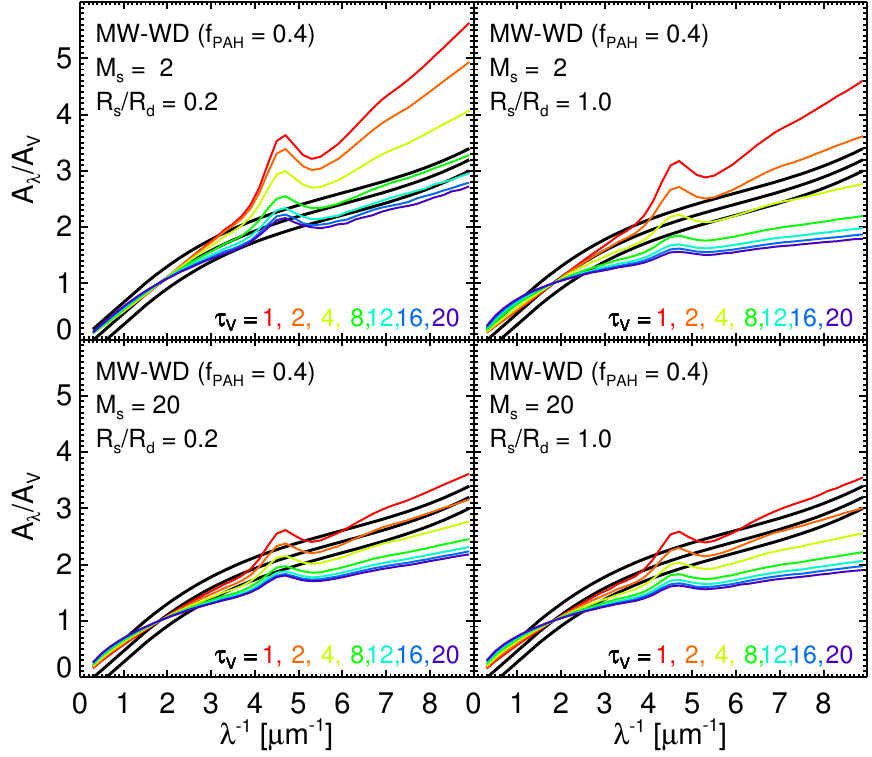}
\par\end{centering}
\caption{\label{fig19}Attenuation curves for a dust type for which the abundance
of PAHs is 0.4 of the MW-WD dust. The curves are shown in three forms:
(top) $E(\lambda-V)/E(B-V)$, (middle) $A_{\lambda}/E(B-V)$, and
(bottom) $A_{\lambda}/A_{V}$. The Calzetti curve and its uncertainty
range are also shown in black.}
\end{figure}

Radiative transfer calculations were performed to compare the attenuation
curves resulting from these two hypothetical dust types. The top four
and bottom four panels in Figure \ref{fig15} show the attenuation
curves obtained for the first and second dust types, respectively.
In the figure, the attenuation curves outside the UV bump wavelength
range are exactly the same as the models with a UV bump strength of
1 shown in Figure \ref{fig8}. The attenuation curves with the same
$R_{{\rm s}}/R_{{\rm d}}$, $M_{{\rm s}}$, and $\tau_{V}$, but with
a different dust type appear to be practically indistinguishable except
for the cases with low $M_{{\rm s}}$ and $R_{{\rm s}}/R_{{\rm d}}$
values ($M_{{\rm s}}=2$ and $R_{{\rm s}}/R_{{\rm d}}=0.2$ in the
figure). Minor differences found for the cases with low $M_{{\rm s}}$
and $R_{{\rm s}}/R_{{\rm d}}$ parameters are attributable to higher
effective extinction in the models with a scattering bump, with scattering
increasing the probability that the photon will be subsequently absorbed.
This example clearly demonstrates that the attenuation curve is mainly
determined not by the extinction curve but by the absorption curve.
Differences in the assumed scattering curves yield relatively minor
changes in attenuation curves.

Here, it should be emphasized that the second dust type, in which
the UV bump feature is assumed to be caused by scattering, should
not be regarded as a physical dust model, but a hypothetical model
arbitrarily devised to illustrate the importance of the underlying
absorption curve. We showed that variation of the scattering curve
produces only minor changes in attenuation curves, provided that the
same absorption curve is used and scattering is strongly (but realistically)
forward-directed. As $g$ is decreased, multiple scattering becomes
more important and the probability for scattered photons to be absorbed
somewhere in the medium increases, resulting in higher attenuation.
However, if $g$ is a constant over the wavelength range (for instance
$g(\lambda)=0$), the effect is to increase $A_{\lambda}$ by amounts
that are almost independent of wavelength; this resulted in no substantial
change in the normalized $A_{\lambda}/A_{V}$ curves unless the optical
depth is too high. We obtained only slightly shallower attenuation
curves ($A_{\lambda}/A_{V}$) than the original ones when isotropic
scattering was assumed over all wavelengths. However, such isotropic
scattering could lead to significantly shallower attenuation curves
at higher optical depths than those we considered in this paper.

\begin{figure}[tp]
\begin{centering}
\medskip{}
\includegraphics[clip,scale=0.92]{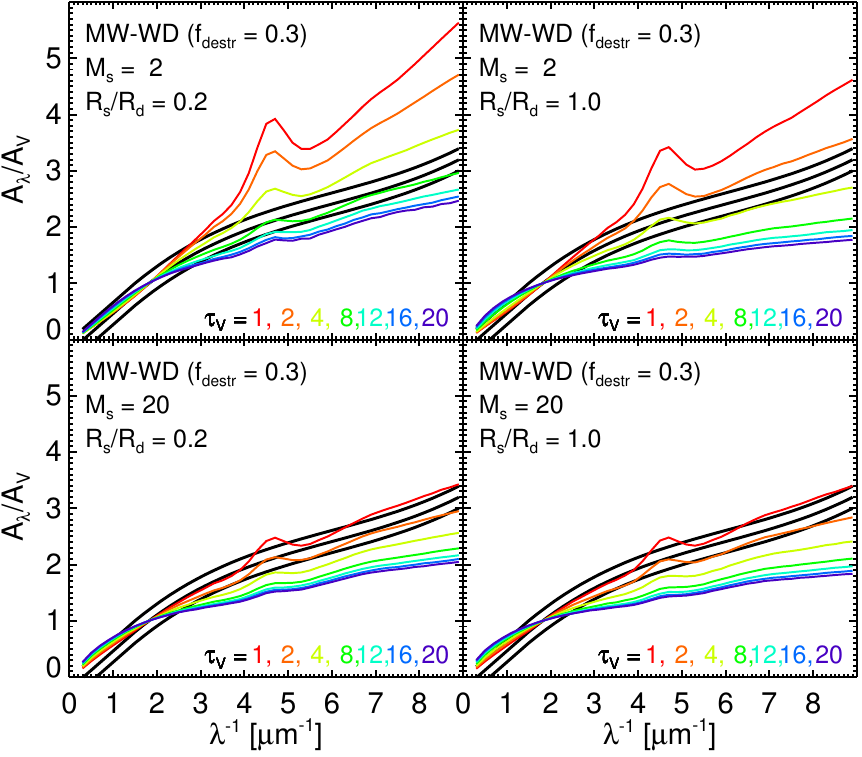}
\par\end{centering}
\begin{centering}
\medskip{}
\par\end{centering}
\begin{centering}
\includegraphics[clip,scale=0.92]{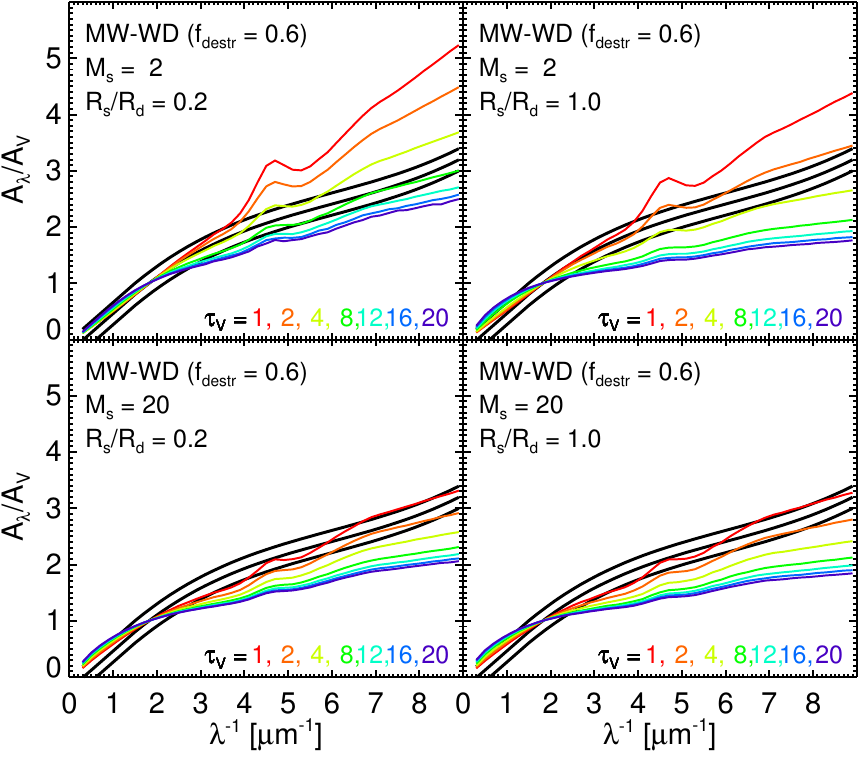}
\par\end{centering}
\caption{\label{fig20}Attenuation curves ($A_{\lambda}/A_{V}$) for the case
in which PAHs are destroyed in low-density regions having (top) 30\%
and (bottom) 60\% of dust mass. The Calzetti curve and its uncertainty
range are also shown in black.}
\medskip{}
\end{figure}

\begin{figure*}[tp]
\begin{centering}
\medskip{}
\includegraphics[clip,scale=0.92]{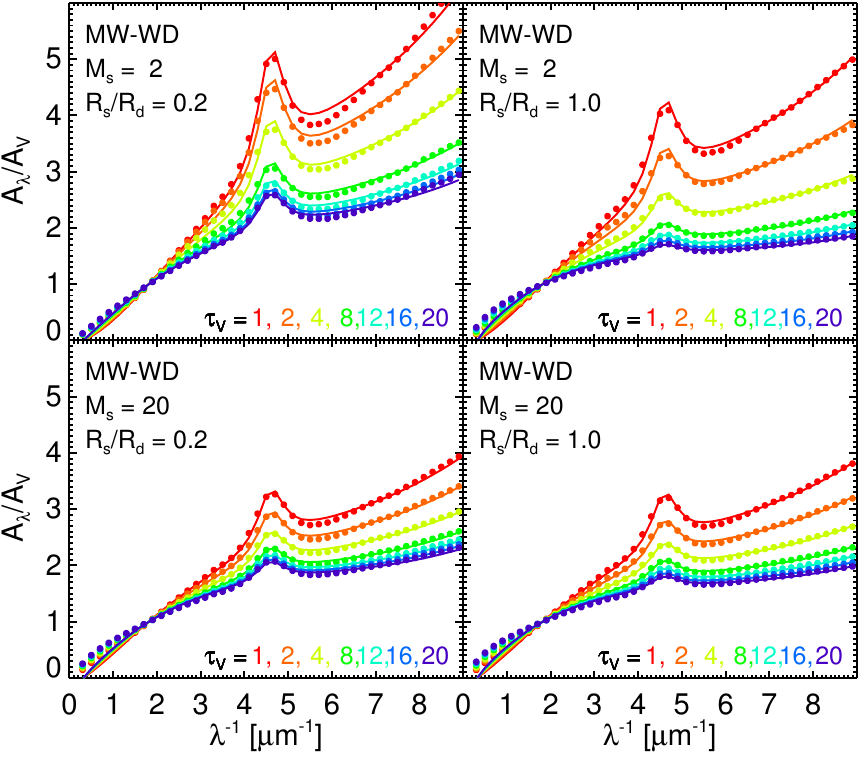}\ \ \includegraphics[clip,scale=0.92]{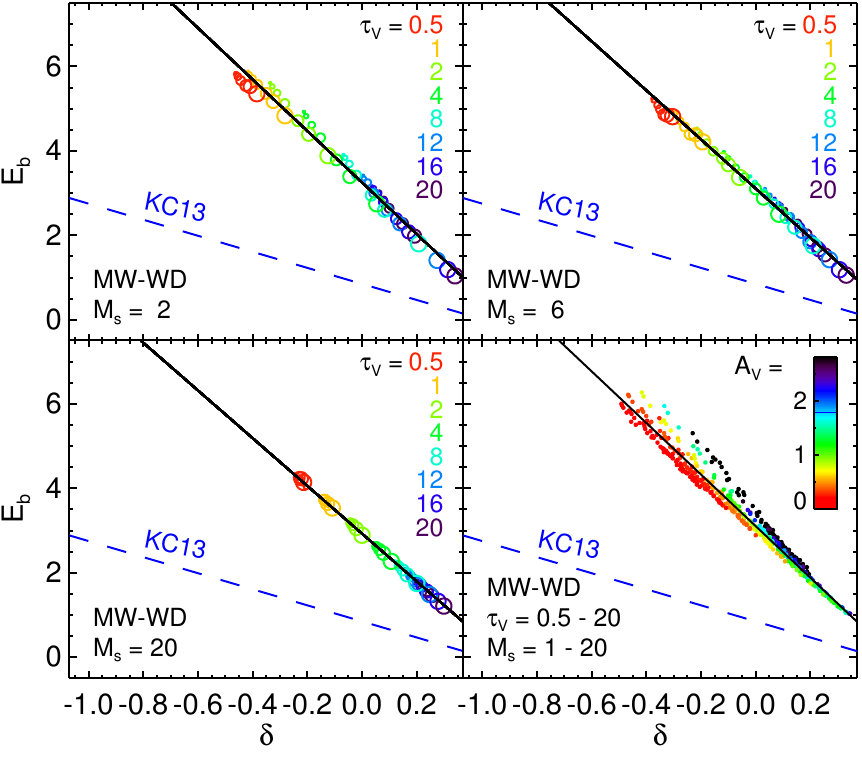}\medskip{}
\par\end{centering}
\begin{centering}
\includegraphics[clip,scale=0.92]{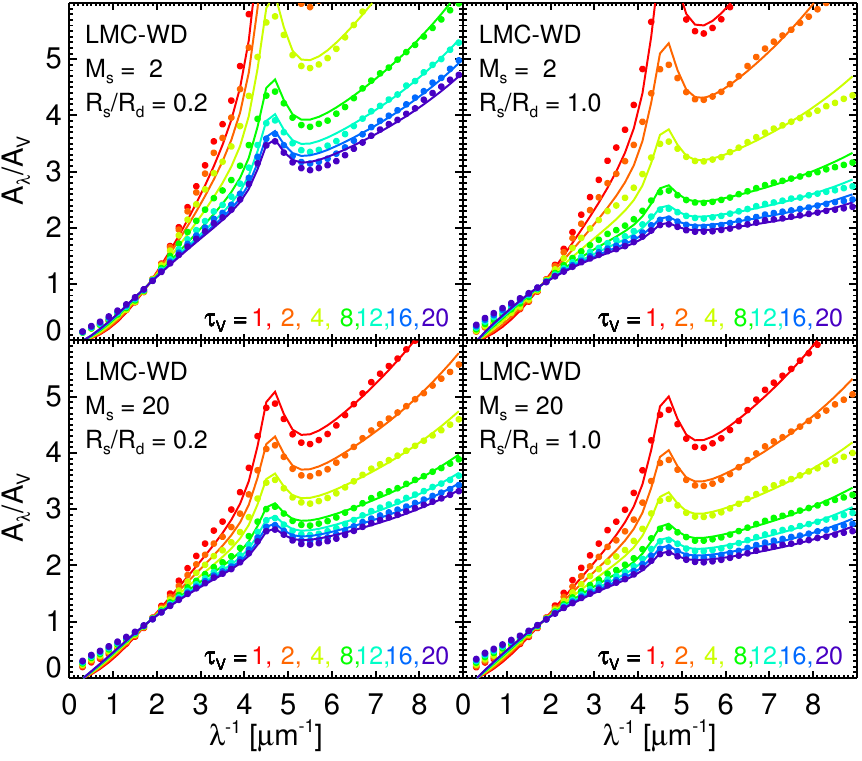}\ \ \includegraphics[clip,scale=0.92]{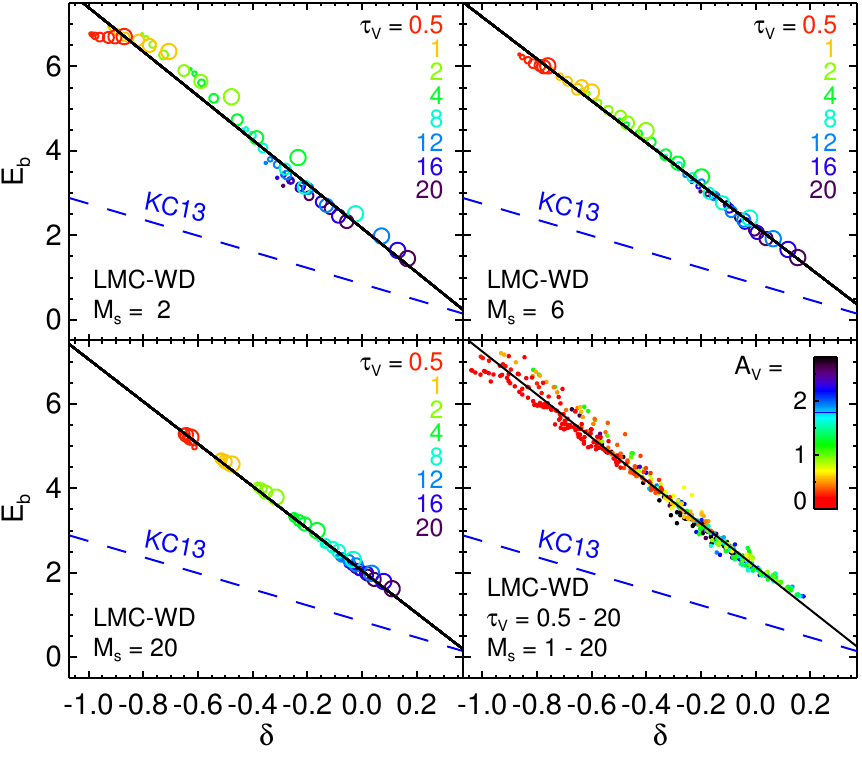}
\par\end{centering}
\caption{\label{fig21}Best-fit results to the modified Calzetti curve. The
top and bottom panels show the fitting results for the attenuation
curves calculated with the MW-WD dust and the LMC-WD dust, respectively.
In the left panels, the attenuation curves for each dust type are
shown as filled circles and the best-fitting modified Calzetti curves
are overplotted as lines with the same colors as the circles. In the
right panels, the best-fit bump strength $(E_{b})$ and slope $(\delta)$
are compared for the models with three Mach numbers ($M_{{\rm s}}=$
2, 6, and 20). The black lines in the right panels are the best-fit
linear functions for the $E_{b}$ vs $\delta$ relationships; the
linear relationship given in \citet{2013ApJ...775L..16K} is shown
as blue dashed lines. The color of the circles denotes the homogeneous
optical depth ($\tau_{V}$), except in the last panels for each dust
type. The size of the open circles in the right panels reflects the
size of the stellar distribution ($R_{{\rm s}}/R_{{\rm d}}=$ 0.0,
0.2, 0.4, 0.6, 0.8, 0.9, and 1.0). The last panel in each dust type
shows the relationship between the best-fit parameters ($E_{b}$ and
$\delta$) estimated from all attenuation curves and the color of
the dots now indicates $A_{V}$ values for individual attenuation
curves. The best-fit lines shown in last panels, which were obtained
by equally weighting the $(E_{b},\ \delta)$ values for all attenuation
curves, are $E_{b}=(3.11\pm0.01)-(6.10\pm0.04)\delta$ for the MW-WD
dust models and $E_{b}=(2.14\pm0.01)-(5.10\pm0.03)\delta$ for the
LMC-WD dust models.}
\medskip{}
\end{figure*}

\begin{figure*}[tp]
\begin{centering}
\medskip{}
\includegraphics[clip]{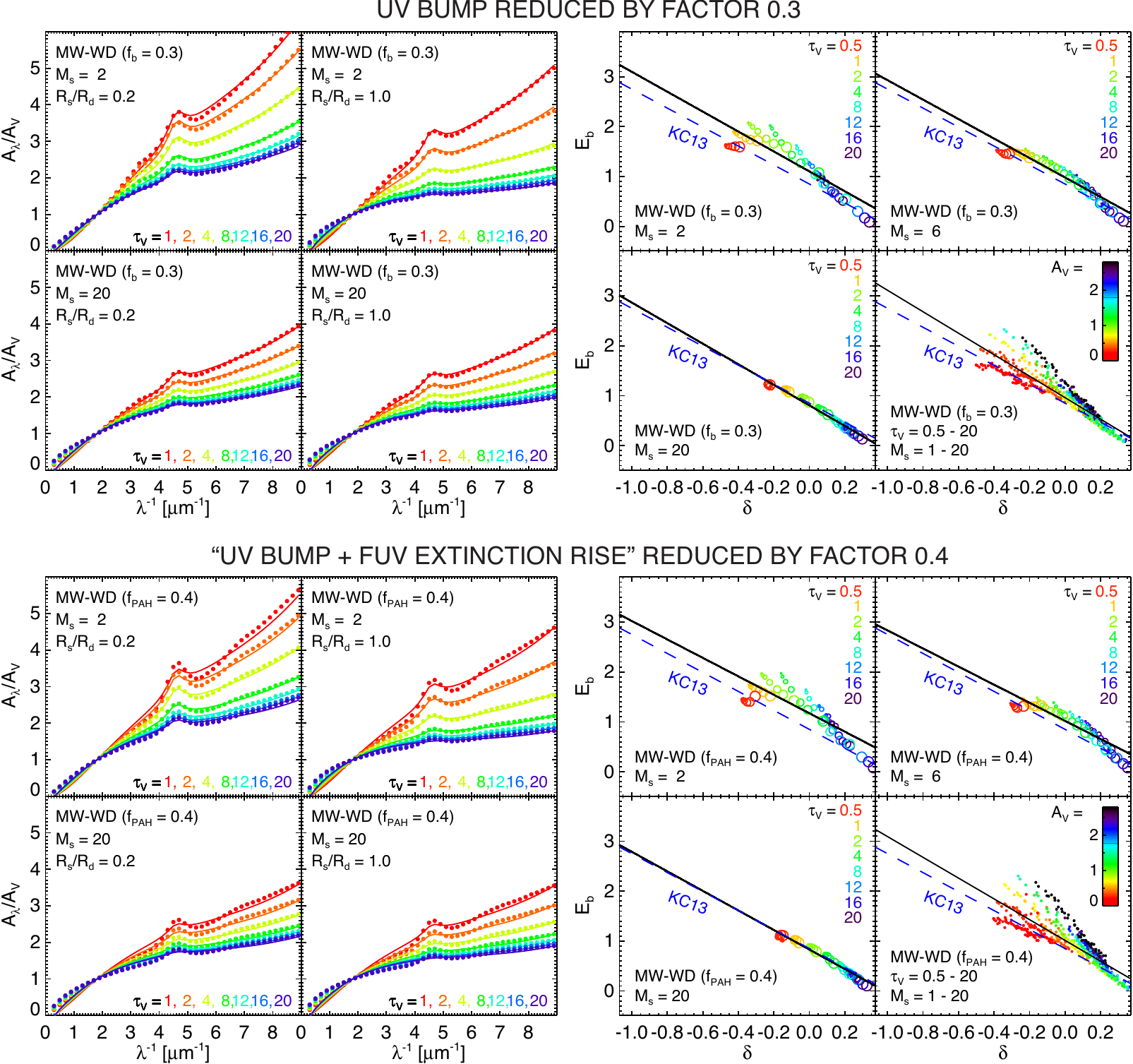}\medskip{}
\par\end{centering}
\caption{\label{fig22}Best-fit results to the modified Calzetti curves. The
top panels show the results for the dust type with a UV bump strength
of 0.3 ($f_{b}=0.3$) relative that of the MW-WD dust. The bottom
panels show the results for the dust type in which the contribution
of the PAH component is reduced by a factor 0.4 ($f_{{\rm PAH}}=0.4$).
In the last panels of each dust type, the best linear fits between
the bump strength and slope are $E_{b}=(0.96\pm0.01)-(2.15\pm0.04)\delta$
for the model with $f_{b}=0.3$ and $E_{b}=(1.01\pm0.01)-(2.08\pm0.05)\delta$
for the model with $f_{{\rm PAH}}=0.4$. The blue dashed line labeled
KC13 is the empirical linear relationship found by \citet{2013ApJ...775L..16K}.
See also the caption of Figure \ref{fig21} for further explanation.}
\medskip{}
\end{figure*}

\begin{figure*}[tp]
\begin{centering}
\medskip{}
\includegraphics[clip,scale=0.92]{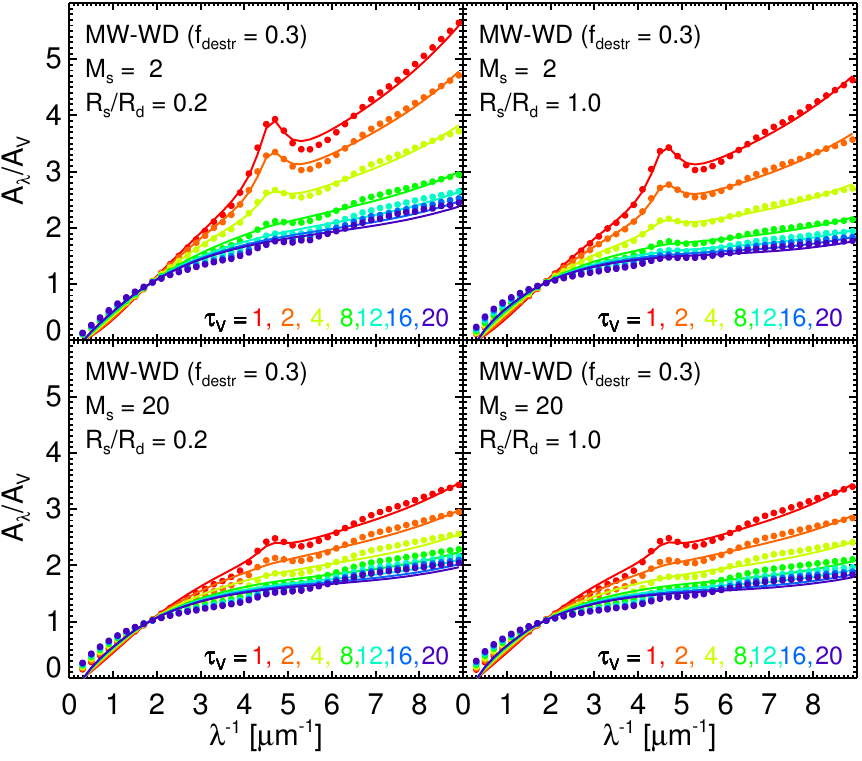}\ \ \ \includegraphics[clip,scale=0.92]{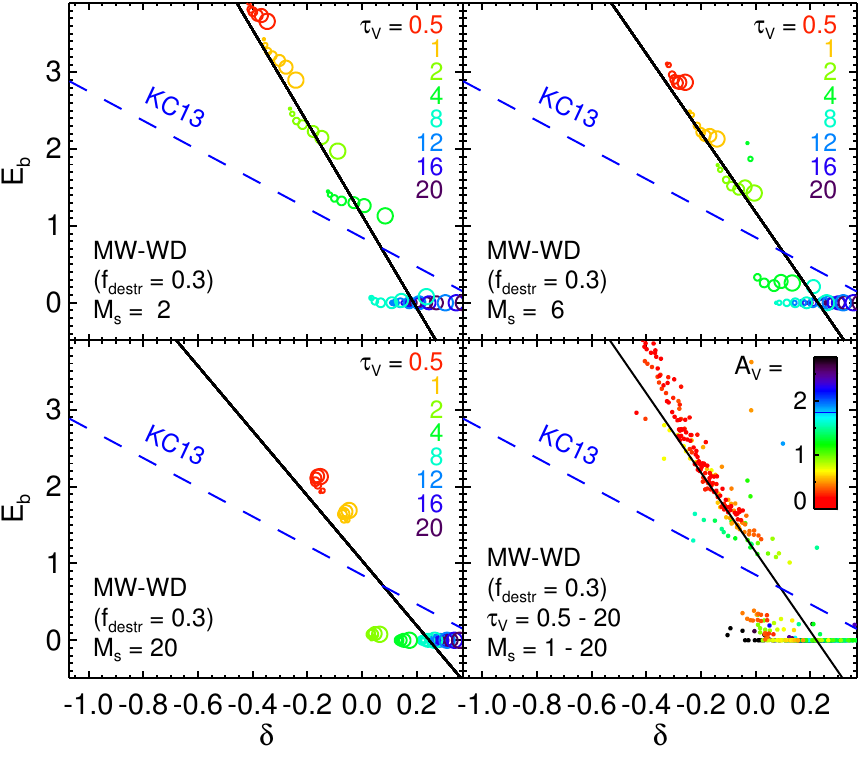}
\par\end{centering}
\begin{centering}
\medskip{}
\includegraphics[clip,scale=0.92]{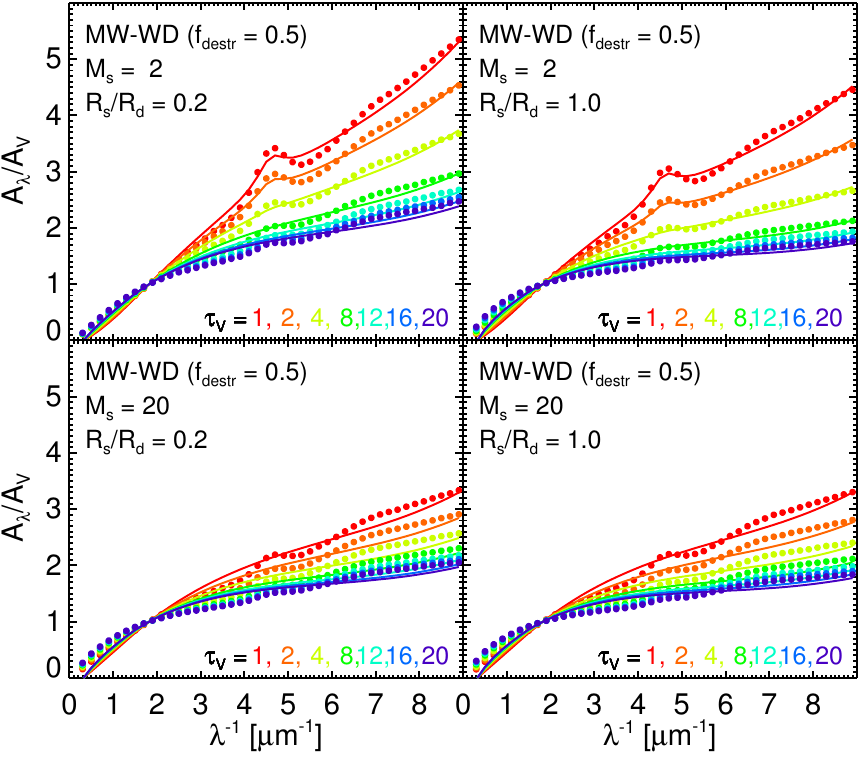}\ \ \ \includegraphics[clip,scale=0.92]{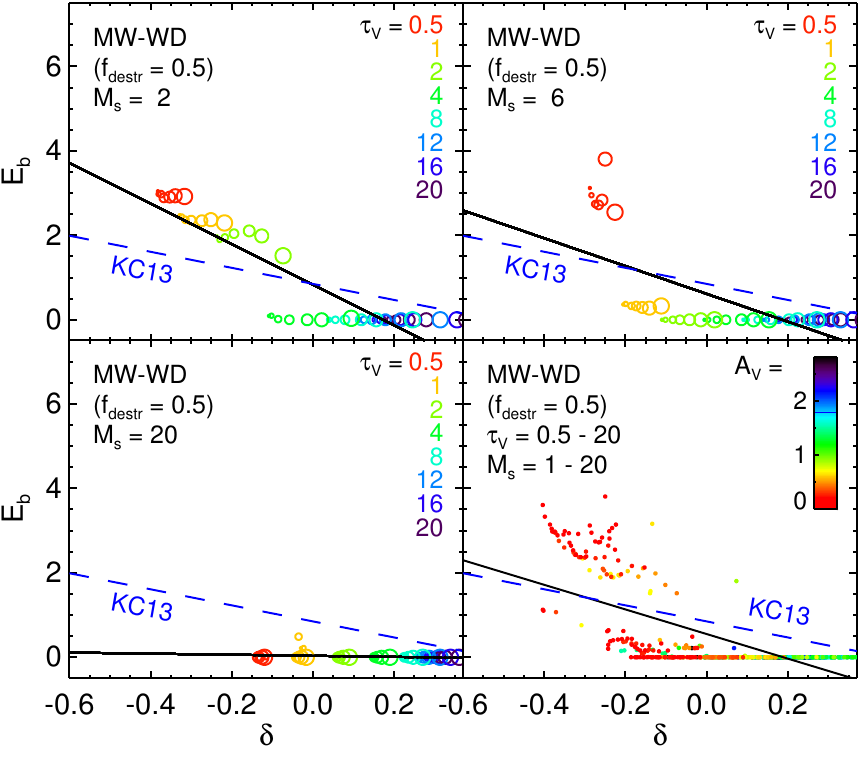}
\par\end{centering}
\caption{\label{fig23}Best-fit results to the modified Calzetti curves for
the case in which PAHs are destroyed in low-density regions of which
fractional dust mass is (top) 30\% and (bottom) 50\%. The best linear
fits between the bump strength and slope are (top) $E_{b}=(0.98\pm0.03)-(4.57\pm0.14)$
for $f_{{\rm destr}}=0.3$ and (bottom) $E_{b}=(0.39\pm0.04)-(2.00\pm0.19)$
for $f_{{\rm destr}}=0.5$. See also the caption of Figure \ref{fig21}
for further explanations.}
\medskip{}
\end{figure*}

\subsection{Variation in Strengths of the UV bump and PAH features}

\label{subsec:3.4}

The polycyclic aromatic hydrocarbon (PAHs) and/or very small carbonaceous
dust grains, which are most likely responsible for the UV absorption
bump feature, are susceptible to destruction in hot gas or strong
UV radiation field \citep{1986A&A...164..159O,1994ApJ...433..797J,1996A&A...305..602A,2010A&A...510A..36M}.
There is evidence for PAH destruction in the M17 \ion{H}{2} region
\citep{2007ApJ...660..346P}. Since the Calzetti curve for local starburst
galaxies lacks a UV bump feature, it is of interest to vary the UV
bump strength and examine resulting variations in the attenuation
curves. As discussed in Section \ref{subsec:3.3}, attenuation curves
were calculated by varying the UV bump absorption strength from 0.0
to 0.9 in steps of 0.1 relative to that of the MW-WD extinction curve.
In the left panels of Figure \ref{fig16}, the absorption curves with
a UV bump strength varying from 0.0 to 0.8 in steps of 0.2 relative
to that of the MW-WD dust are shown in grayscale. Some representative
results are shown in Figure \ref{fig17} for the MW-WD dust type with
a UV bump strength of 0.1 and 0.4. A dust extinction curve with no
UV absorption bump was shown in Section \ref{subsec:3.3} (see the
bottom panels of Figure \ref{fig15}). Although a detailed analysis
on the detectability of a weak UV bump is beyond the scope of the
present study, a UV bump feature with a strength of $\lesssim0.1$
appears to be undetectable at present, especially in the low signal-to-noise
ratio (SNR) IUE spectra used to derive the Calzetti curve and in broad-
or narrow-band photometric SED data. Even with a stronger UV bump
strength, for instance 0.4, the bump feature in the attenuation curves
with $M_{{\rm s}}\sim20$, $R_{{\rm s}}/R_{{\rm d}}\sim1$, and $\tau_{V}\gtrsim4$
would hardly be noticeable in low-SNR spectra or photometric data.

Here, it should be noted that not only the UV absorption bump but
also some part of ``the FUV extinction rise'' in the MW extinction
curve may be produced by PAHs or small carbonaceous dust grains \citep[e.g.,][]{2001ApJ...554..778L,2015ApJ...809..120M}.
If PAHs or small carbonaceous dust grains are responsible for both
the UV bump and the FUV extinction rise, it may be unphysical to arbitrarily
reduce only the UV bump strength while keeping the FUV rise the same.
Therefore, the strength of the whole model PAH contribution to the
optical-UV absorption \textendash{} including the UV bump and absorption
at $\lambda<2000$\AA\ \textendash{} was also varied from 0.0 to
0.9 relative to that of the MW-WD dust to examine its effect on the
attenuation curve. The right panels in Figure \ref{fig16} show variations
of the absorption curve as the strength of the PAH feature, including
both the UV bump and FUV extinction rise features, decreases from
0.8 to 0.0. Decreasing the strengths of the UV bump and PAH features
was done by scaling down the absorption cross-section of PAHs given
in \citet{2001ApJ...554..778L}. Figures \ref{fig18} and \ref{fig19}
show the attenuation curves calculated assuming the PAH contribution
to be 0.0 and 0.4, respectively, relative to the standard MW-WD model.
Unlike the cases (Figure \ref{fig17}) in which only the UV bump strength
is reduced, the overall shape of the attenuation curves as well as
the bump strength is changed. In general, the $E(\lambda-V)/E(B-V)$
curves become slightly steeper at UV wavelengths and the $R_{V}$
values in the $A_{\lambda}/E(B-V)$ curves become higher. This is
because the PAHs absorb not only UV photons but also optical photons
\citep{2002ApJ...572..232L}. The absorption cross section of PAHs
at $B$-band is higher than that at $V$-band and thus the reduction
of the PAHs feature results in less absorption at $B$-band than at
$V$-band. Consequently, $E(B-V)$ is decreased and $R_{V}$ is increased.
This leads the $E(\lambda-V)/E(B-V)$ curves to become steeper in
the UV and the $A_{\lambda}/E(B-V)$ curves to fail to reproduce the
Calzetti curve. On the other hand, the $A_{\lambda}/A_{V}$ curves
become in general shallower. The $A_{\lambda}/A_{V}$ curves without
PAHs in Figure \ref{fig18} are consistent with the Calzetti curve
outside the UV bump wavelength range, but show a wide dip-like feature
near the UV bump. This apparent dip is mainly due to a steep increase
in the absorption (a drop in the albedo) of the silicate grains with
decreasing wavelength near 2000\AA.The $A_{\lambda}/A_{V}$ curves
with the PAH feature strength of 0.4 in Figure \ref{fig19} are very
similar to the Calzetti curve except for the presence of a weak UV
bump. The attenuation curves have slightly higher $R_{V}$ values
compared to the Calzetti curve. We note that the shapes of the $E(\lambda-V)/E(B-V)$
and $A_{\lambda}/E(B-V)$ curves in Figures \ref{fig18} and \ref{fig19}
are direct consequences of the $[A_{\lambda}^{{\rm abs}}/(A_{B}^{{\rm abs}}-A_{V}^{{\rm abs}})]_{\ast}$
curves shown in the bottom right panel of Figure \ref{fig16}. The
$[A_{\lambda}^{{\rm abs}}/(A_{B}^{{\rm abs}}-A_{V}^{{\rm abs}})]_{\ast}$
curves with weaker strength of PAHs have steeper slopes in the UV
and higher values of $[A_{V}^{{\rm abs}}/(A_{B}^{{\rm abs}}-A_{V}^{{\rm abs}})]_{\ast}$.
It can also be found that the properties of the $A_{\lambda}/A_{V}$
curves are mainly caused by the shape of the $[A_{\lambda}^{{\rm abs}}/A_{V}^{{\rm abs}}]_{\ast}$
curves shown in the top right panel of Figure \ref{fig16}. 

In the above models, the strength of the UV bump feature was assumed
to be uniformly reduced over the whole dusty medium. However, the
carriers of the features may be preferentially destroyed only in hot
gas or near strong UV stars. To examine this possibility, we consider
a case where PAHs are completely destroyed in regions with less than
a certain critical density but remain unaffected in higher-density
regions. The critical density $\rho_{{\rm c}}$ was selected such
that the fraction of cumulative mass defined by
\begin{equation}
f_{{\rm destr}}\equiv\left.\sum_{\rho_{i}\le\rho_{{\rm c}}}\rho_{i}\middle/\sum_{i}\rho_{i}\right.,\label{eq:5}
\end{equation}
 varies from 0.1 to 0.9 in steps of 0.1, where $\rho_{i}$ is density
of each cell. Thus a fraction $f_{{\rm destr}}$ of the PAHs are assumed
to have been destroyed. Figure \ref{fig20} shows the attenuation
curves for $f_{{\rm destr}}=0.6$ and 0.3. As can be noted in the
figure, the UV bump feature is much weaker and the shape is grayer
than the former models in which the same amount of PAHs was uniformly
depleted over the entire volume (e.g., the models with $f_{{\rm PAH}}=1-f_{{\rm destr}}$).
This is because the low-density regions with no PAHs take up most
of the volume although the regions contain only a small fraction of
total dust mass, as shown in Figure \ref{fig2}; in other words, high-density
regions containing PAHs and most of dust mass occupy only a relatively
small volume. Consequently, most photons pass through locations with
no PAHs and the resulting attenuation curves have much weaker PAH
features compared to the case with uniformly mixed PAHs.

\subsection{Starlight Originating from Dense Regions}

\label{subsec:3.5}

One more thing to be considered is that bright early-type stars from
which most UV light originates are mainly located near spatially-localized
dense regions. We therefore examined the attenuation curves of the
case in which the stellar source is uniformly distributed but only
in regions where the dust density is above some critical value. The
fraction of cumulative mass measured from the highest density cell,
as opposed to Equation (\ref{eq:5}), was varied from 0.1 to 0.9 in
steps of 0.1. However, we found no significant differences from the
models with a uniform stellar distribution shown in Figure \ref{fig8};
the attenuation curves showed virtually no change in the slope ($\delta$)
defined in next Section and the UV bump strength ($E_{b}$) was only
slightly lowered, by less than 5\%. For test purpose, we also calculated
models in which photons originate only from low-density regions. But,
no significant differences were found; the slope $\delta$ slightly
increased, by $\sim4$\% (shallower attenuation curve). This is because
most of starlight attenuation occurs while photons pass through dense
regions. The chance of passing through high-density regions is not
significantly altered by the distribution of stellar locations if
one is mainly concerned with attenuation curves averaged over all
directions.

Therefore, the UV absorption bump strength is mainly determined by
the abundance of the UV bump carriers. As can be noticed in Figure
\ref{fig8} and next Section (Figure \ref{fig21}), the UV bump can
be almost completely suppressed through radiative transfer effects
at high optical depths ($\tau_{V}\gtrsim20$); the suppression of
the bump is prominent in the $A_{\lambda}/A_{V}$ attenuation curves,
but is less clear in forms of $A_{\lambda}/E(B-V)$ and $E(\lambda-V)/E(B-V)$.
However, the attenuation curves with a strongly-suppressed UV bump
are much shallower than the Calzetti curve. The result is consistent
with that of \citet{2000ApJ...528..799W} in that the Calzetti curve
can be reproduced only with an extinction curve with no, or very weak,
UV bump. This point will be discussed further in Section \ref{subsec:4.1}.

\begin{figure*}[tp]
\begin{centering}
\medskip{}
\par\end{centering}
\begin{centering}
\includegraphics[clip,scale=0.95]{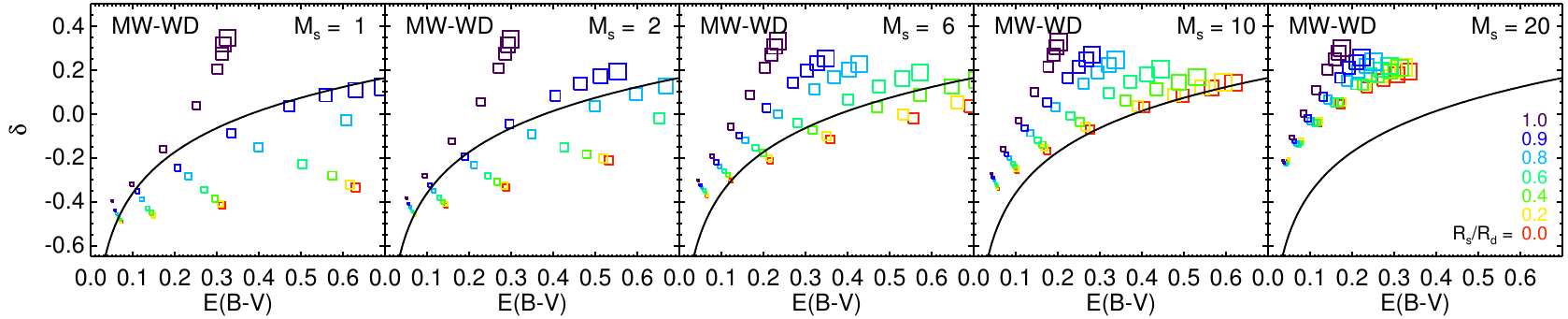}
\par\end{centering}
\begin{centering}
\medskip{}
\par\end{centering}
\begin{centering}
\includegraphics[clip,scale=0.95]{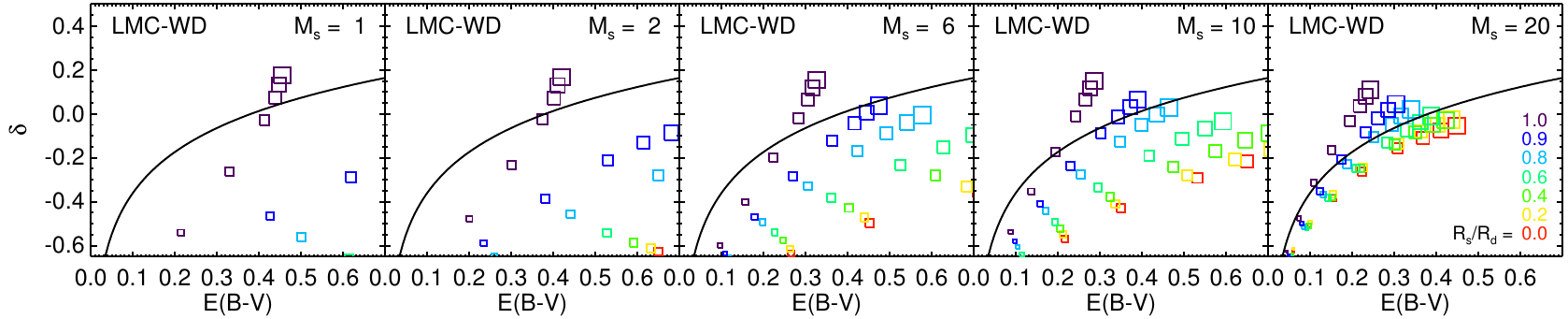}
\par\end{centering}
\begin{centering}
\medskip{}
\par\end{centering}
\begin{centering}
\includegraphics[clip,scale=0.95]{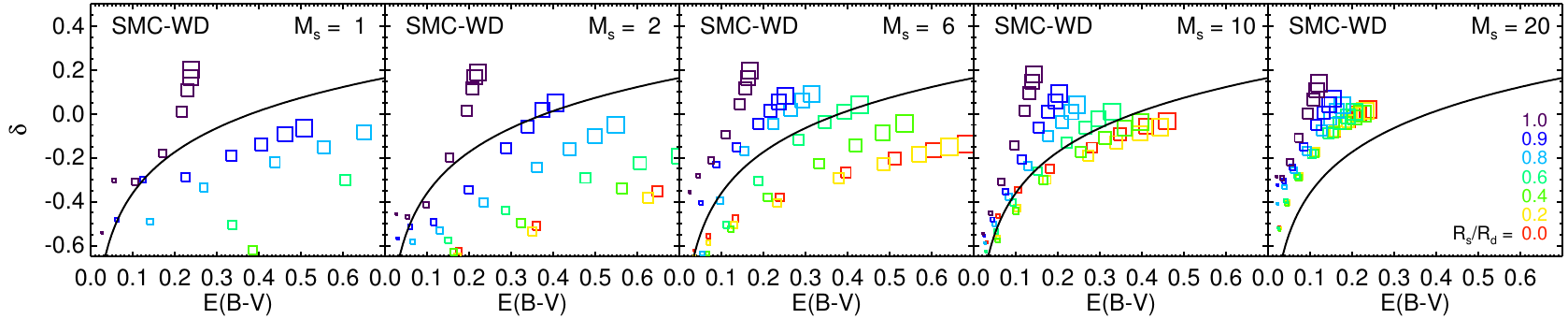}
\par\end{centering}
\caption{\label{fig24}Correlation between the attenuation curve slope $\delta$
and $E(B-V)$ for (top) the MW, (middle) LMC, and (right) SMC dust
types of \citet{2001ApJ...548..296W}. The color of the squares denotes
the size of the stellar distribution ($R_{s}/R_{d}$ = 0.0, 0.2, 0.4,
0.6, 0.8, 0.9, and 1.0), as shown in the rightmost panels. The size
of the squares reflects the homogeneous optical depth $\tau_{V}$
(= 0.5, 1, 2, 4, 8, 12, 16, and 20). Smaller square corresponds to
lower $\tau_{V}$. The black curve denotes the relationship for $\delta$
and $E(B-V)$ given by \citet{2015arXiv151205396S}. Note that color
change in this figure shows the variation of $R_{{\rm s}}/R_{{\rm d}}$
whereas it represents the variation of $\tau_{V}$ in Figures \ref{fig21},
\ref{fig22}, and \ref{fig23}.}
\medskip{}
\end{figure*}

\begin{figure*}[tp]
\begin{centering}
\medskip{}
\includegraphics[clip,scale=0.95]{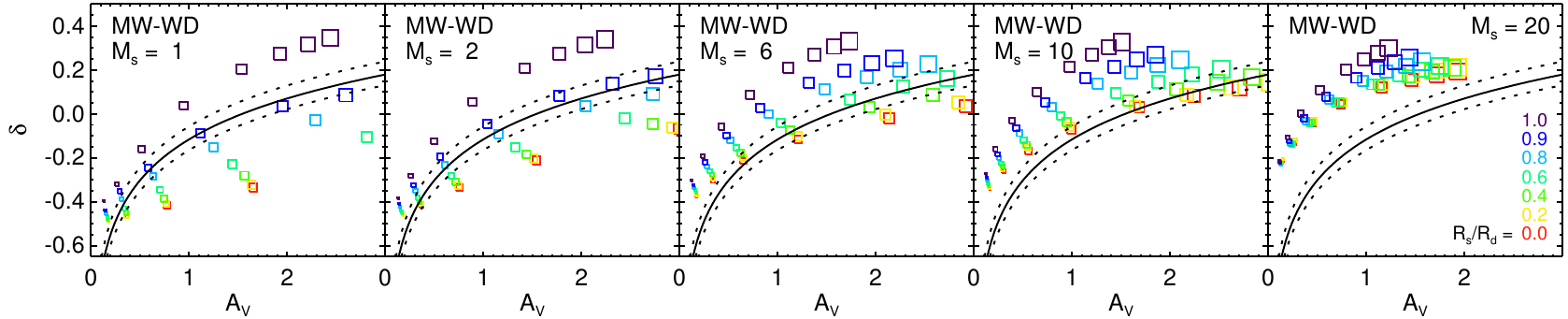}
\par\end{centering}
\begin{centering}
\medskip{}
\par\end{centering}
\begin{centering}
\includegraphics[clip,scale=0.95]{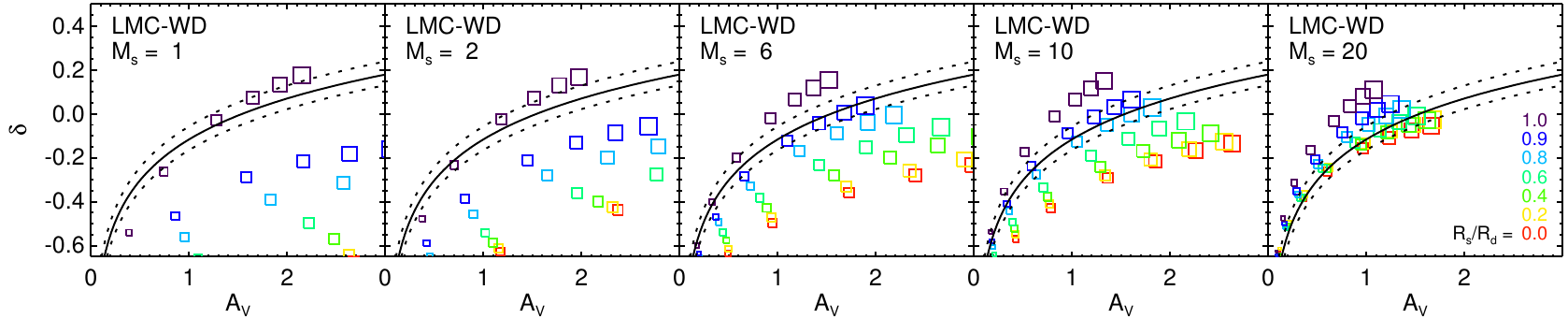}
\par\end{centering}
\begin{centering}
\medskip{}
\par\end{centering}
\begin{centering}
\includegraphics[clip,scale=0.95]{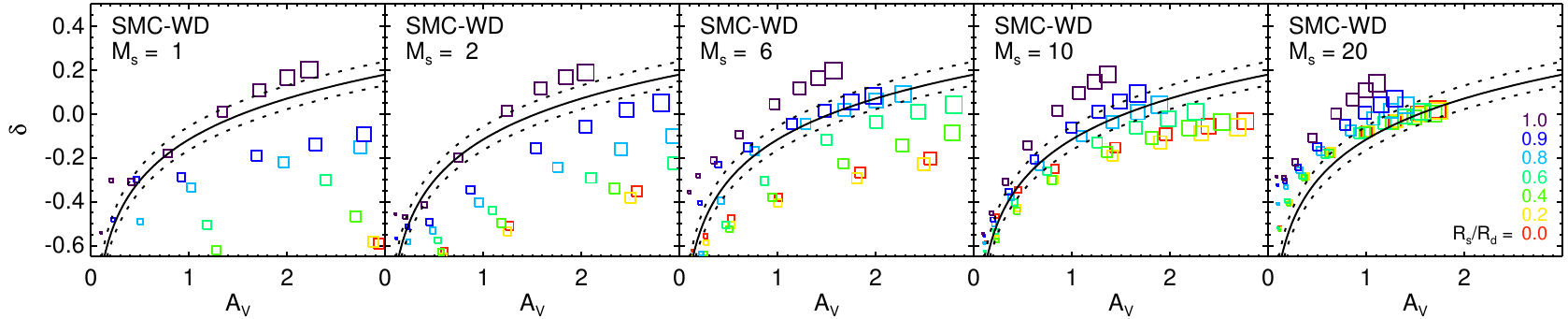}
\par\end{centering}
\caption{\label{fig25}Correlation between the attenuation curve slope $\delta$
and $A_{V}$ for (top) the MW, (middle) LMC, and (right) SMC dust
types of \citet{2001ApJ...548..296W}. The solid curve shows the $\delta$
vs $A_{V}$ curve calculated from Equation (\ref{eq:salmon}) assuming
$A_{V}=4.05E(B-V)$ as in the Calzetti curve. The dotted curves are
upper and lower bounds defined by $R_{V}=4.05\pm0.80$. See also the
caption of Figure \ref{fig24} for further explanation.}
\medskip{}
\end{figure*}

\subsection{Correlation between Slope and UV Bump Strength}

\label{subsec:3.6}

The model attenuation curves reveal a large variation both in overall
shape and UV bump strength; the slope of the attenuation curve tends
to correlate with the UV bump strength. In other words, the UV bump
can be suppressed through radiative transfer effects, but at the expense
of a gray attenuation \citep{2000ApJ...528..799W}. \citet{2013ApJ...775L..16K}
found a strong correlation between the attenuation curve slope and
UV bump strength from a large sample of star-forming galaxies, which
is qualitatively consistent with radiative transfer calculations.
To quantitatively compare with our radiative transfer model results,
we adopt a modified Calzetti function proposed by \citet{2009A&A...507.1793N},
which is represented by

\begin{equation}
A_{\lambda}=\frac{A_{V}}{k(V)}\big[k(\lambda)+D(\lambda)\big]\left(\frac{\lambda}{\lambda_{V}}\right)^{\delta},\label{eq:modified}
\end{equation}
where $k(\lambda)$ indicates the Calzetti curve and $\delta$ is
employed to allow variation of the slope of the attenuation curve:
$\delta>0$ corresponds to shallower attenuation. The Calzetti curve
\citep{2000ApJ...533..682C} is given by
\begin{eqnarray}
k(\lambda) & = & 2.659(-1.857+1.040/\lambda)+R_{V},\nonumber \\
 &  & \ \ \ \ \ \ \ \ \ \ \ \ \ \ \ \ \ \ \ \ \ \ \ \ \ \ \ \ \ \ \ \ 0.63\mu{\rm m}\le\lambda\le2.20\mu{\rm m};\nonumber \\
 & = & 2.659(-2.156+1.509/\lambda-0.198/\lambda^{2}+0.011/\lambda^{3})+R_{V},\nonumber \\
 &  & \ \ \ \ \ \ \ \ \ \ \ \ \ \ \ \ \ \ \ \ \ \ \ \ \ \ \ \ \ \ \ \ 0.12\mu{\rm m}\le\lambda\le0.63\mu{\rm m},\label{eq:calzetti}
\end{eqnarray}
where $R_{V}=4.05\pm0.80$. The UV bump feature is represented by
Drude profile defined by

\begin{equation}
D(\lambda)=\frac{E_{b}(\lambda\Delta\lambda)^{2}}{\left(\lambda^{2}-\lambda_{0}^{2}\right)^{2}+\left(\lambda\Delta\lambda\right)^{2}},\label{eq:Drude}
\end{equation}
where $\lambda_{0}$, $\Delta\lambda$, and $E_{b}$ are the central
wavelength, width, and strength (peak height) of the profile. The
central wavelength of the UV bump is assumed to be $\lambda_{0}=0.2175$
$\mu$m and the width of the bump is $\Delta\lambda=0.035$ $\mu$m,
which were found by \citet{2009A&A...507.1793N} from a stack of spectra
of high-redshift galaxies. Using this profile, \citet{2013ApJ...775L..16K}
derived a linear relation between $\delta$ and $E_{b}$ given by
\begin{equation}
E_{b}=(0.85\pm0.09)-(1.9\pm0.4)\delta.\label{eq:KC13}
\end{equation}

We fitted our attenuation curves with the modified Calzetti function
(Equation \ref{eq:modified}). Figure \ref{fig21} shows the best-fits
for (top) the MW-WD and (bottom) LMC-WD dust type models. Left panels
show the attenuation curves (dot symbols) together with the best-fit
curves (connected lines). It is clear that the modified Calzetti curve
(Eq. \ref{eq:modified}) fairly well reproduces the attenuation curves
calculated with both dust types.

The right panels in Figure \ref{fig21} show the correlations between
the two best-fit parameters ($E_{b}$ and $\delta$), in which the
black and blue lines represent the best linear-fit functions of $E_{b}$
vs $\delta$ from our results and the linear relationship of \citet{2013ApJ...775L..16K},
respectively. The best-fit line shown in the last panel of each dust
type was obtained by equally weighting all the $(E_{b},\ \delta)$
pairs estimated from attenuation curves. The basic properties discussed
in the previous sections are confirmed in the figure. Shallowness
of the attenuation curve anti-correlates with the UV bump strength.
The attenuation curves are shallower (more positive $\delta$) and
have weaker UV bumps (smaller $E_{b}$) for higher $R_{s}/R_{d}$,
$\tau_{V}$ and $M$. The attenuation curves derived from the LMC-WD
dust are slightly steeper than those from the MW-WD dust and have
slightly stronger UV bump strength for a given set of parameters ($R_{s}/R_{d}$,
$\tau_{V}$ and $M$). This is simply because the extinction curve
$[A_{\lambda}/A_{V}]_{\ast}^{{\rm LMC}}$, relative to $[A_{V}]_{\ast}^{{\rm LMC}}$,
of the LMC dust is steeper and has a stronger UV bump feature than
for the MW curve $[A_{\lambda}/A_{V}]_{\ast}^{{\rm MW}}$, as shown
in the left bottom panel of Figure \ref{fig3}. However, note that
the LMC extinction curve $[A_{\lambda}/N_{{\rm H}}]_{\ast}^{{\rm LMC}}$
per hydrogen column density has a weaker UV bump than the MW curve
$[A_{\lambda}/N_{{\rm H}}]_{\ast}^{{\rm MW}}$. The slopes for the
LMC-WD dust range from $\sim-1.5$ to $\sim0.2$, which are on average
smaller (steeper) than those of \citet{2013ApJ...775L..16K} ranging
from $\sim-0.5$ to $\sim0.3$. On the other hand, the MW-WD dust
produces slopes that are consistent with the observational result
of \citet{2013ApJ...775L..16K}. The UV bump strength in both dust
types is, however, much higher than the observational result. We also
note that the UV bump strength ($E_{b}$) for the LMC-WD dust is higher
than for the MW-WD for a given slope ($\delta$); in other words,
$E_{b}$ for the LMC-WD dust increases faster than that of the MW-WD
dust as $\delta$ decreases.

The linear relationships between $E_{b}$ and $\delta$ in Figure
\ref{fig21} are steeper than the result of \citet{2013ApJ...775L..16K},
indicating that a weaker UV bump strength is required to explain their
result. Figure \ref{fig22} shows the best-fit results for the models
with (top) a UV bump strength of 0.3 and (bottom) a PAH feature strength
of 0.4, which both appear to well reproduce the relation of \citet{2013ApJ...775L..16K}.
Note that the equivalent optical depth of $\tau_{V}=20$ mostly corresponds
to $A_{V}$ values less than 3 except the cases with the lowest $R_{s}/R_{d}$
and $M_{s}$ values, as seen in Figure \ref{fig5}. This upper limit
of $A_{V}$ in our results is consistent with the range of the best-fit
$A_{V}$ estimated in \citet{2013ApJ...775L..16K}. We also note that
there is a correlation between $\delta$ and $A_{V}$ in their Figure
2(a) if a single red data point at $(\delta,\;E_{b})\sim(-0.44,\;2.3)$
is ignored, implying a correlation of the slope of the attenuation
curve with the effective optical depth $\tau_{V}$. This trend is
consistent with that shown in the last (lower right) panels for each
dust type in Figure \ref{fig22}, in which every dot corresponds to
a single radiative transfer model and colors of the dots reflect $A_{V}$
of the individual models. Overall correlation between $A_{V}$ and
$\delta$ is clearly shown in the figure. Therefore, our results strongly
suggest that the anti-correlation between $E_{b}$ and $\delta$ found
in \citet{2013ApJ...775L..16K} may be due to radiative transfer effects.
The abundance of the UV bump carriers or PAHs in the sample of \citet{2013ApJ...775L..16K}
would be 30\% or 40\% of that of the MW-WD dust, if the carriers are
assumed to be uniformly mixed over the dusty ISM.

We also fitted the attenuation curves obtained for the cases in which
the PAHs are destroyed only in low-density regions. Figure \ref{fig23}
shows the results for two cases in which the PAHs are destroyed in
low-density regions of which the accumulated mass fraction ($f_{{\rm destr}}$)
is 0.3 and 0.5. In this model type, the relation between $E_{b}$
and $\delta$ could not be uniquely quantified because of a bi-modality
as shown in Figure \ref{fig23}. The best-fit $E_{b}$ values for
most of the models with $f_{{\rm destr}}\gtrsim0.5$ were found to
be approximately zero, indicating no bumps, while most of the models
with $f_{{\rm destr}}\lesssim0.2$ gave steeper linear relationships
between $E_{b}$ and $\delta$ than that of \citet{2013ApJ...775L..16K}.
For the intermediate cases of $f_{{\rm destr}}=0.3$ and 0.4, the
attenuation curves with high $M_{{\rm s}}$ and $\tau_{V}$ gave best-fit
values of $E_{b}\approx0$. On the other hand, relationships between
$E_{b}$ and $\delta$ steeper than that of \citet{2013ApJ...775L..16K}
were found for low $M_{{\rm s}}$ and $\tau_{V}$, as shown for $M_{{\rm s}}=2$
in Figure \ref{fig23}. The models that gave $\delta\sim0$ seem to
indicate that galaxies having PAHs strongly depleted in low-density
regions would lack the UV bump feature in their attenuation curves.
However, these models with $\delta\sim0$ actually showed wavy structures
in the attenuation curves that slightly deviates from the modified
Calzetti curve. We also note that most of the volume has no PAHs in
these models even for $f_{{\rm destr}}=0.3$, as shown in Figure \ref{fig2}.
This model for the PAH abundance does not appear to be consistent
with observations.

\subsection{Correlation between Slope and Color Excess}

\citet{2015arXiv151205396S} applied a Bayesian analysis to a photometric
sample of galaxies at $z\sim2$ and found an average relation between
the attenuation curve slope ($\delta$) and the $E(B-V)$ color excess
given by 
\begin{equation}
\delta=(0.62\pm0.05)\log\left[E(B-V)\right]+0.26\pm0.02.\label{eq:salmon}
\end{equation}
The equation was derived by binning the joint posterior distribution
$P(E(B-V),\delta)$ of $E(B-V)$ and $\delta$ and then fitting the
median $\delta$ for each bin of $E(B-V)$ as a function of $E(B-V)$.
They also showed that radiative transfer models of \citet{2000ApJ...528..799W}
are consistent with this equation.

In Sections \ref{subsec:3.1} and \ref{subsec:3.2}, we noted that
attenuation curves become grayer as the homogeneous optical depth
$\tau_{V}$ increases. Therefore, the observed correlation between
$\delta$ and $E(B-V)$ may be a natural consequence of radiative
transfer effects. To compare Equation (\ref{eq:salmon}) with our
results, Figure \ref{fig24} shows the $\delta$ vs $E(B-V)$ relations
for (top) the MW-WD, (middle) LMC-WD, and (bottom) SMC-WD dust types.
In the figure, clumpiness increases from left ($M_{{\rm s}}=1$) to
right ($M_{{\rm s}}=20$) panels. Variation of the UV absorption bump
strength did not significantly alter the best-fit slope $\delta$;
therefore, the $\delta$ vs $E(B-V)$ relations with a varying UV
bump strength in the MW-WD models were virtually the same as shown
in Figure \ref{fig24}. In Figure \ref{fig24}, Equation (\ref{eq:salmon})
is best matched by the MW-WD models with $M_{{\rm s}}\sim2$ and $R_{{\rm s}}/R_{{\rm d}}\sim0.8$,
LMC-WD models with $M_{{\rm s}}\sim20$ and $R_{{\rm s}}/R_{{\rm d}}\sim0.8$,
and/or SMC-WD models with $M_{{\rm s}}\sim6$ and $R_{{\rm s}}/R_{{\rm d}}\sim0.6$.
In our models, the correlation between $\delta$ and $E(B-V)$ is
simply attributed to their dependence on the optical depth $\tau_{V}$;
attenuation curves become shallower (larger $\delta$) and output
SEDs become redder as the optical depth increases. The left panels
of Figures 6 and 14 in \citet{2015arXiv151205396S} show that relatively
massive galaxies have shallow attenuation curves (denoted by ``starburst-dust''
in the figures) while less massive galaxies have steeper attenuation
curves (``SMC92-dust'') if a data point with stellar mass of $\sim10^{11.5}M_{\odot}$
is ignored. This is consistent with our results that the slope $\delta$
increases with increase of $\tau_{V}$.

\begin{figure*}[t]
\begin{centering}
\medskip{}
\par\end{centering}
\begin{centering}
\includegraphics[clip,scale=0.5]{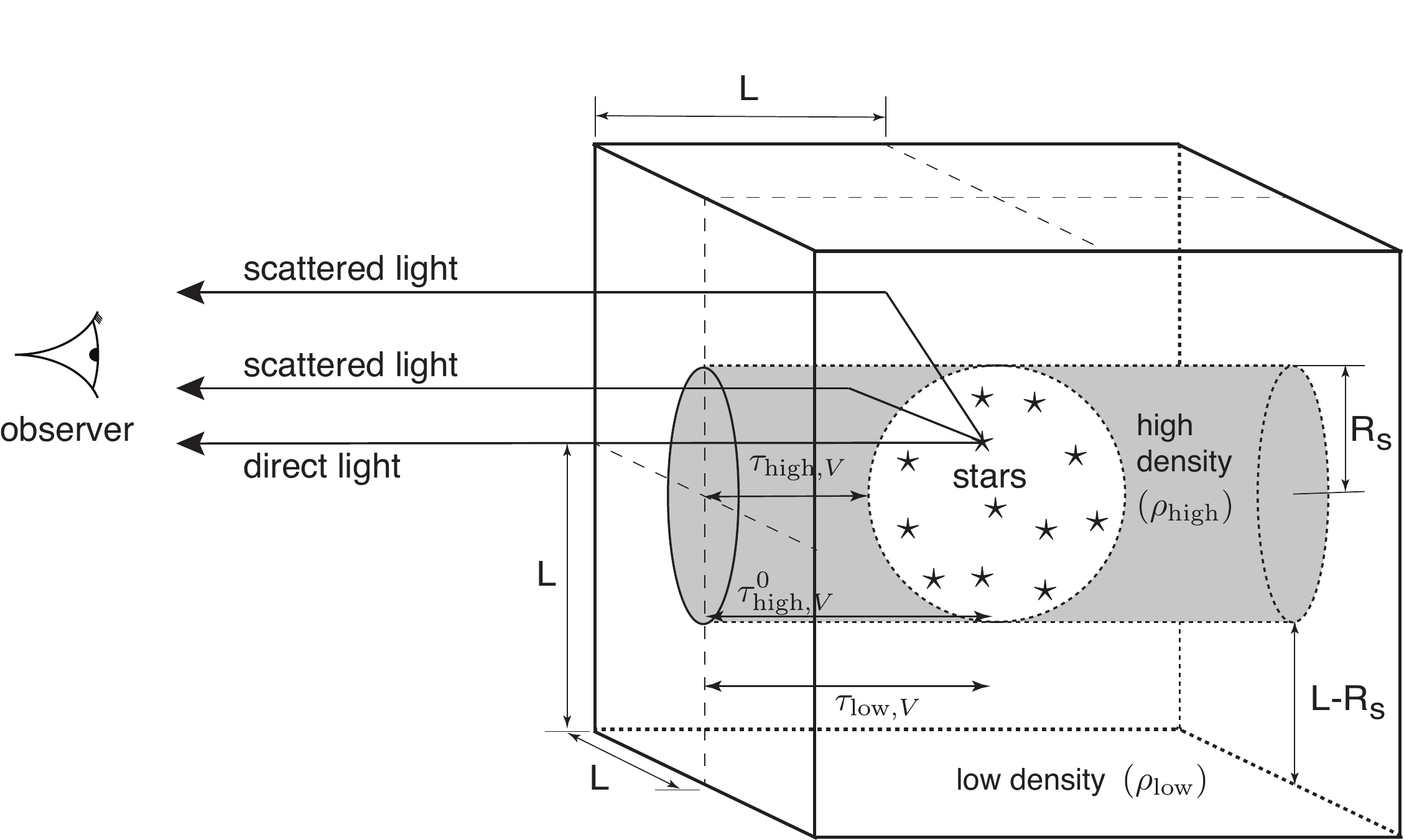}
\par\end{centering}
\medskip{}

\caption{\label{fig26}Scattering-dominant geometry. Direct starlight from
a spherical region with a radius of $R_{{\rm s}}$ is attenuated by
a dense cylinder with the same radius $R_{{\rm s}}$. Starlight is
scattered not only by the high-density cylindrical region but also
by a low-density region outside of the high-density region and measured
by an observer. The optical depths of the low- and high-density media
are defined by $\tau_{{\rm low},V}=\kappa_{V}\rho_{{\rm low}}L$ and
$\tau_{{\rm high},V}=\kappa_{V}\rho_{{\rm high}}(L-R_{{\rm s}})$,
respectively, where $\rho_{{\rm low}}$ and $\rho_{{\rm high}}$ are
densities of the media and $\kappa_{V}$ is extinction coefficient
at V-band. An additional variable $\tau_{{\rm high},V}^{0}$ is also
defined by $\kappa_{V}\rho_{{\rm low}}L$ to represent the optical
depth in the high-density medium at the limit of $R_{{\rm s}}=0$.
There is no dust in the source region.}
\medskip{}
\end{figure*}

Figure \ref{fig24} also shows large variations in the relation between
$\delta$ and $E(B-V)$ as a result of variation of Mach number ($M_{{\rm s}}$)
and/or the stellar distribution size ($R_{{\rm s}}/R_{{\rm d}}$).
In general, $\delta$ becomes larger (shallower) and $E(B-V)$ becomes
smaller as $M_{{\rm s}}$ and/or $R_{{\rm s}}/R_{{\rm d}}$ increase.
The properties were already noted in Sections \ref{subsec:3.1} and
\ref{subsec:3.2}. Therefore, large variations in the $\delta$ vs
$E(B-V)$ diagram shown in Figures 8 and 10 of \citet{2015arXiv151205396S}
may be attributable to diversity of $M_{{\rm s}}$ and $R_{{\rm s}}/R_{{\rm d}}$
rather than to variation in the intrinsic dust properties.

Figure \ref{fig25} shows the $\delta$ vs $A_{V}$ relation, which
is also convenient in interpreting observational results. In the figure,
we also compare with the relation derived from Equation (\ref{eq:salmon})
for $R_{V}=A_{V}/E(B-V)=4.05\pm0.80$ as in the Calzetti curve. The
$\delta$ vs $A_{V}$ relation leads to conclusions similar to those
described for Figure \ref{fig24}, except that the LMC-WD dust type
(with $M_{{\rm s}}\sim10$ and $R_{{\rm s}}/R_{{\rm d}}\sim0.8$)
now best matches the relation inferred from Equation (\ref{eq:salmon}).
However, we note that intrinsic dust properties are not well constrained
only with the observational data of $\delta$ vs $E(B-V)$ or $\delta$
vs $A_{V}$. Both the MW-WD and SMC-WD dust types can explain the
observational data of \citet{2015arXiv151205396S} equally well, but
the attenuation curves derived with the SMC-WD dust type are not consistent
with the overall shape of the Calzetti curve as shown in Figure \ref{fig10}.
The MW-WD model with a UV bump strength reduced by a factor of $\sim0.3-0.4$
is preferred to explain the observational attenuation curves of star
forming galaxies, as described in Sections \ref{subsec:3.2} and \ref{subsec:3.6}.

\section{DISCUSSION}

\label{sec:4}

We showed that the primary determinant of the attenuation curve is
not the underlying extinction curve but the absorption curve. \citet{2006MNRAS.370..380I}
studied the UV color variation by varying the wavelength dependence
of the scattering albedo while keeping the MW extinction curve, finding
a strong effect of the adopted albedo on the UV color. This result
implies that the attenuation curve in galaxies would strongly depend
on the underlying albedo curve. This accords well with our result
in that the variation of albedo is in fact equivalent to a variation
of absorption. We also note that no scattering effect was considered
in the turbulent media foreground screen models of \citet{2003ApJ...599L..21F}
and \citet{2005ApJ...619..340F,2011A&A...533A.117F}. \citet{2003ApJ...599L..21F}
state that it is somewhat surprising that their attenuation model
works well even without the inclusion of scattered light. In fact,
this is not surprising because absorption is the most important factor
in producing attenuation curves.

As noted in \citet{2012A&A...545A.141B}, radiative transfer models,
including ours and those of \citet{2000ApJ...528..799W}, indicate
that SED fitting codes must allow for variations of the attenuation
curve to fit the UV continuum of galaxies. Our radiative transfer
results for a wide range of configurations considered in the present
paper indicate that the modified Calzetti curve (Equation \ref{eq:modified})
is versatile enough to reproduce theoretical attenuation curves as
well as to model galactic SEDs.

In the following, we further discuss three topics that are relevant
to our results.

\subsection{Correlations with Galaxy Properties}

\label{subsec:4.1}

It appears that the UV bump feature is present in typical star-forming
galaxies with generally weaker strengths than in the MW extinction
curve. The UV strength seems to be associated with galaxy activity.
\citet{2009A&A...499...69N} found that galaxies with evidence of
a UV bump feature host older stellar populations than galaxies lacking
an evident bump. It was also shown that more active galaxies with
higher specific star formation rates (SFRs) tend to have weaker UV
bumps \citep{2011MNRAS.417.1760W,2012A&A...545A.141B,2013ApJ...775L..16K}.

This trend can be explained by one of the following scenarios or by
their combinations: (1) variations in the grain size distribution,
(2) age-dependent extinction, and (3) radiative transfer effects.
First of all, variations in the grain size distribution can cause
the trends. The balance between grain formation, growth, and destruction
would be altered by star formation activity; small dust grains can
be easily destroyed in active galaxies \citep[e.g., ][]{2003ApJ...594..279G}.
Lack of small dust grains or PAHs in active galaxies would yield a
weak UV bump as well as a shallower attenuation curve. This is in
qualitative agreement with the observational findings. On the other
hand, an increased rate of grain shattering in grain-grain collisions
could conceivably augment the population of very small grains in active
galaxies.

Second, the observed trend might also be ascribed to the age-dependent
extinction model, in which old stars are attenuated only by the diffuse
ISM, while younger stars experience a higher attenuation due to molecular
clouds in addition to the diffuse ISM \citep{2000ApJ...542..710G,2007MNRAS.375..640P}.
In the models of \citet{2000ApJ...542..710G} and \citet{2007MNRAS.375..640P},
the attenuation curves for young stars in dense molecular clouds showed
very weak UV bumps. However, we found no significant change in attenuation
curves when stars are located only within high-density regions, as
described in Section \ref{subsec:3.5}. In the age-dependent extinction
model (set B) of \citet{2007MNRAS.375..640P}, the optical depth of
molecular clouds was assumed to be $\tau_{{\rm MC}}=20$ at 1 $\mu$m,
corresponding to $\tau_{{\rm MC}}\thickapprox50$ at V-band (for the
MW dust). As noted in Section \ref{subsec:3.5} and \citet{2000ApJ...528..799W},
radiative transfer models with such high optical depths can produce
attenuation curves with virtually no UV bump. However, the attenuation
curves will be much shallower than the Calzetti curve. It is not clear
that the slope of the attenuation curves derived in \citet{2000ApJ...542..710G}
and \citet{2007MNRAS.375..640P} are consistent with that of the Calzetti
curve. Moreover, as noted in \citet{2013MNRAS.432.2061C}, \citet{2000ApJ...542..710G}
and \citet{2007MNRAS.375..640P} assume isotropic scattering in solving
the radiative equation for a spherically symmetric molecular cloud
of uniform density \citep[see][]{1994MNRAS.268..235G}. The approximation
of isotropic scattering produces even shallower attenuation curves
with a much weaker UV bump at high optical depths as considered in
their models.

Finally, we note that radiative transfer effects can also explain
the trend. The ISM in more active star-forming galaxies would likely
be more clumped, leading to shallower attenuation curves with weaker
UV bumps. The observational data in Figure 2 of \citet{2013ApJ...775L..16K}
show evidence of a correlation between the slope ($\delta$) and $A_{V}$.
This is also an understandable consequence of radiative transfer effects,
as shown in Figure \ref{fig22}. In this interpretation, weaker UV
bump strengths and shallower attenuation curves in more active galaxies
do not necessarily indicate more effective destruction of the UV bump
carriers, but could instead be explained by having a more clumpy ISM
and larger dust optical depths in these systems. However, the abundance
of the UV bump carriers in the galaxy samples of \citet{2013ApJ...775L..16K}
should be $\sim$ 30\% or 40\% of that of the MW-WD dust. The reduced
strength of the UV bump should be useful in constraining the size
distribution of dust grains and the dust chemical composition.

\begin{figure}[t]
\begin{centering}
\medskip{}
\includegraphics[clip,scale=0.57]{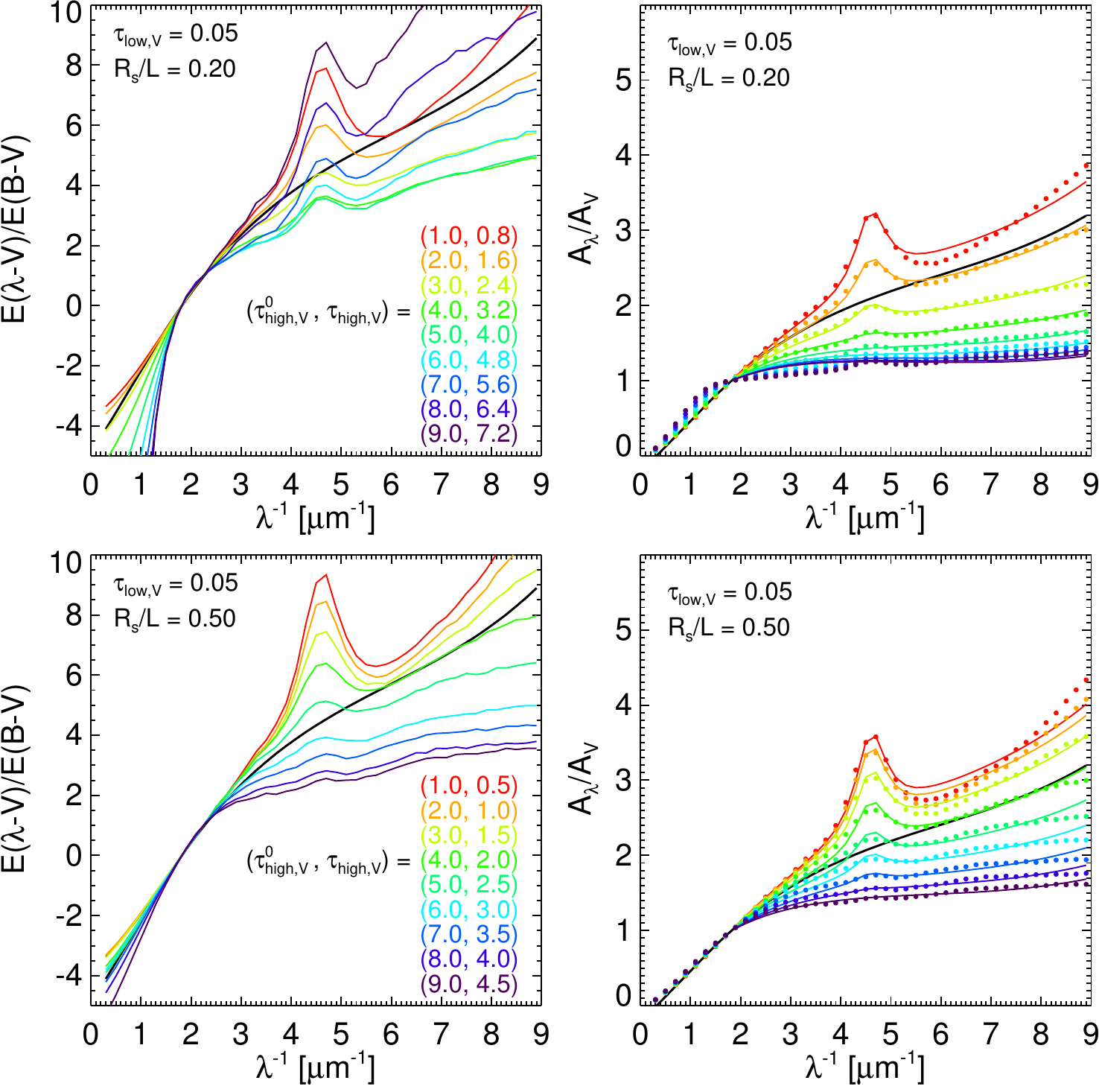}
\par\end{centering}
\caption{\label{fig27}Attenuation curves calculated for the scattering-dominant
geometry shown in Figure \ref{fig26}. The optical depth of the high-density
region ($\tau_{{\rm high},V}$) was varied such that $\tau_{{\rm high},V}^{0}(=\kappa_{V}\rho_{{\rm high}}L)=1,2,\cdots,9$
while the optical depth of the low-density region ($\tau_{{\rm low},V}$)
was fixed to 0.05. Both $\tau_{{\rm high},V}^{0}$ and $\tau_{{\rm high,V}}$
are shown in parentheses for convenience. The top and bottom panels
show the attenuation curves for $R_{{\rm s}}/L$ = 0.2 and 0.5, respectively.
The left and right panels show the results in two forms, e.g., $E(\lambda-V)/E(B-V)$
and $A_{\lambda}/A_{V}$, respectively. In the right panels, the attenuation
curves are shown as filled circles and the best-fitting modified Calzetti
curves are overplotted as lines with the same colors as the circles.
The MW-WD dust is assumed for the model calculations. The black curves
are the Calzetti Curves.}
\medskip{}
\medskip{}
\end{figure}

\begin{figure}[t]
\begin{centering}
\medskip{}
\includegraphics[clip,scale=0.57]{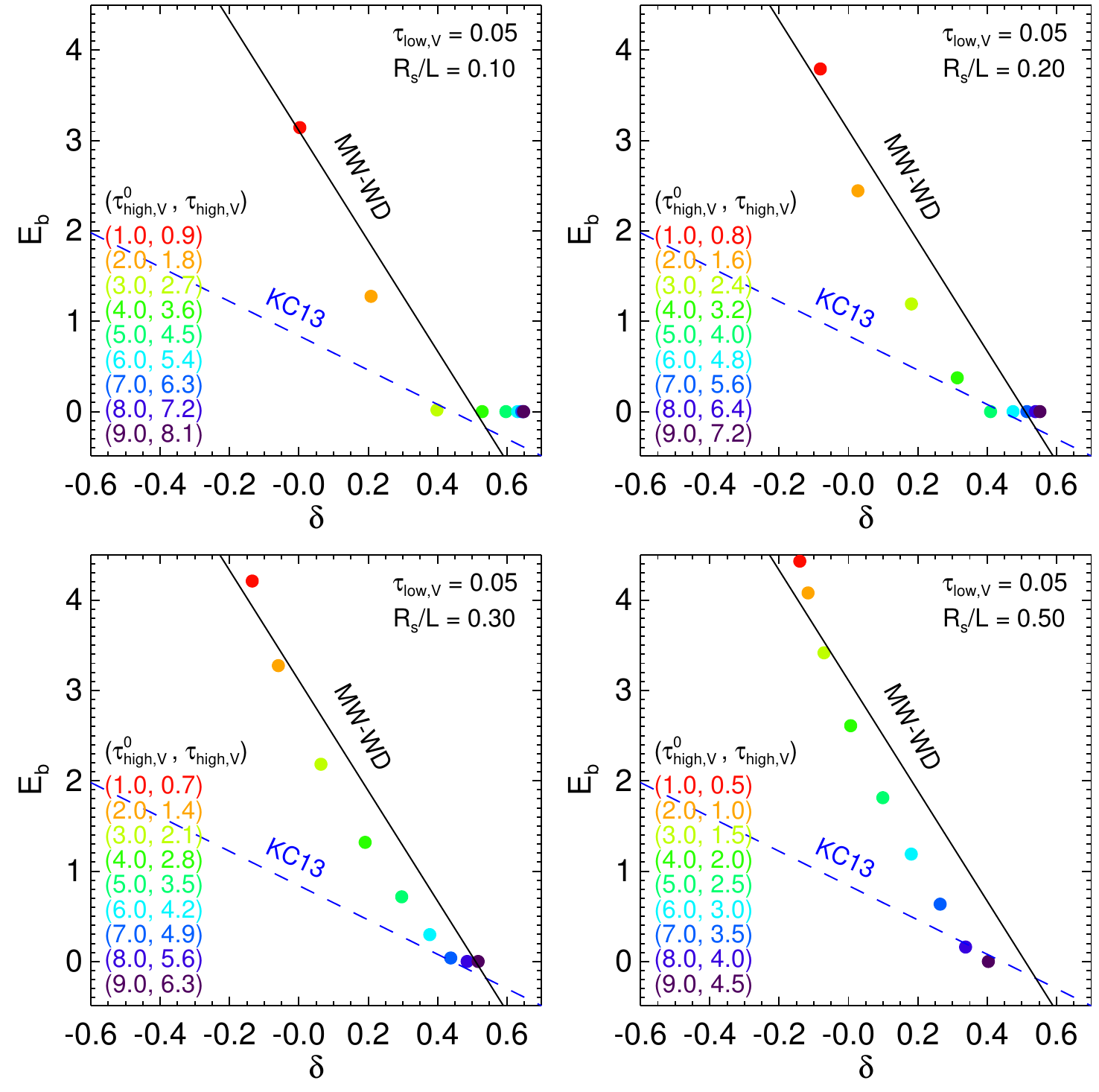}
\par\end{centering}
\caption{\label{fig28}Best-fit results to the modified Calzetti curves for
the attenuation curves calculated in the scattering-dominant geometry
(Figures \ref{fig26} and \ref{fig27}). The best-fit bump strength
($E_{b}$) and slope ($\delta$) for the models with four radii of
the blocking cylinder ($R_{{\rm s}}/L$ = 0.1, 0.2, 0.3, and 0.5)
are compared. As in Figure \ref{fig27}, $\tau_{{\rm high},V}^{0}$
was varied from 1 to 9 in steps of 1 and $\tau_{{\rm low},V}$ was
fixed to 0.05. The blue dashed line labeled KC13 is the empirical
linear relationship found by \citet{2013ApJ...775L..16K}. The black
line labeled MW-WD is the best-fit linear function for the MW-WD dust
found in Figure \ref{fig21}.}
\medskip{}
\medskip{}
\end{figure}

Dependence of the attenuation curves on galaxy inclination angle has
been observed. Highly inclined galaxies or edge-on galaxies tend to
have stronger UV bumps \citep{2011MNRAS.417.1760W,2013ApJ...775L..16K}.
However, there seem to be conflicting results with regard to the shape
of attenuation curves. \citet{2011MNRAS.417.1760W} found grayer or
shallower attenuation curves for highly inclined galaxies. In radiative
transfer calculations for a model late-type galaxy, \citet{2004ApJ...617.1022P}
found shallower curves with weaker UV bumps for highly inclined galaxies.
On the contrary, \citet{2013ApJ...775L..16K} observed a weak tendency
of steeper attenuation curves for edge-on galaxies. \citet{2015arXiv151205396S}
found no correlation between the attenuation curve shape and the axis
ratios (inclinations) of galaxies.

We can expect a dependence of attenuation curves on inclination angle
from our results, even though the present study did not consider detailed
geometries of disk galaxies. The vertical scaleheight of dust in disk
galaxies is comparable to that of young stars and smaller than that
of old stars, but the dust radial scalelength parallel to the major
axis is larger than those of stars \citep[e.g., ][]{1997A&A...325..135X,2007A&A...471..765B,2014ApJ...785L..18S}.
Therefore, in edge-on galaxies, stars, especially young stars responsible
for UV light, will be hidden behind the radially extended dust layer.
This configuration is similar to a geometry with $R_{{\rm s}}/R_{{\rm d}}<1$
in our model. On the other hand, the star/dust geometry for face-on
galaxies will be close to the case with $R_{{\rm s}}/R_{{\rm d}}\ge1$.
The geometry with $R_{{\rm s}}/R_{{\rm d}}=1$ for face-on galaxies
would give rise to slightly shallower attenuation curves with weaker
UV bumps than that with $R_{{\rm s}}/R_{{\rm d}}<1$ for edge-on galaxies
if the optical depths viewed face-on and edge-on were the same. However,
the increase of the optical depth at high inclination angles would
yield much stronger (opposite) effects on the attenuation curves than
due to the difference in the star/dust geometry, as shown, for instance,
in Figure \ref{fig22}. In the right panels of Figure \ref{fig22},
variation of $E_{b}$ and $\delta$ due to change in $R_{{\rm s}}/R_{{\rm d}}$
(represented by symbol size) is relatively small compared to that
due to the $\tau_{V}$ change (represented by color). The same trend
was also noted in Section \ref{subsec:3.2}. Consequently, edge-on
galaxies would have shallower attenuation curves with weaker UV bumps.
This expectation is consistent with the conclusion obtained with radiative
transfer calculations for a disk galaxy in \citet{2004ApJ...617.1022P},
but seems not to accord with the observational results of \citet{2011MNRAS.417.1760W}
and \citet{2013ApJ...775L..16K}. Note that our models and those of
\citet{2004ApJ...617.1022P} are complementary in the sense that we
assumed a spherical, clumpy ISM, but \citet{2004ApJ...617.1022P}
assumed a smooth disk + bulge geometry. For the same reason, both
models have limitations in explaining the inclination angle dependence
of attenuation curve. We, therefore, need to investigate more realistic
models incorporating both the clumpiness of ISM and the disk + bulge
geometry of galaxies to better understand the observed dependence
of the attenuation curve on inclination angle. More systematic observational
studies on the inclination angle dependence of attenuation curve are
also required.

\begin{figure*}[t]
\begin{centering}
\medskip{}
\includegraphics[clip,scale=0.8]{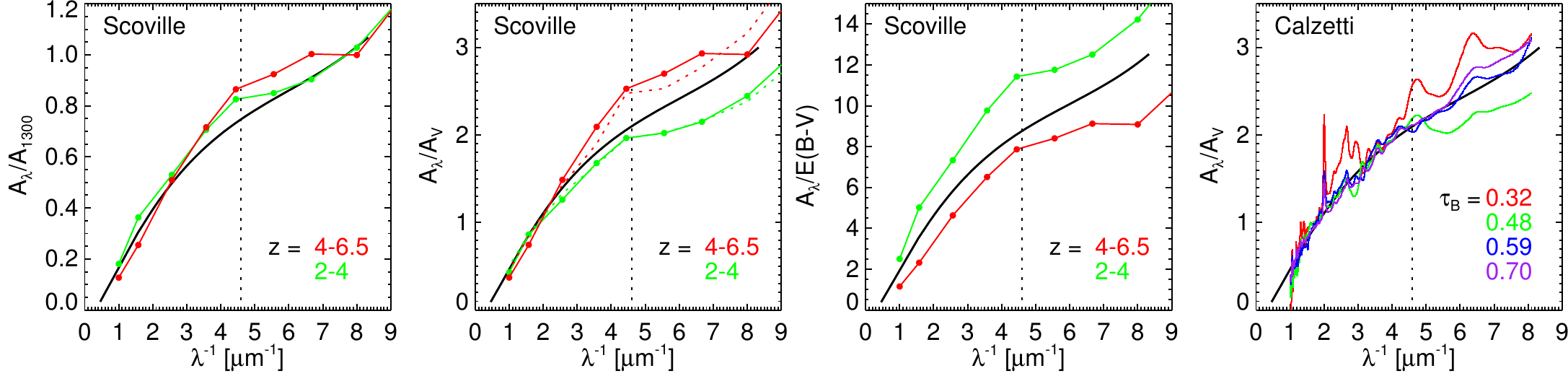}
\par\end{centering}
\caption{\label{fig29}Comparison of the two attenuation curves of high-redshift
galaxies in \citet{2015ApJ...800..108S} and those derived from the
four template spectra of local starburst galaxies in \citet{1994ApJ...429..582C}.
The first to third panels show the Scoville et al. attenuation curves
normalized to $A_{1300}$, $A_{V}$, and $E(B-V)$, respectively.
The last panel shows the attenuation curves derived from the template
spectra of local starburst galaxies, after being smoothed with a Gaussian
function. The features in the last panel, except the 2175\AA\ bump,
are atomic lines. Dashed lines in the second panel are for the modified
Calzetti law (Equation (\ref{eq:modified})). In each panel, the black
line is the Calzetti curve.}
\medskip{}
\end{figure*}

\subsection{Scattering-Dominant Geometry}

\label{subsec:4.2}

A geometrical configuration in which single-scattered light dominates
direct starlight or scattered light is selectively observed could
produce a UV bump-less attenuation curve. One example of the scattering-dominant
geometry can be found in halos of edge-on galaxies \citep[e.g., ][]{2014ApJ...785L..18S,2015ApJ...815..133S}.
Figure \ref{fig26} illustrates a simplified geometry to represent
a scattering-dominant configuration, in which the direct starlight
from a sphere with a radius $R_{s}$ is attenuated by a cylindrical
dense medium with the same radius and the scattered light from a low-density
medium outside the cylinder is substantial. The spherical volume containing
stars is carved out to contain no dust and starlight is not absorbed
within the source region. Height of the cylindrical medium is defined
by $2L$, as shown in Figure \ref{fig26}. Optical depths of the low-
and high-density medium at V-band are defined by $\tau_{{\rm high},V}\equiv\kappa_{V}\rho_{{\rm {\rm high}}}(L-R_{{\rm s}})$
and $\tau_{{\rm low},V}\equiv\kappa_{V}\rho_{{\rm low}}L$, respectively.
We also define an additional variable $\tau_{{\rm high},V}^{0}\equiv\kappa_{V}\rho_{{\rm high}}L=\tau_{{\rm high},V}(R_{{\rm s}}=0)$,
which is more appropriate to represent the dust density of the high-density
medium. Here, $\rho_{{\rm high}}$ and $\rho_{{\rm low}}$ are the
densities of high- and low-density medium, respectively, and $\kappa_{V}$
is extinction opacity at V-band. Monte Carlo radiative transfer modeling
has been done for this geometry by assuming the MW-WD dust model.
Figure \ref{fig27} shows the resulting attenuation curves obtained
by fixing $\tau_{{\rm low},V}$ to be 0.05 and varying $\tau_{{\rm high},V}^{0}$
from 1 to 9 in steps of 1. The top and bottom panels show the results
for $R_{{\rm s}}/L$ = 0.2 and 0.5, respectively. The best-fit modified
Calzetti curves for the attenuation curves are also shown in the right
panels.

In Figure \ref{fig27}, both the $E(\lambda-V)/E(B-V)$ and $A_{\lambda}/A_{V}$
curves are close to the overall shape of the Calzetti curve or the
modified Calzetti curve. Note that these attenuation curves for the
scattering dominant geometry were calculated by measuring the flux
at a single (edge-on) direction along the cylinder axis, while all
the other attenuation curves presented in this paper were obtained
by averaging over all directions. Therefore, the attenuation optical
depths in Figures \ref{fig27} and \ref{fig28} were calculated as
$\tau_{{\rm \lambda}}^{{\rm att}}=-\ln(4\pi\tilde{f}_{\lambda}^{{\rm esc}})$,
where $\tilde{f}_{\lambda}^{{\rm esc}}$ is the output SED per unit
solid angle divided by source SED. As $\tau_{{\rm high},V}$ increases,
starlight from the high-density region is more blocked and the contribution
of singly-scattered light from the low-density region increases. Because
of no bump feature in the scattering cross-section and very low optical
depth of the low-density scattering medium, the scattering-dominant
SEDs lack a UV bump feature at highest $\tau_{{\rm high},V}$. However,
the UV absorption bump feature appears unless $\tau_{{\rm low},V}\ll1$.
As the size of source distribution $(R_{{\rm s}})$ decreases, starlight
is more attenuated and shallower attenuation curves with weaker UV
bump strengths are obtained. Therefore, the scattering-dominant geometry
would reduce at least partly the UV bump strength.

Figure \ref{fig28} shows the relations between $E_{b}$ and $\delta$
obtained by fitting the attenuation curves for $R_{{\rm s}}/L=$ 0.1,
0.2, 0.3, and 0.5 to the modified Calzetti curves, together with the
empirical relationship found by \citet{2013ApJ...775L..16K} and the
best-fit linear function for the MW-WD dust models found in Figure
\ref{fig21}. The same densities as in Figure \ref{fig27} were adopted.
As optical depth $\tau_{{\rm high},V}$ increases, $E_{b}$ and $\delta$
start to deviate from the relation derived from the MW-WD dust models
and become closer to the empirical equation found by \citet{2013ApJ...775L..16K}.
It might be possible to further decrease the UV bump strength by adopting
the turbulent ISM density distribution for the high-density cylindrical
medium.

\citet{2014ApJ...785L..18S} and \citet{2015ApJ...815..133S} modeled
the scattered FUV light observed above the galactic disk of edge-on
galaxies by assuming the MW-WD dust properties. On the other hand,
\citet{2014ApJ...789..131H} claimed that the the UV colors measured
in halos of edge-on galaxies are consistent with the SMC-type dust,
lacking a UV bump. However, the SED shape of the scattered flux is
mainly determined by the wavelength dependence of the scattering cross-section,
which has no UV bump feature \citep{1995ApJ...446L..97C,1977A&A....54..841A}.
The geometry in Figure \ref{fig26} is not for edge-on galaxies, but
appropriate to demonstrate a case in which substantial scattered starlight
is observed. Therefore, the results obtained from the geometry illustrate
that edge-on galaxies with only the scattered UV SED data may not
be useful for determination of dust type in the halos without a detailed
radiative transfer model.

\begin{figure}[t]
\begin{centering}
\medskip{}
\includegraphics[clip,scale=0.9]{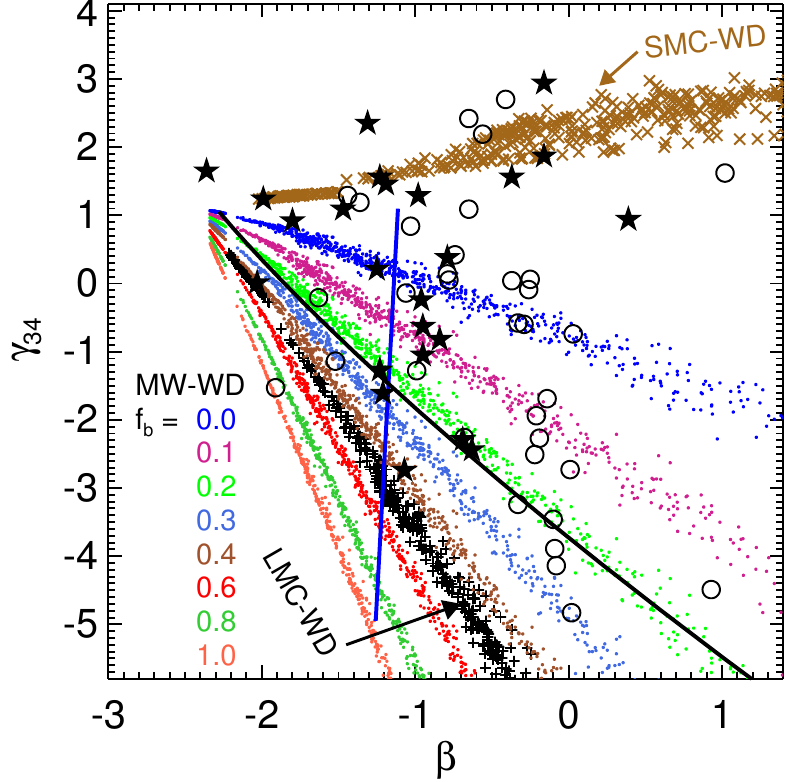}
\par\end{centering}
\caption{\label{fig30}Relation between the slope parameters $\gamma_{34}$
and $\beta$. The UV bump strength ($f_{b}$) for the MW-WD dust type
was varied from 0 to 1.0. The results for the SMC-WD and LMC-WD dust
types are also shown in brown cross ($\times$) and black plus ($+$)
symbols, respectively. The blue vertical line denotes the result calculated
for the Calzetti curve + UV bump with a varying strength. The black
line shows the result for the modified Calzetti curve, adopting the
$E_{b}-\delta$ relation of \citet{2013ApJ...775L..16K}. In both
cases, the bump strength $E_{b}$ increases from 0 ($\gamma_{34}\approx1$)
to 2 ($\gamma_{34}\approx-5$). The IUE data for 24 local starburst
galaxies and the Focal Reducer and Spectrograph Deep Field (FDF) data
for the galaxies at $z\sim2$ were taken from \citet{2005A&A...444..137N}
and are shown as stars and circles, respectively.}
\medskip{}
\end{figure}

\subsection{UV bumps in local starbursts}

\label{subsec:4.3}

It is generally believed that the UV bump feature is very weak or
absent in local starburst galaxies because of lack of the feature
in the Calzetti curve. The presence of a UV bump in typical star-forming
galaxies has been regarded as a significant difference from the starburst
galaxies. However, we will now argue that the local starburst galaxies
may not be so different from more normal star-forming galaxies.

\citet{2015ApJ...800..108S} carefully selected high-redshift galaxies
which are supposed to resemble local starburst galaxies. They presented
average attenuation curves in two redshift bins ($z=2-4$ and $4-6.5$)
normalized to attenuation at $\lambda=0.13$ $\mu$m. The attenuation
curves for the two redshift bins are reproduced in the first panel
of Figure \ref{fig29}. In order to compare their results with the
Calzetti curve and our results, we renormalized the attenuation curves
to the values at $V$-band. The renormalized $A_{\lambda}/A_{V}$
curves are shown in the second panel of Figure \ref{fig29}. From
the figure, it is now clear that the curve resembles the attenuation
curves shown in Figures \ref{fig8} and \ref{fig17}. The red ($z=4-6.5$)
and green ($z=2-4$) curves are quite similar to the attenuation curves
calculated for low and high optical depths, respectively, in the figures.
To derive the UV bump strength, the modified Calzetti curve (Equation
\ref{eq:modified}) was applied to the data. The lower redshift data
was well fitted by the parameters of $(\delta$, $E_{b}$) = (0.13,
0.73), which is consistent with the relation between ($\delta$, $E_{b}$)
given by \citet{2013ApJ...775L..16K}. The higher redshift data was
also found to be comparable with a modified Calzetti curve with ($\delta$,
$E_{b}$) $\sim$ ($-0.06$, 1.4). This implies that the Scoville
et al. attenuation curves are consistent with those of \citet{2013ApJ...775L..16K}.
The two modified Calzetti curves that represent the Scoville et al.
attenuation curves are shown as dotted lines in the second panel.
The Scoville et al. curves are plotted in the $A_{\lambda}/E(B-V)$
form in the third panel. The lower-redshift curve has a lower $E(B-V)$
color excess than the higher-redshift curve, which led to higher $A_{\lambda}/E(B-V)$
in the lower-redshift curve. The property that the order of $A_{\lambda}/A_{V}$
curves is reversed in the $A_{\lambda}/E(B-V)$ curve was found in
Section \ref{subsec:3.2}.

To derive an average attenuation curve for the \emph{IUE} sample of
local starburst galaxies, \citet{1994ApJ...429..582C} sorted the
sample galaxies according to Balmer decrement bins, averaged the galactic
spectra $F_{\nu}$ in each bin, and then divided the spectra by the
spectrum showing the lowest extinction. Then, the log ratios were
divided by the color excess estimated from the Balmer decrement. The
resulting reddening curves were used to derive a final attenuation
curve. The template spectra for each Balmer decrement bin\footnote{\href{http://www.stsci.edu/hst/observatory/crds/cdbs_kc96.html}{http://www.stsci.edu/hst/observatory/crds/cdbs\_kc96.html}}
were downloaded to compare with the present results. The four template
spectra were smoothed with a Gaussian function and are shown in the
fourth panel of Figure \ref{fig29}, in which spiky and/or broad features
except the one at $\lambda^{-1}\sim4.6$ $\mu$m$^{-1}$ are strong
atomic absorption lines, such as H$\alpha$ 6564\AA\ and \ion{C}{4}
$\lambda\lambda$1551\AA. We note that the two curves denoted by
red and green colors are fairly similar to the Scoville et al. curves
with same colors in the second panel, suggesting that the two template
curves have a UV bump strength similar to that of the Scoville et
al. curves, although the other template curves show no evidence for
a UV absorption bump. Thereby, we can conclude that at least two of
the four template spectra of the local starburst galaxies have (weak)
UV bumps comparable to other star-forming galaxies, even though there
is no UV bump in the \citet{1994ApJ...429..582C} overall curve.

UV continuum slopes are also useful in testing for the presence of
UV bumps in spectra of local starburst galaxies. \citet{2005A&A...444..137N}
divided the UV wavelength range $1200-2500$\AA\ into several sub-ranges
and parameterized the UV SEDs of their own sample galaxies observed
at $z\sim2$ and the \emph{IUE} spectra of the local starburst galaxies
analyzed in \citet{1994ApJ...429..582C} with a power-law ($F_{\nu}\propto\lambda^{\beta}$)
in each sub-range. They compared the observational data with the radiative
transfer models of \citet{2000ApJ...528..799W}. In particular, the
slopes $\beta$, $\gamma_{3}$, and $\gamma_{4}$ defined in wavelength
ranges of $1268-1740$, $1920-2175$, $2175-2480$\AA, respectively
(see their Table 4 for the precise definitions of the wavelength windows),
were used to identify the presence of the UV bump feature. A derived
parameter $\gamma_{34}\equiv\gamma_{3}-\gamma_{4}$ measures the curvature
of the UV continuum across the UV bump and is very sensitive to the
presence of the UV bump feature.

Figure \ref{fig30} shows the plot of $\gamma_{34}$ versus $\beta$
estimated from our radiative transfer calculations for the MW-WD dust
with varying strength of the UV bump ($f_{b}$, defined relative to
the MW extinction curve). In the figure, change in the UV bump strength
is denoted by different colors. The models with varying strength of
the whole PAH feature (including both the UV bump and FUV extinction
rise) showed no significant difference from those with varying only
the UV bump strength and thus are not displayed in this paper. This
is because the slope parameters are less sensitive to variation of
the FUV extinction rise component. The results for the LMC-WD and
SMC-WD dust types are also overplotted with plus ($+$) and cross
($\times$) symbols, respectively. The loci of the Calzetti curve
with a varying UV bump strength are denoted by the near-vertical blue
line. The result for the modified Calzetti curve is also shown by
the black line adopting the relation between the bump strength and
the attenuation curve slope of \citet{2013ApJ...775L..16K}. Both
lines show variation as the bump strength ($E_{b}$) increases from
0 (high $\gamma_{34}$) to 2 (low $\gamma_{34}$). Note that the result
obtained for the modified Calzetti curve is consistent with that for
MW-WD dust with bump strength suppressed by $f_{b}\approx0.2-0.3$,
consistent with the results in Figure \ref{fig22}. In calculating
$\gamma_{34}$ and $\beta$ values for the models, we assumed the
stellar continuum slope parameters to be $\gamma_{34}=1.29$ and $\beta=-2.44$
estimated from the stellar population synthesis model of \citet{2005MNRAS.362..799M}
with solar metallicity, standard Salpeter initial mass function, and
continuous star formation during an age of 500 Myr. The \emph{IUE}
sample of local starburst galaxies and the samples of $z\sim2$ galaxies
are denoted by triangles and circles, respectively. The loci occupied
by the MW-WD dust type ($f_{b}=1$) are slightly shifted to higher
$\beta$ values than those calculated with the MW-WG dust in \citet{2005A&A...444..137N}.
For the SMC-WD dust type, $\gamma_{34}$ increases with $\beta$.
However, for the SMC-WG dust, $\gamma_{34}$ remains somewhat constant
($\sim1$) as $\beta$ increases. It is clear that about half of the
\emph{IUE} sample galaxies can be described by the MW-WD dust models
with weak UV bumps, rather than by the SMC-WD dust. This strengthens
the previous conclusion that at least some of the \emph{IUE} starburst
galaxies have a UV bump, though weaker than that of the MW-WD dust.
The mean slopes of $\left\langle \beta\right\rangle =-1.05$ and $\left\langle \gamma_{34}\right\rangle =0.27$
are closer to the values obtained from a UV bump-less model of the
MW-WD dust than with the SMC-WD dust model. However, the slope parameter
$\gamma_{34}$ ($\sim1$) for the Calzetti curve with no UV bump is
more consistent with those of the models obtained from the SMC-WD
dust. Note also that no significant difference is found between the
distributions of the slopes for the local starburst galaxies and the
galaxies at $z\sim2$, implying the similarity between the local starburst
galaxies and the high-redshift galaxies at $z\sim2$. There are minor
changes in $\beta$ among the various synthesis models considered
in \citet{2005A&A...444..137N}, but this does not alter our conclusion.

In Figure \ref{fig30}, the slope parameters for the LMC-WD dust are
similar to those of the MW-WD dust models with a bump strength of
$\sim0.5$. This suggests that the overall distribution of the slope
parameters for the galaxies in the figure may be equivalently reproduced
by the LMC-WD dust if its UV bump strength was varied. However, the
overall shape of the $A_{\lambda}/A_{V}$ attenuation curves derived
from the LMC-WD dust are in general steeper than the Calzetti curve,
as shown in Figure \ref{fig9}. As noted in Section \ref{subsec:3.2},
the shape of the Calzetti curve was better consistent with those of
the MW-WD attenuation curves over a wider range of model parameters
than those of the LMC-WD dust. Therefore, most of the calculations
in the present study were based on the MW-WD dust extinction curve.

Using the \emph{IUE} and \emph{IRAS} data for local starburst galaxies,
\citet{1999ApJ...521...64M} proposed a formula to estimate dust attenuation
from a UV spectral slope $\beta$ (this is different from $\beta$
in \citealt{2005A&A...444..137N}), based on the tight relation between
the IR to UV flux ratio (referred to as the infrared excess, IRX)
and $\beta$. However, it has been observed that star-forming galaxy
samples selected at optical and UV wavelengths are located significantly
below the IRX-$\beta$ relation of \citet{1999ApJ...521...64M} \citep[e.g., ][]{2005ApJ...619L..51B,2007ApJS..173..524B}.
\citet{2012ApJ...755..144T} found that the UV fluxes in \citet{1999ApJ...521...64M}
were significantly underestimated because of the small aperture of
the \emph{IUE} ($10''\times20''$) and consequently, their IRX values
were overestimated. The aperture size effect led to the IRX-$\beta$
relation being shifted upward on the plot. \citet{2012ApJ...755..144T}
calibrated the UV fluxes for the local starburst galaxies with the
\emph{GALEX} data and derived a revised IRX-$\beta$ relation, which
is shifted downward compared to that of \citet{1999ApJ...521...64M}.
The new relation was then found to be consistent with the star-forming
galaxy samples. This seems to suggest that the attenuation properties
of local starburst galaxies are in fact similar to those of other
star-forming galaxies. However, this does not directly indicate the
existence of a UV bump in local starburst galaxies. In a future study,
it may be worthwhile to apply the modified Calzetti curve to the \emph{IUE}
spectral data of local starburst galaxies together with the recent
photometric data of \emph{GALEX} and the \emph{Swift} UVOT to better
understand the attenuation curves in local starburst galaxies.

\section{SUMMARY}

\label{sec:5}

The dust attenuation curves of starlight in galactic environments
were investigated through Monte-Carlo radiative transfer models, which
employ dust density distributions appropriate for a clumpy ISM. The
full numerical results of the models are made available to the community
via the authors' WWW site \footnote{\href{http://kwangilseon.github.io}{http://kwangilseon.github.io/}}.
The principal conclusions of this paper are as follows: 
\begin{itemize}
\item It is confirmed that the absolute amount of dust is not well determined
by the color excess $E(B-V)$ due to the saturation of $E(B-V)$.
\item Scattering at $\lambda^{-1}\gtrsim2\ \mu{\rm m}^{-1}$ can lead to
either reddening or blueing depending on the wavelength dependence
of the albedo.
\item Theoretically derived attenuation curves show large variations both
in the curve slope and the UV bump strength, depending on the star/dust
geometry, clumping, and total amount of dust.
\item The attenuation curves become grayer (or shallower) and the strength
of the 2175\AA\ absorption bump becomes weaker as the optical depth,
the clumping, and the size of source distribution increase.
\item The shape of the attenuation curves is primarily determined by the
wavelength dependence of absorption rather than by the extinction
(absorption + scattering) curve.
\item Attenuation curves consistent with the original Calzetti curve \citep{1994ApJ...429..582C,2000ApJ...533..682C}
are obtained using the MW dust model of \citet{2001ApJ...548..296W},
provided the 2175\AA\ absorption bump is partially or totally suppressed.
\item The discrepancy between our results and previous studies that claim
SMC-type dust to be the most likely origin of the Calzetti curve is
mainly attributed to the differences in the adopted dust albedo and
the strength of the 2175\AA\ bump; this study uses the theoretically
calculated albedos and suppresses the bump strength whereas the previous
ones used empirically estimated albedos and the bump strength fixed
to that of the MW extinction curve.
\item The model attenuation curves are generally well represented by the
modified Calzetti curve (Equation \ref{eq:modified}) proposed by
\citet{2009A&A...507.1793N} to allow variation of the attenuation
curve slope and UV bump strength.
\item The correlation between the slope and UV bump strength (Equation \ref{eq:KC13}),
with steeper curves having stronger bumps, as found in a large sample
of star-forming galaxies \citep{2013ApJ...775L..16K}, is well reproduced
if the abundance of the UV bump carriers or PAHs is assumed to be
30\% or 40\% of that of the MW dust.
\item The correlation between the slope and UV bump strength found by \citet{2013ApJ...775L..16K}
is explained by radiative transfer effects which lead to shallower
attenuation curves with weaker UV bumps as the ISM is more clumpy
and has larger dust mass.
\item The correlation between the slope and the $E(B-V)$ color as found
in \citet{2015arXiv151205396S} can be explained by radiative transfer
effects; as optical depth increases, attenuation curves become shallower
and redder.
\item It is demonstrated that a scattering-dominant geometry in which direct
starlight is mostly absorbed and single-scattered light dominates
over the direct starlight can produce attenuation curves with very
weak UV bump and with shallow slope.
\item It is also shown that at least some of the local starburst galaxies
appear to have a UV bump feature in their attenuation curves, albeit
much weaker than that of the MW dust.
\end{itemize}

\acknowledgements{Numerical simulations were partially performed by using a high performance
computing cluster at the Korea Astronomy and Space Science Institute.
BTD was supported in part by NSF grant AST-1408723.}

\end{document}